\documentstyle[preprint,aps,psfig]{revtex}

\begin{document}
\draft

\title{ Protein Folding Kinetics: Time Scales, Pathways, and Energy
Landscapes in Terms of Sequence Dependent Properties 
}
\author{T. Veitshans\(^{1}\), D. K. Klimov\(^{2}\), 
D. Thirumalai\(^{2}\) } 
\address{\(^{1}\)
Laboratoire de Spectrom\'{e}trie Physique, associ\'{e} au CNRS\\
Universit\'{e} J.\ Fourier, Grenoble I; B.P.\ 87; 38402 
Saint--Martin d'H\`{e}res Cedex, France \\ 
and \\
\(^{2}\) Institute for Physical Science and Technology and
Department of Chemistry and Biochemistry\\
University of Maryland, College Park, Maryland 20742, USA}

\maketitle

\begin{flushleft}
{\small 
{\bf Background: } Recent experimental and theoretical studies have
revealed that protein folding kinetics can be quite complex and
diverse depending on various factors such as size of the protein sequence and
external conditions. For example, some proteins fold apparently in a
kinetically two state manner whereas others follow complex routes to
the native state. We have set out to provide the theoretical basis for
understanding the diverse behavior seen in the refolding kinetics of
proteins in terms of properties that are intrinsic to the sequence. 
\vspace{0.5cm}

{\bf Results:} The folding kinetics of a number of sequences for 
off-lattice continuum
models of proteins is studied using Langevin simulations at two
different values of the friction coefficient. We show for these models
that there is a remarkable correlation between folding times, \(\tau
_{F}\), and \(\sigma = (T_{\theta } - T_{F})/T_{\theta } \), where
\(T_{\theta }\) and \(T_{F}\) are the equilibrium 
collapse and folding transition
temperatures, respectively. The microscopic dynamics reveals that
several scenarios for the kinetics of refolding 
arise depending on the range of values of
\(\sigma \). For relatively small \(\sigma \) the chain reaches the
native conformation by a direct native conformation  nucleation collapse
(NCNC) mechanism without being trapped in any detectable
intermediates. For moderate and
large values of \(\sigma \) the kinetics is described by the kinetic
partitioning mechanism (KPM)  according to which a fraction of molecules \(\Phi
\) (kinetic partition factor)  reaches the native conformation via 
the NCNC mechanism. The remaining
fraction attains the native state by off-pathway processes that
involve trapping in several misfolded structures. The rate determining
step in the off-pathway processes is the transition from the misfolded
structures to the native state.  The partition factor
\(\Phi \) is also determined by \(\sigma \): smaller the value of
\(\sigma \) larger is \(\Phi \). The qualitative aspects of our results
are found to be independent of the friction coefficient. The
simulation results and theoretical arguments are used to obtain
estimates for time scales for folding via the NCNC 
mechanism in small proteins, those with less than about 70
amino acid residues. 
\vspace{0.5cm}

{\bf Conclusions: } We have shown that the various scenarios for
folding of proteins, and possibly other biomolecules, can be
classified solely in terms of \(\sigma \). Proteins with small values
of \(\sigma \) reach the native conformation via a nucleation collapse
mechanism and their energy landscape is characterized by having one
dominant native basin of attraction (NBA). On the other hand proteins
with large \(\sigma \) get trapped in competing basins of attraction
(CBA) in which they adopt misfolded structures. Only a small fraction
of molecules access the native state rapidly, when \(\sigma \) is
large. For these sequences the
majority of the molecules approach the native state by a three stage
multipathway mechanism, in which the rate determining step involves a
transition from one of the CBA's to the NBA. 
\vspace{0.5cm}

{\bf Key words: } collapse and folding transition
temperature, kinetic partitioning mechanism,  native conformation  nucleation
collapse, protein folding, three-stage multipathway mechanism. }

\end{flushleft}

\section{Introduction}

It has become clear over the last few years that the study of minimal 
models has given rise to a novel theoretical understanding of the 
kinetics of protein folding 
\cite{Dill,Bryn95,Wol,Chan94,Thirum94,Abk1,Socci95,Thirum96}. 
The general scenarios that have emerged from 
these studies are starting to be confirmed
experimentally \cite{Otzen,Fer,Rad,Englander,Schindler,Kief,Itzhaki}. 
In particular, 
there is now some experimental support \cite{Rad,Kief} for the kinetic
partitioning mechanism (KPM) first described using minimal off-lattice
models\cite{Thirum94,Guo,ThirumGuo}. The principles 
emerging from these studies have also been used to predict the 
folding pathways and the nature of  
kinetic intermediates in specific proteins. For example, it was shown 
that the single disulfide intermediate 14-38 in bovine pancreatic 
trypsin inhibitor, which denotes 
that the structure of this intermediate contains a covalent  
disulfide bond between cysteines at location 14 and 38, 
forms early and decays before other more stable single intermediates 
start to form \cite{CamProx}. This 
theoretical prediction has been \underline{subsequently} verified 
experimentally \cite{Dadlez}. The 
theoretical studies
\cite{Dill,Bryn95,Wol,Chan94,Thirum94,Abk1,Socci95,Thirum96,Mirny} 
have in fact provided, perhaps for the first time, a 
firm basis for understanding and predicting the overall scenarios that can 
arise in 
{\em in vitro} refolding kinetics of proteins. Since the general scenarios 
for 
refolding kinetics have been understood from a qualitative viewpoint it 
is of interest to correlate in a quantitative manner the dependence 
of folding 
times for a number of sequences in terms of parameters that can be 
{\em experimentally measured}. In a recent paper theoretical arguments were
used to provide quantitative estimate of some of the important time
scales that arise naturally according to the KPM \cite{Thirum95}. In this paper
computational studies are used to complement the previous work. We
should note that Onuchic {\em et al.}  
have also initiated complementary approaches
to understand in a quantitative manner the folding kinetics of small
\(\alpha \)-helical proteins \cite{Onuchic95}.

In our earlier work we showed using lattice models that the foldability of 
proteins (namely, the ability of a sequence with a unique native  state 
to access it in finite time scales under folding conditions) can be 
understood in terms of 
two characteristic thermodynamic temperatures which are intrinsic to the sequence 
\cite{Klim,Cam93}. One of them 
is \(T_{\theta }\), the collapse transition 
temperature, at which there is a transition from a random coil to an 
almost compact state. The other is the folding transition temperature, 
\(T_{F}\), below which the polypeptide chain is predominantly in the 
native conformation. In the biochemical literature \(T_{F}\) is 
roughly the melting temperature, \(T_{m}\). It was established for a 
variety of 
sequences (in both two and three dimensions) that the folding time 
correlates extremely well with \(\sigma \) \cite{Klim,Cam93}, where
\begin{equation}
\sigma = \frac{T_{\theta } - T_{F}}{T_{\theta }}. 
\label{sigma}
\end{equation}
Based on very general arguments \cite{Thirum95} it can be shown that 
\(T_{F} \le 
T_{\theta }\), hence \(0 \le \sigma \le 1\).  
Both \(T_{F}\) and \(T_{\theta }\) are sensitive functions of not 
only the sequence but also the external conditions. This can be 
verified experimentally and data in the literature in fact support this 
obvious result. For example, Alexander 
{\em et al.} have shown for the IgG binding protein that \(T_{F}\) (in 
their notation \(T_{m}\)) varies 
linearly with pH \cite{Alex}. These authors have also determined
\(T_{\theta }\) for two forms of IgG binding protein.  
Thus, foldability of sequences 
and the associated 
kinetics for a given sequence can be altered by changing the 
external conditions. The major purpose of this article is to explore the 
folding kinetics as a function of \(\sigma \) 
using off-lattice simplified models of polypeptide chains. In
addition, we provide detailed analysis of folding kinetics for a
number of sequences at two values of the friction coefficient to
assess the role of viscosity on the qualitative aspects of the 
folding scenarios.

The physical reason for expecting that \(\sigma \) would control the 
folding rates in proteins is the following. If \(\sigma \) is small, then 
\(T_{\theta } \approx T_{F}\) and hence all the conformations 
that are sampled  at \(T \lesssim T_{F}\) have relatively high 
free energy. Any barrier that 
may exist between these high free energy mobile conformations can  
be overcome easily provided the temperature at which folding 
occurs is not too low. Thus, for small \(\sigma \) one can in principle 
fold a polypeptide chain at a relatively high temperature (in the
range where collapse and the acquisition of the native conformation
are almost synchronous)  and access the 
native conformation rapidly. For these cases the folding process would 
appear to be kinetically two state like \cite{Jackson1}. 
Furthermore, studies based on
lattice models suggest that for sequences with small and moderate
values of \(\sigma \) the kinetic accessibility of the native
conformation together with its thermodynamic stability can be achieved
over a relatively broad temperature range \cite{Klim}. 
On the other hand when \(\sigma 
\approx 1\), then \(T_{F} \ll T_{\theta }\) and in this case the 
folding process would inevitably be affected by kinetic traps and 
misfolded structures. Since some of these misfolded structures can have 
many elements in common with the native structure they can be fairly 
stable \cite{Guo,ThirumGuo}. Since \(T_{F}\) is low for sequences 
with large \(\sigma \) 
these stable structures could have long lifetimes even if the free
energy barriers 
separating them and the native state are only moderate. Thus, it is likely 
that sequences with \(\sigma \approx 1\) are in general not foldable on 
biologically relevant time scales. These expectations are borne out in this 
study and a quantitative 
relationship between folding rates and \(\sigma \) is given. 
The energy landscape perspective can be used to argue that small values of
\(\sigma \) correspond to the native state having a 
large native basin of attraction (NBA) \cite{Honey90} or funnel
\cite{Bryn95,Wol,Leop}.

Before we close this introduction a brief comment on the use of minimal
models to understand folding kinetics is pertinent. 
This is especially important because their utility in
getting insights into protein folding kinetics has been 
questioned \cite{Honig}.  The minimal  models 
do not explicitly contain all the
features that are known to be important in imparting stability to
proteins.  However, there are many aspects of the minimal models that
mimic the dominant interactions in proteins \cite{Dill}. These involve chain
connectivity, hydrophobicity as the driving force, and sequence
heterogeneity. In addition, off-lattice models studied in this paper
and elsewhere \cite{Rey,Kastella} 
which use a realistic representation of the potentials for \(\alpha
\)-carbons of a polypeptide chain yield
\((\phi,\psi)\) values consistent with the Ramachandran plot
\cite{Helix}. The aspects
of real proteins that are not faithfully represented here are side chains
and hydrogen bonds. Straub and Thirumalai have argued that lower order
effects like stability arising from hydrogen bonds are included in the
simplified 
off-lattice models of the sort considered here in a coarse grained
manner \cite{Straub}. This is achieved by 
suitably renormalizing the dihedral angle potentials. 
Despite these important limitations the studies 
based on minimal models of proteins have been the only source of concrete 
testable theoretical predictions in the field of protein folding 
kinetics \cite{Dill,Bryn95,Wol,Chan94,Thirum94,Abk1,Socci95,Thirum96}. 
Insights based on the energy landscape picture of 
folding have lead, for example, to the microscopic picture of native
conformation nucleation 
collapse (NCNC) mechanism in refolding of proteins
\cite{Guo,ThirumGuo,Mirny}. 
Recently, experimentalists have begun interpreting their data on 
certain proteins \cite{Itzhaki,Sosnick} using the concept of
NCNC mechanism. Thus, despite certain 
limitations these studies have 
already offered considerable insights into the folding kinetics of 
biomolecules \cite{Dill,Bryn95,Wol,Thirum96} in both {\em in vitro} 
and {\em in vivo}. 

Since this article is rather lengthy we provide a brief roadmap to the
rest of the article. The details of the model and simulation methods
are given in Sec. (II) and (III), respectively. The main results are
presented in Sec. (IV) and (V). The paper is concluded in Sec. (VI). The
appendix contains useful formulae for obtaining time scales for the
dominant nucleation collapse process for small proteins. Readers not
interested in technical details can skip Sec. (II) and (III) entirely.

\section{Description of the Model}

The model used in our simulations is a variant of the one 
introduced in 
our previous studies \cite{Guo,ThirumGuo}. 
We use continuum minimal model representation of a 
polypeptide chain. In these classes of models only the principle features 
of proteins responsible for imparting stability are retained. These 
include hydrophobic forces, excluded volume interactions, bond angle 
and dihedral angle degrees of freedom. The simplified model can be 
thought of as a coarse grained representation containing only 
the \(\alpha \)-carbons of the protein 
molecule. The polypeptide is modeled as a chain consisting of \(N\) 
connected beads with each corresponding to a set of particular \(\alpha \)-carbons 
in a real protein. 
In order to simplify
the force field 
we assume that sequence is essentially built 
from residues of three types, hydrophobic (\(B\)), hydrophilic (\(L\)), and 
neutral (\(N\)). Our previous studies have established that the three
letter code 
can be used to construct the basic structural motifs in proteins, namely, 
\(\alpha \)-helix and \(\beta \)-turns \cite{Guo,Helix,Honey92}.  
In this study we mimic the 
diversity in the 
hydrophobic residues in proteins  using a dispersion in the
interactions between  \(B\) residues (see below). 

The potential energy of a conformation, which is specified by the set of 
vectors \(\{\vec{r}_{i}\}\), \(i=1,2...N\), is taken to be of the following 
form 
\begin{equation}
E_{p}(\{\vec{r}_{i}\}) = V_{BL} +  V_{BA} +  V_{DIH} +  V_{NON}
\label{E}
\end{equation}
where \(V_{BL}, V_{BA}, V_{DIH}\), and \(V_{NON}\) correspond to 
bond-length potential, bond-angle potential, dihedral angle potential, and 
non-bonded potential, respectively. A brief summary of these interactions 
is given below. 

{\bf (a) Bond-length potential.} In our previous studies we assumed 
that the length of 
the covalent bond connecting the successive beads to be fixed. The 
constraint of fixed bond length, which was enforced using the RATTLE 
algorithm \cite{Andersen}, proves to be computationally demanding. In the present study 
we use a stiff harmonic potential between successive residues, which keeps 
the bond length approximately fixed, i.e., 
\begin{equation}
V_{BL} = \sum_{i=1}^{N-1}\frac{k_{r}}{2}(|\vec{r}_{i+1} - \vec{r}_{i}| - 
a)^{2},
\label{V_BL}
\end{equation}
where \(k_{r} = 100\epsilon_{h}/a^{2}\), \(a\) is the average 
bond length between two beads, and \(\epsilon_{h}\), the average 
strength of the hydrophobic interaction, is the unit of energy in our 
model. We have verified that using the potential in Eq. (\ref{V_BL}) gives 
the same results for the sequence that has been previously studied \cite{Guo}. 

{\bf (b) Bond-angle potential.} The potential for the bending 
degrees of freedom, describing
the angle between three successive beads \(i\), \(i+1\), \(i+2\) is taken
to be
\begin{equation}
V_{BA}  = \sum_{i=1}^{N-2} \frac{k_{\theta }}{2}(\theta _{i} - 
\theta _{0})^{2}, 
\label{V_BA}
\end{equation}
where \(k_{\theta }=20\epsilon_{h}/(rad)^{2}\) and 
\(\theta _{0}=1.8326 \ \ rad \) or \(105^{\circ}\). 

{\bf (c) Dihedral angle potential.} This potential describes the ease of 
rotation around the angle formed between four consequent  beads. This 
degree of freedom is largely responsible in determining secondary
structures  in a polypeptide 
chain \cite{Creightonbook}. The \(i^{\text{th}}\)  dihedral angle 
\(\phi _{i}\) is formed  
between vectors \(\vec{n}_{i} = (\vec{r}_{i+1,i} \times \vec{r}_{i+1,i+2})\) 
and   \(\vec{n}_{i+1} = (\vec{r}_{i+2,i+1} \times \vec{r}_{i+2,i+3})\), 
i.e., it is the angle between the plane  defined by beads \(i\), \(i+1\), 
\(i+2\) and the one spanned by beads \(i+1\), \(i+2\), \(i+3\). The 
vector \(\vec{r}_{i,i+1} = \vec{r}_{i+1} - \vec{r}_{i}\). The general form of 
the potential describing the dihedral angle degrees of freedom is well 
known \cite{McCammon} and can be represented as 
\begin{equation} 
V_{DIH} =\sum_{i=1}^{N-3} [ A_{i}(1+\cos\phi_{i}) + B_{i}(1+\cos 3\phi_{i})]
\label{V_DIH}
\end{equation}
If two or more of the four beads in defining \(\phi_{i}\) are neutral 
(\(N\)) the \(A_{i}\) and \(B_{i}\) are taken to be \(0\epsilon_{h}\) and 
\(0.2\epsilon_{h}\), respectively. For all other cases \(A_{i} = B_{i} 
=1.2\epsilon_{h}\). For the larger values of \(A_{i}\) and \(B_{i}\) the 
trans state is preferred and this leads to the formation of extended 
conformation. The presence of neutral residues, which are introduced so 
that loop formation is facilitated, has the effect of decreasing the 
barrier and energetic differences between the trans and gauche states 
\cite{Honey92}. 

{\bf (d) Non-Bonded potential.} The non-bonded potentials arise between 
pairs of 
residues that are not covalently bonded. These forces 
together with those arising from the dihedral angle degrees of 
freedom  (which provide favorable local interactions for the formation of
secondary structures) are 
responsible for the overall 
formation of the three dimensional topology of the polypeptide chain. 

We take simple forms to represent the non-bonded  
interaction terms. We assume that the effective potential describing the 
interaction between the residues \(i\) and \(j\) (\(|i-j| \ge 3\)) depends 
on the type of residues involved. The total non-bonded  potential is 
written as
\begin{equation}
V_{NON} = \sum_{i=1}^{N-3} \sum_{j=i+3}^{N} V_{ij} (r), 
\label{V_NON}
\end{equation}
where \( r = |\vec{r}_{i} - \vec{r}_{j}|\). 
The potential between two \(L\)-beads or between a (\(L,B\)) pair is taken 
to be 
\begin{equation}
V_{L\alpha} (r) = 4\epsilon_{L}\left[\biggl(\frac{a}{r}\biggr)^{12} + 
\biggl(\frac{a}{r}\biggr)^{6}\right]\ \ (\alpha = L \ \ \text{or} \ \ B),
\label{V_L}
\end{equation}
where \(\epsilon_{L} = \frac{2}{3} \epsilon_{h}\). This potential is purely
repulsive with a value of \(2\epsilon_{h}\) at \(r = 2^{1/6}a\), which
is the location of the minimum in the hydrophobic potential (see Eq.
(\ref{V_BB})). The presence of \(r^{-6}\) term gives rise to a
potential that is longer ranged than the usual \(r^{-12}\) term. The additional
term may be interpreted to arise from the hydration shells around the
hydrophilic residues. 

The interaction between the neutral residues and the others is expressed 
as 
\begin{equation}
V_{N\alpha}(r) = 4\epsilon_{h}\biggl(\frac{a}{r}\biggr)^{12} \  \ (\alpha = N,\ \ L, 
\ \ \text{or} \ \ B).   
\label{V_N}
\end{equation}
If both the residues are hydrophobic (\(B\)) the potential of interaction is 
taken to be 
\begin{equation}
V_{BB}(r)  =  4\lambda \epsilon_{h}\left[\biggl(\frac{a}{r}\biggr)^{12} - 
\biggl(\frac{a}{r}\biggr)^{6}\right],
\label{V_BB}
\end{equation}
where \(\epsilon_{h}\) determines the strength of the hydrophobic 
interaction. The above form for \(V_{BB}(r)\)  
can be thought of as approximate 
representation (capturing the primary minimum) of the potential of mean 
force between spherical  
hydrophobic spheres in water \cite{Pangali}. 

The dimensionless parameter \(\lambda \) is 
assumed to have a Gaussian distribution 
\begin{equation} 
P(\lambda ) = \frac{1}{(2\pi \Lambda ^{2})^{1/2}}\exp\biggl(-
\frac{(\lambda - \lambda _{0})^{2}}{2\Lambda ^{2}}\biggr), 
\label{P(lambda)}
\end{equation} 
The mean value of \(\lambda _{0} = 1\). 
The introduction of the distribution in the strength of the hydrophobic 
interaction creates diversity among hydrophobic species, and hence provides a 
better caricature of proteins. The standard deviation \(\Lambda \) controls 
the degree of heterogeneity of the hydrophobic interactions and if its 
value becomes too large then the unambiguous division of residues into three 
distinct types becomes difficult. Consequently, we keep the value of 
\(\Lambda \) 
small enough so that the prefactor \(4\lambda \epsilon_{h}\) is in 
general greater than \(4\epsilon_{L}\) (see Eqs. (\ref{V_L},\ref{V_BB})). 
This also assumes 
that the interaction between hydrophobic residues remains attractive at 
the separation corresponding to Lennard-Jones minimum. For large values of 
\(\Lambda \) (not used in our present study) the distribution function 
in Eq. (\ref{P(lambda)}) has to be truncated at some positive value of 
\(\lambda 
\) so that the \(\lambda  \epsilon_{h}\) does not become negative. 

In our earlier studies \cite{Guo,ThirumGuo,Honey92} we used \(\Lambda =0\) and hence all \(B\) beads were 
identical. Thus, our previous studies correspond exactly to a three letter 
code. The current potential function with random hydrophobic interaction 
gives more specificity to the interactions, and yet preserves the overall 
hydrophobic interactions as the driving force for structure formation. 

\section{Simulation methods} 

\subsection{Langevin dynamics}

Following our earlier work we have used Langevin dynamics  for
simulating  folding kinetics \cite{Guo,Honey92}. We include
a damping term in the equation of  motion with a properly chosen
friction coefficient \(\zeta \) and the  Gaussian  random force to
balance the energy dissipation caused by  friction.  The equation of
motion written for the generalized coordinate \(x\) is  given by 
\begin{equation}
m\ddot{x}  =  -\zeta\dot{x} + F_{c} + \Gamma \equiv F
\label{mx}
\end{equation}
where \(F_{c} = -\frac{\partial E_{p}}{\partial x}\) is the 
conformation force, which is a negative gradient of potential energy with 
respect to the coordinate \(x\), \(\Gamma \) is the random force 
having a 
white noise spectrum, and \(m\) is the mass of a bead.  The equation of 
motion Eq. (\ref{mx}) is numerically integrated 
using the velocity form of the Verlet algorithm \cite{Verlet}. If the 
integration step is \(h\), the position  of a bead 
at the time \(t+h\) is expressed through the second order in \(h\) as 
\begin{equation}
x(t+h)  =  x(t) + h\dot{x}(t) + \frac{h^{2}}{2m} F(t). 
\label{x(t+h)}
\end{equation}
Similarly, the velocity \(\dot{x}(t+h)\) at the time \(t+h\) is 
given by 
\begin{eqnarray}
\dot{x}(t+h) & = &
\biggl(1-\frac{h\zeta}{2m}\biggr)\biggl(1-\frac{h\zeta}
{2m}+ \biggl(\frac{h\zeta}{2m}\biggr)^{2}\biggr)\dot{x}(t) + 
\frac{h}{2m}\biggl(1-\frac{h\zeta}{2m} 
+\biggl(\frac{h\zeta}{2m}\biggr)^{2}\biggr) \nonumber \\
 & & \times \biggl(F_{c}(t) + \Gamma (t) + F_{c}(t+h) + \Gamma
(t+h)\biggr) + o(h^{2}). 
\label{v(t+h)}
\end{eqnarray}
Because we assume that the random force \(\Gamma \) has a white noise 
spectrum, the autocorrelation function \( <\Gamma(t)\Gamma(t')>\) 
is expressed in the form 
\begin{equation} 
<\Gamma(t)\Gamma(t')> =  2\zeta k_{B}T \delta (t-t'). 
\end{equation}
Since the equation of motion Eq. (\ref{mx}) is discretized and solved 
numerically, this formula can be rewritten as 
\begin{equation}
<\Gamma(t)\Gamma(t+nh)>  =  \frac{2\zeta k_{B}T}{h}\delta_{0,n}, 
\end{equation}
where \( \delta_{0,n} \) is the Kronecker delta and \(n=0,1,2...\).  Thus,
in the context of this model changing the temperature of the system
essentially means changing the standard variance in the Gaussian
distribution of the random force \(\Gamma \). 

Temperature is measured in the units of \(\epsilon
_{h}/k_{B}\). In the underdamped limit, i.e. when  \(\zeta\dot{x} \)
is negligible compared to the inertial term in the equation  of
motion (\ref{mx}), a natural choice of the unit of time is 
\(\tau _{L} =  (m
a^{2}/\epsilon _{h})^{1/2}\). The simulations have been done  in
low to moderate friction limit which in the rate theory of reactions
would correspond to the energy diffusion regime. The
integration step  used  in the equation of motion is taken to be \(h=
0.005 \tau _{L}\). All the  sequences were studied at two values of
the friction coefficient \(\zeta  _{L} = 0.05 m \tau _{L}^{-1}\)
and \(\zeta _{M} = 100\zeta_{L}=  5 m \tau _{L}^{-1}\). The
relation between the  time unit \(\tau _{L} \)  and the folding time
scales in real proteins as well as  the range of \(\zeta \) used in
this study  are  discussed in the Appendix.  The Appendix also 
gives estimates for certain time
scales in the folding kinetics of proteins using the 
simulation results and theoretical
arguments.  In our simulations the
mass of residue \(m\), the bond length \(a\), the hydrophobic
energy constant \(\epsilon_{h}\), and the Boltzmann constant \(k_{B}\)
are set to unity. 

\subsection{Determination of native conformation}

For each sequence we determined the native conformation by adapting the 
procedure similar to that used in our recent work on lattice model of 
proteins \cite{Klim}. The database of sequences generated is described in Sec. 
(III.D). As 
in our earlier works \cite{Guo,Honey92} we have used a combination of 
slow cooling and 
simulated annealing to determine the native conformation. The chain is 
initially heated to \(T=5.0\) and equilibrated at this temperature for 
\(2000\tau _{L}\). The temperature is then quenched to \(T=1.0\)  and the 
chain is reequilibrated for an additional \(2000\tau _{L}\). This process 
of quenching the chain from \(T=5.0\) to  \(T=1.0\)  was repeated several 
times so that we generated a set of independent conformations at  
\(T=1.0\). These structures are used as starting conformations for the 
slow cooling process. In order to ascertain that the starting 
conformations are independent the overlap between a pair of these
conformations  
(see Eq. (\ref{chi})) 
averaged over all distinct pairs denoted by \(\overline{\chi } \) is
calculated. This yields 
\(\overline{\chi } \sim 0.9\) that roughly corresponds to the value of
\(\chi \) for 
a pair of randomly generated conformations.   The temperature of the 
system in the simulations starting from one of the well 
equilibrated conformations at
\(T=1.0\), is 
slowly decreased to \(T=0.0\) (see Sec. (III.C) for details). In the process 
of reaching \(T=0.0\) the energies of the conformations are recorded. This 
process is repeated for several (typically 10) initial conformations. The 
conformation with the lowest energy is assumed to be the native state for 
the sequence. After determining the native conformation 
by this method we 
raised the temperature from \(0.0\) to \(0.2\) in \(1000 \tau _{L}\) and 
then 
lowered it to \(T=0.0\), i.e. we performed a simple simulated annealing 
procedure. In all instances the resulting structure and the energy coincided 
with those obtained by the slow cooling protocol. It 
should be emphasized that this  method cannot  guarantee that the 
structures are indeed global energy minima. 
However, the determination of native structures for these sequences 
by other 
optimization techniques leads to the same structures \cite{Amara}. 
Thus, we are fairly 
certain that the  structures found by this method indeed are the lowest 
energy structures for our model. 

\subsection{Thermodynamic properties}

We and others have shown that each foldable sequence is 
characterized by two natural temperatures \cite{Dill,Chan94,Socci95,Guo,Cam93}. 
One of them is \(T_{\theta }\) 
below which the chain adopts more or less compact conformation. The 
transition at \(T_{\theta }\) is (usually) second order in character suitably 
modified by a finite size effects. Following our earlier studies 
\(T_{\theta }\) is located by determining the temperature dependence of 
the heat capacity \cite{Guo,Klim,Cam93}
\begin{equation}
C_{v}=\frac{<E^{2}>-<E>^{2}}{T^{2}}. 
\label{Cv}
\end{equation}
The location of the peak in \(C_{v}\) is taken to be \(T_{\theta }\). 
Previous studies have shown that at \(T \approx T_{\theta }\) the radius 
of gyration changes dramatically reaching a value roughly coinciding
with that for  
a compact conformation \cite{Guo,Honey92}. This, of course, is 
usually taken to be a signature of "collapse" transition in
homopolymers \cite{DeGenes}. 

The second crucial temperature is the folding transition temperature 
\(T_{F}\). There are several ways of calculating \(T_{F}\) all of which 
seem to give roughly similar estimates \cite{Socci95,Cam93,Sali94b}. 
We use the fluctuations in the 
structural overlap function to estimate \(T_{F}\). The structural 
overlap function is defined as 
\begin{equation}
\chi = 1 - \frac{2}{N^{2}-5N+6} \sum_{i=1}^{N-3} \sum_{j = i+3}^{N}  
\Theta(\varepsilon - |r_{ij} - r_{ij}^{N}|), 
\label{chi}
\end{equation}
where \(r_{ij}\) is the distance between the beads \(i\) and \(j\) for a given 
conformation, \(r_{ij}^{N}\) is the corresponding distance  in the native 
conformation, and \(\Theta (x)\) is the Heavyside function. If  \( 
|r_{ij} - r_{ij}^{N} | \le \varepsilon \) then the beads \(i\) and \(j\) 
are assumed to form a native contact. In our simulations we take 
\( \varepsilon = 0.2 a\). 

It follows from the definition of \(\chi \) that at finite temperatures 
\(<\chi (T)>\), the thermal average, is in general non-zero. The folding 
transition temperature is obtained from the temperature dependence of the 
fluctuations in \(\chi \)
\begin{equation}
\Delta \chi   = <\chi ^{2}(T)> - <\chi (T) >^{2}
\label{dchi}
\end{equation}
For sequences with a unique ground state \(\Delta \chi \) exhibits a peak 
at \(T \simeq T_{F}\) \cite{Cam93}. It has been shown that for these
simple off-lattice models this 
transition is a finite size first order phase transition \cite{Guo}. 
Our previous 
lattice model studies have shown that \(T_{F}\) obtained from the
temperature dependence of  \(\Delta 
\chi \) is in general slightly smaller than that calculated from the 
midpoint of \(<\chi (T)> \) or other suitable order parameters 
\cite{Klimlong}. 

The thermodynamic properties like \(<\chi (T)>\), total energy
\(<E(T)>\) (the sum of kinetic and potential energies) 
etc. are calculated using time averages over sufficiently long 
trajectories. The trajectories which are generated in the search for 
the native structure can be used to get an approximate estimate of the 
temperature interval \((T_{l},T_{h})\), which includes the temperatures 
\(T_{\theta}\) and \(T_{F}\). In all cases we set 
\(T_{h}=1.0\), while \(T_{l}\) varies from \(0.3\) to \(0.4\). Each 
trajectory starts with the same zigzag initial conformation. The chain is then 
heated at 
\(T=5.0\) and brought to equilibrium at \(T=1.0\) as described in Sec. 
(III.B). 
The method of slow cooling employed for calculating thermodynamic averages is 
identical to that presented in \cite{Klimlong}. The system is periodically  
cooled (starting at the temperature \(T_{h}\))  by an amount   
\(\Delta T \). 
The time \(\tau _{max} \) is the time of running the   
simulations at the fixed temperature, \(T_{i}=T_{h}-i \Delta T \), where 
\(i=0,1,2...\). 
In this study we have set \(\Delta T =0.02\), the
time \(\tau _{max}= 2500\tau _{L}\), and the equilibration time after the
change of the temperature by \(\Delta T \) to be \(\tau _{eq}=250\tau 
_{L} \).
These values are used with most sequences within the entire temperature
interval \((T_{l},T_{h})\). The thermodynamic 
values for one particular initial condition \(i\) is calculated as 
\begin{equation}
\overline{f_{i}}(T) = \frac{1}{\tau _{av}} \int _{\tau _{eq}}^{\tau _{eq} 
+ \tau _{av}} f_{i}(T,t)dt,
\label{fav}
\end{equation}
where \(\tau _{av} = \tau _{max} - \tau _{eq}\). The equilibrium 
thermodynamic value is obtained by averaging over a number of 
initial conditions 
\begin{equation}
<f(T)> = \frac{1}{M} \sum _{i=1}^{M} \overline{f_{i}}(T). 
\label{favav}
\end{equation}
We found that \(M=50\) was sufficient to obtain accurate  results for
equilibrium properties. For most 
sequences the values of parameters (see 
above) used in the course of equilibration were 
large enough to obtain converged results. In some instances the equilibration times had to be increased to 
obtain converged results. The functions \(<E(T)>\), \(C_{v} (T)\), and
\(\Delta \chi (T)\) are obtained by fitting the data by polynomials, and
\(<\chi (T)>\) was fit with two hyperbolic tangents. 

We should note that due to the intrinsic heterogeneity of these 
systems non-ergodicity effects often manifest themselves \cite{Guo}. 
If this is the case, 
we need to do weighted averaging of thermodynamic quantities, as described 
elsewhere \cite{Guo,Str2}, to get converged results. This issue was
not encountered for the sequences that were examined in this study. 

It is obvious that the thermodynamic quantities are independent of the 
underlying dynamics provided the dynamics yields the Boltzmann 
distribution at \(t \rightarrow \infty \). Since the thermodynamics is 
only determined by the Boltzmann factor \(\exp(-E_{p}/k_{B}T)\) it is 
convenient to determine them at low friction where the sampling of 
conformation space appears to be more efficient \cite{Guo,Honey92}.

\subsection{Database of sequences}

For our model one can, in principle,  generate infinite number of sequences 
because of the continuous  distribution of the effective hydrophobic 
interactions (see Eq. (\ref{P(lambda)})). 
The vast majority of such sequences would be random, and
hence would not fold to a unique native state on finite time scale. 
Our goal is to obtain a number of these sequences with 
the 
characteristic temperatures \(T_{\theta }\) and \(T_{F}\) such that 
they span a reasonable range of \(\sigma \) (see Eq. (\ref{sigma})). It is 
clear 
that merely creating random sequences will not achieve this objective.
In general, random sequences would take extraordinarily long times to
fold. It is known that foldable sequences (those that reach the native
state in finite times) are designed to have a relatively smooth energy
landscape. Such sequences may, in fact, be minimally
frustrated \cite{Bryn89} or have compatible long and short-range
interactions \cite{Go83}. 
Thus, in order to generate sequences that span the range of  \(\sigma
\) and which are foldable, 
we used the most primitive design procedure in the inverse protein 
folding problem \cite{Shakh93,Shakh94}. 
Because our objective is not to provide the most optimal solution to the 
inverse folding problem the more reliable methods introduced recently were 
not utilized \cite{Deutsch}. 

In all our studies the number of beads \(N=22\). The composition of all 
sequences is identical, namely, all of them contain \(14\) hydrophobic 
beads, \(5\) hydrophilic beads, and \(3\) neutral beads. The sequences in 
this model differ from each other because of the precise way in which 
these beads are connected. In addition, due to the distribution of 
hydrophobic interactions not all hydrophobic beads are identical. The 
latter condition also introduces diversity among sequences. 

The method for creating the database of sequences is as follows. 
The first sequence A, 
\(B_{9}N_{3}(LB)_{5}\), examined  has been already studied in our 
previous work \cite{Guo}. This allows us to ascertain that our modified model 
(incorporating stiff harmonic bond-length potential instead of RATTLE 
algorithm) yields results consistent with our earlier studies. This 
sequence has zero 
\(\Lambda \), so all hydrophobic residues are identical.   
All other sequences (to be used as starting conditions for 
Monte 
Carlo optimization procedure, see below)  were generated at random with 
different  standard deviations \(\Lambda \), but preserving the same 
composition, i.e. 14 \(B\) beads, 5 \(L\) beads, and 3 \(N\) beads. 
By "random generation" of a sequence we mean that a sequence is randomly
constructed from the beads of three types and the values of the
parameter 
\(\lambda \) specifying non-bonded interactions between hydrophobic
residues in a sequence (Eq. (\ref{V_BB})) were obtained using Gaussian
distribution Eq. (\ref{P(lambda)}). 
Specifically, we used \(\Lambda =0\) (one sequence), \(\Lambda =0.1\) (one 
sequence), \(\Lambda =0.17\) (one sequence), \(\Lambda =0.3\) (five 
sequences).

The next step in creating the database of sequences is the choice of the
target conformation, the precise choice of which is rather arbitrary
because the only natural requirement is that it should be compact. 
However, due to intrinsic propensity toward bend formation near clusters
of \(N\) residues, it seems reasonable to restrict the choice of target
conformation topology to that with single U-turn. Obviously, the number of
residues in a sequence \(N=22\) allows us to define much more complicated
topologies featuring multiple U-turns. In order to avoid the comparison of
folding behavior of the sequences with the native conformations of
different topologies the only criterion for selecting target conformations
is that they must have a single U-turn and be reasonably compact. The role
of the topology of the ground state in determining the folding kinetics
will be addressed in a future paper. Thus, using the conditions described
above we have selected the target conformations from the database of low energy
structures found in the course of slow cooling simulations. 

Once a target conformation is chosen Monte Carlo algorithm
in sequence space \cite{Shakh93,Shakh94} is used to obtain an optimal sequence by means of the
primitive inverse design procedure, i.e., the sequence 
that has the lowest energy and is
compatible with the target conformation. There is no guarantee that this
procedure is by any means the best way of designing optimal sequences as has
been pointed out by Deutsch and Kurosky \cite{Deutsch}. However, for our 
purposes this naive
procedure suffices. The main idea is to perform small random permutations
of a sequence while keeping its composition fixed and accepting (or
rejecting) new sequences with respect to the Boltzmann factor \(P=\exp(-
\Delta E/k_{B}T)\), where \(\Delta E \) is the energy variation due to
permutation and \(T\) is the temperature of Monte Carlo optimization
scheme \cite{Shakh93,Shakh94}. Hence, this algorithm is aimed at lowering the energy of the
target conformation.  The sequence providing the lowest energy at the
target conformation is chosen as the desired sequence and used for further
analysis.  The control parameter which specifies the degree (or "quality")
of optimization is the temperature \(T\). For most sequences we have used
low value of \(T=0.2\). We have made several attempts to run Monte
Carlo algorithm in a sequence space by gradually decreasing the
temperature from high values of \(T \gtrsim 1.0\) up to the very low
values \(T \lesssim 0.01\). This method, however, does not often provide
lower energies of target conformations than that based on quenching the
temperature at a certain value, and this is probably due to moderate
ruggedness of the energy landscape in sequence space.  It must be
emphasized that the optimization scheme does not guarantee that the target
conformation is actually the native state of the optimized sequence. This
should be checked in the course of molecular dynamics simulations (see
below). 

The nine sequences obtained using the procedure described
above are listed in Table 1 and are labeled A through I. Of the nine 
sequences eight (B-I) were
generated using Monte Carlo method in sequence space. 
All sequences have different native conformations, except the pairs 
of sequences A,D and B,C, which share the same native state. It must 
also be emphasized that although sequences F-H have 
identical distribution of beads they differ from each other with respect to 
the strength of hydrophobic interactions, i.e. they have different 
sets of prefactors \(\lambda \) (Eq. (\ref{P(lambda)})).

\subsection{Simulation temperature}

        In order to compare the rates of folding for different sequences
it is desirable to subject them to identical folding conditions. The
equilibrium value of \(<\chi (T)>\) measures the extent to which the
conformation at a given temperature \(T\) is similar to the native
state. At sufficiently low temperature \(<\chi (T)>\) would approach zero,
but the folding time may be far too long. We chose to run our folding
simulations at a sequence dependent simulation temperature \(T_{s}\)
which 
is subject to two conditions: (a) \(T_{s}\) \underline{be less than}
\(T_{F}\) for a specified sequence so that the native conformation has the
highest occupation probability; (b) the value of \(<\chi(T=T_{s}) >\) be a
constant for all sequences, i.e. 
\begin{equation} 
<\chi(T=T_{s}) >=\alpha. 
\label{026}
\end{equation} 
In our simulations we choose \(\alpha =0.26\) and for all
the sequences studied \(T_{s}/T_{F} < 1 \). This general procedure for
selecting the simulation temperatures has been already used in 
recent studies of folding kinetics using lattice models 
\cite{Klimlong,Cam95,Sali94b}.  

An alternative way of choosing \(T_{s}\) is to assume that the
probability of occupation of the native state be the same for all
sequences. In our previous study \cite{Klimlong} on lattice models 
we have used this
method for a small number of sequences. The trends in the folding
times at the resulting simulation temperatures (as well as the kinetics)
were very similar to those found when Eq. (\ref{026}) is used to
determine \(T_{s}\). 

It is also possible to keep the simulation temperature constant for all
sequences. Because \(T_{F}\) and \(T_{\theta }\) vary greatly
depending on sequence such a choice would not ensure that the
probability of being in the native conformation is roughly the same
for all sequences or that all sequences are qualitatively similar to
the same extent. In other words the folding conditions for the
sequences would effectively be different if the temperature is held
constant. This argument also implies that the statistical trends of
folding kinetics with respect to intrinsic sequence dependent
properties are expected to hold only over an optimal range of folding
conditions which in simulations are entirely determined by
temperature.

\subsection{Monitoring folding kinetics}

	The simulation procedure for obtaining folding kinetics resembles
the slow cooling method apart from the one principal difference that after
heating the chain it is quenched to the temperature \(T_{s}\), defined
from the condition \(<\chi(T=T_{s}) >=0.26\).  The 
temperature is held 
constant after the quench to \(T=T_{s}\). The
duration of the folding simulations depends on the rate of folding of a
particular sequence and is typically on the order of \(10^{4}\tau _{L} \). 
For 
each sequence we generated  between \(100-300\) independent trajectories.  
The folding
kinetics is monitored using the fraction of trajectories \(P_{u}(t)\)
which does not reach the native conformation at time \(t\)
\begin{equation}
P_{u}(t)=1 - \int_{0}^{t} P_{fp}(s)ds,
\label{Pu}
\end{equation}
where \(P_{fp}(s)\) is the distribution of first passage times, 
\begin{equation}
P_{fp}(s) = \frac{1}{M} \sum_{i=1}^{M} \delta (s - \tau 
_{1i}).
\label{Pfp}
\end{equation}
In Eq. (\ref{Pfp}) \(\tau _{1i}\) denotes the first passage time for the
\(i^{\text{th}}\) trajectory, i.e. the time when a sequence adopts
native state for the first time. It is easy to show that the mean first
passage time \(\tau _{MFPT}\) to the native conformation (which is
roughly the folding time, \(\tau _{F}\)) can be calculated
as
\begin{equation}
\tau _{MFPT} = \int_{0}^{\infty}tP_{fp}(t)dt =
\int_{0}^{\infty}P_{u}(t)dt. 
\label{tfp}
\end{equation}
The mean first passage time \(\tau _{MFPT}\) can also be calculated
using \(\tau _{1i}\)  for the M trajectories so that
\begin{equation}
\tau _{MFPT} = \frac{1}{M}\sum_{i=1}^{M} \tau_{1i}. 
\label{tfpdirect}
\end{equation}
We find that for all the sequences \(P_{u}(t)\) 
can be adequately fit with several exponentials of the form 
\begin{equation}
P_{u} (t) = \Phi \exp\biggl(-\frac{t}{\tau _{FAST}}\biggr) + 
\sum _{k} a_{k} \exp \biggl(-\frac{t}{\tau _{k}}\biggr),  
\label{Pufit}
\end{equation}
where the sum is over the dominant misfolded structures, \(\tau _{k}\)
are the time scales for activated transition from one of the misfolded
structures to the native state, and \(\sum _{k} a_{k} = 1 - \Phi
\). 
In this study we report  \(\tau _{MFPT}\) using Eq. (\ref{tfp}) by
fitting \(P_{u}(t)\) according to Eq. (\ref{Pufit}). We have
explicitly verified that Eq.  (\ref{tfp}), with \(P_{u}(t)\) given 
by Eq. (\ref{Pufit}), and Eq. (\ref{tfpdirect}) 
yield practically identical results. 

It has been shown in a series of articles that multiexponential fit of the
function \(P_{u}(t)\) can be understood in terms of the kinetic
partitioning mechanism  \cite{Thirum96,Guo,ThirumGuo,Mirny} and is
indicative of the distribution of time scales in the refolding of
biomolecules. 
According to KPM a fraction of molecules
\(\Phi \) folds to the native conformation very rapidly, while the remainder 
(\(1 - \Phi \)) approaches the native state via a complex three stage
multipathway mechanism (TSMM).  Therefore, the time 
constants \(\tau _{FAST}\) and
\(\tau _{k}\) can be interpreted as the characteristic folding times of
the fast and slow phases, respectively.  When appropriate fit of the
function \(P_{u}(t)\) with exponentials is performed, the calculation of
the mean first passage time \(\tau _{MFPT}\) becomes straightforward.

An alternative way to calculate folding times is based on the analysis of
the time dependence of the overlap function. The overlap function \(\chi
\) is constantly calculated during simulations and its average time
dependence is obtained as
\begin{equation}
<\chi (t)> = \frac{1}{M} \sum_{i=1}^{M} \chi _{i}(t) ,
\label{chiav}
\end{equation}
where \(\chi _{i}(t)\) is the value of \(\chi \) for the \(i^{\text{th}}\)
trajectory at time \(t\). We find that \(<\chi (t)>\) can also be fit
by a sum of exponentials (usually by one or two, see \cite{Klim,Cam93}) 
\begin{equation}
<\chi (t)> = a_{1} \exp\biggl(-\frac{t}{\tau_{1}}\biggr)+a_{2}
\exp\biggl(-\frac{t}{\tau_{2}}\biggr)
\label{chifit}
\end{equation}
Here \(\tau_{1} \) gives the estimate for the time scale of 
native conformation 
nucleation collapse  process  \cite{Guo,ThirumGuo}, while the 
largest time 
constant in the fit \(\tau
_{2}\) serves as an estimate of the folding time \(\tau _{F}\).  
For most sequences
biexponential fit provides the most accurate results. However, several
sequences (typically ones with relatively small values of \(\sigma \))
demonstrate a clear single exponential behavior of \(<\chi
(t)>\).  In some cases there are additional slow components present on
larger time scales as well. It has been shown in our previous papers that 
defining folding times using the functions \(P_{u}(t)\) or \(<\chi
(t)>\) yields
qualitatively similar results \cite{Guo,Klimlong}. The same 
conclusion is valid for this model
as well. Thus, \(\tau _{FAST}\) and the largest 
\(\tau _{k}\) in Eq. (\ref{Pufit}) are 
roughly proportional to \(\tau_{1} \) and \(\tau_{2} \), respectively.

\section{Results}
\subsection{Thermodynamic properties}

In this section we present the results on thermodynamics and kinetics of
folding. Using the methodology described above, we studied 9
sequences, 8 of which were generated by performing Monte Carlo 
simulations in sequence space. The native conformation of each sequence was
determined using the procedure described in Sec. (III.B). The example of the 
native conformation for the
sequence G is given in Fig. (1).
This picture demonstrates that all three neutral residues shown in grey  
are concentrated in the turn region. It is also clearly seen that
hydrophobic residues shown in blue tend to be in close contact to each
other due to their inherent attractive interaction, while hydrophilic
residues shown in red point outwards. 

For
each sequence we calculated the two characteristic equilibrium 
temperatures, collapse
transition temperature \(T_{\theta }\) and folding transition temperature
\(T_{F}\) from the temperature dependence of \(C_{v}\) and \(\Delta
\chi \), respectively. The plots of \(<\chi >\) and \(<E>\) were also 
obtained.  Fig. 
(2) displays these functions for the sequence G.  The plot of
\(<\chi (T)>\) in Fig. (2a)  indicates that at high temperatures \(0.8 
\lesssim T \lesssim 1.0\) the value of \(<\chi > \approx 0.8\), so the 
chain 
has negligible amount  of the native structure.  It was
already shown in \cite{Honey92} that under these conditions
polypeptide 
chain is in a random coil conformation. The overlap function 
gradually decreases with
the temperature and at \(T=0.3\) it reaches the value below \(0.2\). We do
not plot \(<\chi >(T)\) for \(T < 0.3\), because at such low temperatures
it is difficult to obtain reliable
thermodynamic averages due to non-ergodicity problems. Fortunately, this is 
not necessary for the 
aim of our study because all characteristic temperatures \(T_{\theta }\),
\(T_{F}\) and the simulation temperature \(T_{s}\) are relatively high.
The peak of the specific heat \(C_{v}\) (see Fig. (2d)) which corresponds to 
the collapse
transition temperature \(T_{\theta }\) is at \(0.78\). At this
temperature protein undergoes a transition from an extended coil state  to
compact conformation. In fact, we calculated the radius of
gyration \(<R^{2}_{g}>\) as a function of temperature for few sequences and
found that at \(T \approx T_{\theta }\) it shows a sudden drop in
accord with earlier and more recent studies
\cite{Guo,Honey92}. However,
at \(T_{\theta }\) the overlap function is still relatively large (\(<\chi
> \approx 0.7\)). The fluctuation of the overlap function \(\Delta \chi
\) achieves a maximum at \(T=0.62\) and this is taken to be the folding
temperature \(T_{F}\). The value of \(T_{F}\) calculated from the 
midpoint of \(<\chi >\) (i.e., when \(<\chi >\) is about 0.5) is also 
around 0.62. In general, we have found that \(T_{F}\) obtained from the 
peak of \(\Delta \chi \) is slightly lower than that calculated from 
the temperature dependence of similar measures like \(<\chi >\) 
\cite{Klimlong}. 
It was demonstrated that this temperature
corresponds to first order folding transition to the native
conformation \cite{Guo,Cam93}. 
By monitoring \(\chi (t) \) for several individual trajectories under
equilibrium conditions at \(T \approx T_{F}\) we find that the protein
fluctuates between the native and disordered conformations.  
All the nine sequences show similar behavior from which the various 
thermodynamic parameters can be easily extracted. The parameter \(\sigma 
\) (see Eq. (\ref{sigma})) for the nine sequences ranges from 0.14 to 0.65. 
Thus, a 
meaningful correlation between the folding time and \(\sigma \), which is 
one of the major purposes of this study, can be established.

The simulation temperature \(T_{s}\) for the sequence G defined by Eq. 
(\ref{026}) is found to be 0.41. In Fig. (3) we present the dependence of 
\(T_{s}\)
on the parameter \(\sigma \). It is seen that the simulation temperature
\(T_{s}\) is a decreasing function of \(\sigma \). Thus, high values of
\(T_{s}\) are found for the sequences with small values of \(\sigma \),
and this prompts us to anticipate that such sequences are fast folders
(see below). However, it was argued \cite{Klimlong} that this correlation must be
viewed as statistical.  This implies that if two sequences have 
close values of \(\sigma \), then a precise correlation with
\(T_{s}\) is not always expected. On the other hand, if a large number of
sequences spanning a range of \(\sigma \) is generated, then we expect
a statistical correlation to hold. We also expect these conclusions to
hold over a range of temperatures which is favorable for folding. 
The present off-lattice studies and
those based on lattice models \cite{Klim,Klimlong} confirm this expectation. 

\subsection{Dependence of \(T_{\theta }\) and \(T_{F}\) on sequence} 

One of the major results in this study (see Sec. (IV.D)) is
that the folding times for all sequences correlate extremely well
with \(\sigma \) (cf. Eq. (\ref{sigma})). Therefore, it is of interest
to investigate how \(T_{\theta }\) and \(T_{F}\) vary with the
sequence. It seems reasonable to assert that the folding temperature
\(T_{F}\) depends rather sensitively on the precise sequence. In fact,
it has been argued that to a reasonable approximation \(T_{F}\) is
determined by the nature of the low energy spectrum (a sequence
dependent property), at least in
lattice models \cite{Camacho96}. The sensitive dependence of \(T_{F}\)
on the sequence is explicitly confirmed in the present paper and in
the previous lattice models as well \cite{Klimlong,Camacho96}. 
In Table 1 we display 
\(T_{F}\) for the nine sequences. The values of \(T_{F}\) range from
0.20 to 0.62. Thus, the largest  \(T_{F}\) is about three times larger
than that of the smallest value. 

It might be tempting to think that \(T_{\theta }\) should be
insensitive to the sequence and should essentially be determined by
the composition of the sequence. This expectation arises especially
from heteropolymer theory \cite{Garel}. According to the
heteropolymer model, which
essentially ignores short length scale details, \(T_{\theta }\) is
determined by the average excluded volume interactions, \(v_{0}\), and
the average strength of hydrophobic interactions, \(\lambda
_{0}\epsilon _{h}\). Both
these values are expected to be roughly constant, especially if the sequence
composition is fixed. The determination of \(T_{\theta }\) for a
polypeptide chain based on these arguments ignores surface terms and
may, in fact, be valid in the thermodynamic limit, i.e. when the
number of beads tends to infinity. However, polypeptide chains are
finite sized and hence the nature of surface residues which depends on
the precise sequence are critical in the determination of \(T_{\theta
}\). This, in fact, is borne out in our simulations. In Table 1 we
also display \(T_{\theta }\) for the nine sequences. Although the
largest \(T_{\theta }\) is only approximately \(1.5\) times (as oppose
to a factor of three for \(T_{F}\)) larger
than that of the lowest 
it is clear that \(T_{\theta }\) is very sequence dependent even
though the composition for all sequence is \underline{identical}. All
the sequences have 14 hydrophobic residues. Thus, both \(T_{\theta }\)
and \(T_{F}\) are determined not only by the intrinsic sequence but
also by external conditions. In fact, \(T_{\theta }\) and \(T_{F}\),
and, consequently, \(\sigma \) can be manipulated by altering the
external solvent conditions (pH, salt, etc.). It therefore follows
that  a single foldable
sequence can have very different values of \(\sigma \) depending on
the solvent conditions and hence can exhibit very different kinetics. 

It is interesting to obtain estimates for \(T_{\theta }\) and
\(T_{F}\) using realistic values of \(\epsilon _{h}\), the average 
strength of hydrophobic interaction. From Table 1 we note that 
the range of \(T_{\theta }\) is
\((0.58 - 0.80)\epsilon _{h}/k_{B}\) with the lower values
corresponding to sequences with larger \(\sigma \). The value of
\(\epsilon _{h}\) ranges from \((1-2) kcal/mol\).  Assuming that
\(\epsilon _{h} \approx 2 kcal/mol\) the range of \(T_{\theta }\) is
\((48-67)^{\circ}C\). It appears that the better designed sequences (ones
with smaller \(\sigma \) values) have more realistic values of
\(T_{\theta }\). Similarly the range of \(T_{F}\) for better designed
sequences is \((33-50)^{\circ}C\). These estimates suggest that
optimized sequences can fold over a moderate range of temperatures
rapidly and with relatively large yield. These expectations are
explicitly demonstrated here.

\subsection{Kinetics of folding: The Kinetic Partitioning Mechanism (KPM)}

We studied the folding kinetics using the function \(P_{u}(t)\),
which gives  the fraction of unfolded molecules (trajectories) at 
time \(t\). We also computed the time dependence of \(<\chi (t)>\) to gain
additional kinetic information concerning the approach to the native
conformation. The function \(P_{u}(t)\) has been
obtained for each sequence at the simulation temperature \(T_{s}\) from
the analysis of large number of individual trajectories 
(\(M=100-300\)) starting with
different initial conditions. The resulting plots of \(P_{u}(t)\) were
fitted with a sum of  exponentials (one, two, or three)  and the mean 
first passage time \(\tau
_{MFPT}\) (taken to be equal to \(\tau _{F}\)) for each sequence was calculated as described in Sec. (III.F). 
In general, it is found that after a short transient time \(P_{u}(t)\) is 
extremely well fit by a sum of exponentials (cf. Eq. (\ref{Pufit})). 
The partition factor, \(\Phi \),  gives the fraction of 
molecules that reaches the native conformation on the time scale \( \tau 
_{FAST} \) by a NCNC mechanism and \(\tau _{k}\) (\(\gg \tau
_{FAST}\)) being the time scales over 
which the remaining fraction \(1-\Phi \) reaches the native state 
\cite{Guo,ThirumGuo,Klimlong}. 

Based on fairly 
general theoretical considerations it has been shown that \(\sigma \) 
(\(=(T_{\theta } - T_{F})/T_{\theta }\)) can be used to discriminate 
between fast and slow folding sequences \cite{Cam93,Thirum95}. This
has been confirmed 
numerically for lattice models \cite{Klim,Klimlong,Camacho96}. 
We classify fast folding sequences 
as those with relatively large values of \(\Phi \) (\( \gtrsim 0.9\)). 
These sequences reach the native conformation without forming any
discernible intermediates and essentially display a two state kinetic 
behavior. The plot of \(P_{u} (t)\) for one of these sequence (sequence 
G) which can be fit with only one exponential in Eq. 
(\ref{Pufit}) is presented in Fig. (4a). 
It is obvious that \(\Phi \) depends on the sequence (via \(\sigma
\)), the temperature, and other external conditions. Four 
sequences out of nine appear to be fast folders 
displaying a two state kinetic approach to the native conformation 
with \(\Phi \gtrsim 0.9\). These sequences have 
\(\sigma \) values less than about 0.4. 

The other five sequences have \(\Phi \) values less than 0.9 and hence 
can be classified as moderate or slow folders. The values of \(\sigma \) for 
these 
sequences exceed 0.4. The discrimination of sequences into slow and fast 
based on \(\Phi \) is arbitrary. An example of the kinetic 
behavior of a slow folder 
(sequence A) probed using \(P_{u}(t)\) is 
shown in Fig. (4b). The generic behavior of \(P_{u}(t)\) as a sum of
several 
exponentials has been argued to be a consequence of the kinetic  
partitioning mechanism  \cite{Guo,Thirum95}. 
Typically for slow folding sequences \(\tau _{FAST}\) varies from 
\(200\tau _{L} \) to \(600\tau _{L}\), whereas the largest value of 
\(\tau _{k}\) (as defined by Eq. (\ref{Pufit})) 
lies in the interval from \(2500\tau _{L} \) to 
\(2.7 \times 10^{6}\tau _{L}\). Slow folding  
trajectories reach the native state  via three stage multipathway 
mechanism  \cite{Guo,Thirum95,Cam93}, according to which 
random collapse of a protein (first stage) is followed by a slow 
search of the native state among compact conformations (second stage)  
that eventually leads the polypeptide chain to one of several misfolded 
structures. These misfolded structures have  many characteristics of the 
native state. Generically the rate determining step in the TSMM involves 
the transition (crossing a free energy barrier) from the misfolded 
structure to the native state (third stage) \cite{Cam93}. 

In order to obtain insights into the microscopic origins of the slow and 
fast phases we have analyzed the dynamic behavior of various 
trajectories. We have found that for sequences that find the native 
conformation essentially in a kinetically two state manner all the 
trajectories reach the native conformation without forming any discernible 
intermediates. Furthermore, for these cases once a certain number of 
contacts is established the native state is reached very rapidly, which 
is reminiscent of a nucleation process \cite{Abk1,Guo,ThirumGuo}. The 
time scale for such nucleation dominated processes is relatively 
short and it has been suggested that in these cases 
the collapse process and the 
acquisition of the native conformation occur almost simultaneously
\cite{Thirum95}. It is for 
this reason we refer to this process as native conformation  
nucleation collapse (NCNC). 
This process has been referred to as nucleation condensation mechanism
by Fersht \cite{Fer}.  These points 
are illustrated by examining the dynamics of the structural overlap 
function \(\chi (t)\) for fast folders. A typical  plot for \(\chi (t)\) 
for a fast folder (sequence G) is shown in Fig. (5). In Fig. (5) (as
well as in Figs. (6-8,15,16)) we plot 
\(\chi (t)\) for a trajectory labeled \(k\) averaged over a few 
integration steps \(h\), i.e., 
\begin{equation}
\overline{\chi _{k}}(t) = \frac{1}{\overline{\tau }} \int 
_{t-\overline{\tau }/2}^{t+\overline{\tau }/2} \chi _{k} (s) ds.
\label{avchi}
\end{equation}
The value of \(\overline{\tau } = 5 \tau _{L}\) which is much less than any 
relevant folding time scales. This figure (Fig. (5a)) shows 
that within \(380\tau _{L}\) (the first passage time) the chain 
reaches the native conformation. After the chain reaches the native state 
there are fluctuations around the equilibrium value of \(<\chi >\)
(\(=  0.26 \)). Another example of folding trajectory for this
sequence is presented in lower panel, which is further analyzed in
Sec. (IV.D). 

The dynamical behavior shown in Fig. (5) for fast trajectories should be
contrasted with the trajectories for other sequences that reach the
native state by indirect  off-pathway processes. An example of such a behavior for
the moderate folder (sequence E, \(\Phi = 0.72\)) is  shown in Fig. (6).  
The behavior
presented  in Fig. (6a) shows  that after an initial rapid collapse
(on the time scale of about  \(100-200\tau _{L}\)) the chain explores 
intermediate state (where  \(\chi (t)\) is
roughly constant for a large fraction \(\tau  _{I}/\tau _{1i}\) of
time, where \(\tau _{I}\) is the life-time of the intermediate state)
before reaching the native conformation at  \(\tau _{1i} = 3026\tau
_{L}\). Fig. (6b) shows another off-pathway trajectory for this sequence,
in which native conformation is reached at  \(  1970\tau _{L}
\). Although these slow trajectories are qualitatively similar, they
clearly demonstrate that the chain samples different misfolded
conformations depending on the initial conditions before it 
finally finds the native state. This fact
further supports the multipathway character of the indirect folding process.  
After the native conformation is reached the  overlap
function fluctuates around the equilibrium value \(<\chi >=0.26\) or makes
sudden jumps to the higher values of \(\chi  \approx 0.4\)
 and fluctuates around these values 
for a finite time. Such dynamics clearly reflects  frequent visits 
to low lying  structures (see Sec. (IV.D).  The
behavior shown in Fig. (6) is very typical of the trajectories that
reach  the native conformation via indirect mechanisms which are
conveniently quantified in terms of TSMM. Fig. (7) presents a typical
indirect trajectory for fast  sequence I, which has the partition 
factor  \(\Phi \) slightly less than unity. 
This trajectory reaches the native conformation at \(2384 \tau
_{L}\). 

It is also instructive to compare the dynamical behavior of the
nucleation   trajectories of fast and slow folding sequences. An
example of  a trajectory that reaches the native conformation via
nucleation collapse  mechanism  for sequence E is shown in
Fig. (8).  It is important to note that the qualitative
behavior of  \(\chi (t)\) presented  in Fig. (8) is very similar to
that shown in Fig. (5). 
This further confirms  that the underlying
mechanism that leads the chain directly to the native  conformation
for sequences with large \(\sigma \) is similar to the   nucleation
process. The only difference is that the partition factor  \(\Phi \)
is less for sequences with large \(\sigma \) than  for
ones with small \(\sigma \). Fig. (8) also indicates that after reaching
native state the  chain makes frequent visits to neighboring misfolded
conformations  and, in some instances, gets trapped in these for
relatively long times.

The kinetic behavior described above  suggests  that the value of
\(\sigma \) can be used to 
classify sequences according to their ability to access the native 
state. It appears that not only does \(\sigma \) correlate well with the
intrinsic kinetic accessibility of the native conformation it also
statistically determines the kinetic partition factor \(\Phi 
\). In Fig. (9a) we show the dependence of \(\Phi \) on \(\sigma \) for 
the nine sequences. The trend which emerges from this plot is that the 
sequences with larger 
values of \(\sigma \) (and consequently with larger \(\tau _{MFPT}\)) have 
smaller values of \(\Phi \). For example, for the slow folding sequence 
labeled A with the largest value of \(\sigma =0.65\) the fraction of fast 
trajectories is \(\Phi =0.43\). In contrast, the fastest folding sequence 
labeled I (\(\sigma =0.14\))
for which biexponential fit of \(P _{u}(t)\) is needed, has the value of  
\(\Phi \gtrsim 0.9\).

\subsection{Probes of kinetic and equilibrium intermediates using
inherent  structures: Roles of NBA and CBA}

The question of the nature and relevance of intermediates in protein
folding is of abiding interest. Our studies here and elsewhere 
\cite{Thirum95,Klimlong} have
demonstrated that the scenarios for folding can be conveniently
classified in terms of \(\sigma \) provided the foldable sequences are
compared in a similar manner. In order to probe the role of
intermediates in the approach to native state we have analyzed three
sequences (E, G, and I) using the kinetic order parameter
profiles. Sequences G and I are classified as fast folders (the
partition factor \(\Phi \) exceeds 0.9) while sequence E is a moderate
folder with the associated \(\sigma \) (\(\Phi =0.72\)) 
lying in the boundary between fast and slow folding sequences. 

We analyze the role of kinetic and equilibrium intermediates (defined
below) using the following methodology. Each trajectory is divided
into a kinetic part and an equilibrium part. The kinetic part of a
trajectory labeled \(i\) includes
the portion from the beginning till the
native state is reached for the first time, namely, the first passage
time, \(\tau _{1i}\). The equilibrium part is taken to be the
remaining portion of the trajectory from \(\tau _{1i}\) till \(\tau
_{max}\). For convenience we take \(\tau _{max}\) to be the same for
all trajectories. In order to characterize the nature of intermediates
we use the overlap function, \(\chi \), which, as described earlier,
gives the degree of similarity to the native conformation. It is
possible that the same value of \(\chi \) may correspond to different
conformations and in some instances to conformations that are even
structurally unrelated to each other. However, by studying the
distribution of overlap function over a range of \(\chi \) 
for several independent initial conditions and by directly
comparing the resulting conformations and calculating \(\chi \) between
them we can ascertain the states
that are visited with overwhelming probability before and after
reaching the native conformation. In order to probe the nature of
kinetic and equilibrium intermediates that the chain samples en route
to the native conformation we have determined the "inherent"
structures \cite{Still} (see below). The inherent structures are obtained from the
time course of \(\chi (t)\) examples of which are shown in
Figs. (5-8). The basins of attractions are obtained before the chain
reaches the native conformation for the first time (i.e., the
"kinetic" basins) and are determined as follows. As the polypeptide
chain approaches the native conformation (but has not yet reached it,
i.e., \(t < \tau _{1i}\)), we
record several (usually about 10) instantaneous conformations which serve as
initial conditions for steepest descent simulations. In this method the
temperature is set to zero and the velocities of all residues are
rescaled to zero after each integration step. This  results in a
"downhill" motion of a sequence on the energy surface. The final
conformations of the steepest descent quench simulations (provided they
are sufficiently long) are the conformations of local energy minima
(inherent structures) which the sequence explores in the folding process.  
These conformations 
obtained at different times and with distinct initial
conditions allow us to map the distribution of folding pathways. The
same technique for getting inherent structures  was used 
after the first
passage time \(\tau _{1i} < t < \tau _{max}\) as well. 
These would give us the "equilibrium"
intermediates. This analysis allows to compare the nature of
intermediates in the kinetic pathways. 

For sequence G we determined the inherent structures using the
instantaneous conformations labeled (1-6) (all of which occur at \(t < 
\tau _{1i}\)) shown in Fig. (5b). The inherent structures for this
particular trajectory (and for others as well) almost always coincide
with the native state. This clearly shows that for sequence G, for
which the native state is reached by the nucleation collapse
mechanism, the various inherent structures directly map into the native
basin of attraction (NBA). The rapid approach to the NBA is the reason
for the two state kinetics displayed. It also follows that the NBA is
relatively smooth, i.e. the energy fluctuations characterizing the
roughness is comparable to \(k_{B}T_{s}\). The roughness associated
with the NBA implies that the polypeptide chain spends a finite amount
of time in close proximity (\(\chi (t) \approx <\chi >\)) to
the native conformation prior to reaching it. It is worth emphasizing
that this sequence (\(\sigma = 0.20\)) fluctuates only around the
native state even for \(t > \tau _{1i}\) for all the trajectories
examined. 

Of the nine sequences we have examined I is the fastest folder, i.e.,
it has the smallest folding time. Nevertheless, the partition factor
\(\Phi \) is slightly (but measurably) less than unity. The amplitude
of the slow component is very small (for this sequence the
biexponential fit to \(P_{u}(t)\) suffices). 
These observations suggest that 
the underlying topography explored could be somewhat different from
that of sequence G which is also a fast folder. Most of the
trajectories reach the native state for sequence I rapidly without
forming any intermediates and resemble the behavior shown in Fig. (5)
for sequence G. However, there are "off-pathway" trajectories for this
sequence an example of which is shown in Fig. (7). The inherent
structures at the kinetic part for this particular trajectory   
(\(t < \tau _{1i}\)) were determined using the conformations labeled
(1-6) (Fig. (7)). In addition, the inherent structures were also
calculated using the conformations (7-12) that the chain samples after
the first passage time for this trajectory. We found that these
inherent structures are all identical and differ very slightly (as
measured by the overlap function). Consequently we 
characterize them as native-like intermediates. This sequence, although is
a fast folder, has at least one 
competing basin of attraction (CBA) in which the
structure is quite similar to the native state. Since there is a
small fraction of molecules that reach the CBA prior to reaching the
NBA the \(\Phi \) value is smaller than unity. The comparison between
sequences G and I, both of which folds very rapidly, shows that there
can be significant differences in the underlying energy surface. This
is further illustrated in Sec. (IV.E). 

According to our classification sequence E is at least a moderate
folder and exhibits the full range of the kinetic partitioning
mechanism (\(\Phi = 0.72\)). 
There is a significant component of initial trajectories
that reach the native state via three stage multipathway
mechanism. Examples of these off-pathway trajectories are shown in
Fig. (6). We have obtained the 
inherent structures using conformations labeled 
(1-6) in Fig. (6a) (that occur before \(\tau _{1i}\)) 
and using the conformations labeled 
(7-20) (that are sampled for times greater than \(\tau _{1i}\)). It is
found that these structures are nearly the same (excluding structure
(6)) indicating that, in
this instance, the polypeptide samples native-like intermediates
en route the native conformation. In this sense the behavior for this
trajectory is no different from that observed for off-pathway
trajectories for sequence I (Fig. (7)). 

The result of a similar analysis using another trajectory shown in
Fig. (6b) is dramatically different. The inherent structures obtained
using the conformations labeled (1) and (2) and (3-7) are completely
different from each other. Furthermore, the equilibrium intermediates
identified with the inherent structures obtained using the
instantaneous conformations (11-17) do not resemble those calculated
during the kinetic portion (1-7). We do find that the equilibrium intermediates
for this trajectory (11-17) are virtually 
identical to those calculated using the
conformations (CBA's) sampled by other trajectories displayed in
Figs. (6a,8). Examination of other off-pathway trajectories reveal the
presence of an exceptionally stable intermediate with \(\chi
 \approx
0.8\). In fact, this intermediate survives for \(90,000\tau _{L}\), while
a typical first passage time is only about \(1000\tau _{L}\). 
Such intermediates
described above 
are never visited again after folding is completed and hence they are
kinetic intermediates. 

These observations imply that for moderate and slow folders there are
several competing basins of attraction. Some of these serve as
equilibrium intermediates, i.e., these have native-like
characteristics and the chain revisits them even after reaching the native
state. Others, which occur relatively early in the folding process,
perhaps during the initial collapse process itself, 
are kinetic intermediates that
are not visited after the native state is reached, at least during the
time course of our simulations. Thus, for
moderate and slow folders one has a distribution of CBA's. The
presence of CBA's provide the entropic barriers to folding
\cite{Cam95} resulting
in slow approach to the native state. In contrast, for fast folders
the only intermediates that are encountered, if at all, are all 
native-like. Thus, for fast folding sequences only the NBA
dominates. In such cases the energy landscape can be thought of as
being funnel-like \cite{Bryn95,Leop}.

\subsection{Free energy profiles}

The analysis in the preceding subsection indicates that the free
energy profile can be quite complex. The shapes of these profiles
depend crucially on the sequence and external conditions (in our
simulations that is specified  only by the temperature). We have
attempted a 
caricature of the free energy surface by computing the histogram of 
states expressed in terms of the potential energy
\(E_{p}\) and \(\chi \). 
The histogram of states, which measures the probability of occurrence
of the state with a given \(E_{p}\) and \(\chi \), is defined as 
\begin{equation}
g(E_{p},\chi ) = 
\frac{1}{M} \sum _{i=1}^{M} \frac{1}{\tau
_{max}-\tau _{1i}} \int _{\tau _{1i}} ^{\tau _{max}} 
\delta (E_{p} - E _{p,i}(t)) 
\delta (\chi  - \chi _{i}(t))ds,
\label{g}
\end{equation}
where \(E _{p,i}(t)\) and \(\chi _{i}(t)\)
are the values of potential energy and overlap function for
the trajectory \(i\) at time \(t\) averaged over a small interval of 
\(5\tau _{L}\). 
We have calculated \(g(E_{p},\chi )\) for
three sequences at the sequence dependent simulation temperature
\(T_{s}\). The values of \(M=100\), a grid size of \(0.1\) is used for
\(E_{p}\) and \(\chi \) is increased in
intervals of 0.01. If \(\tau _{max} >> \tau _{1i}\) then
Eq. (\ref{g}) gives the equilibrium distribution function. A free
energy profile may be illustrated using the potential of mean force 
defined as 
\begin{equation}
W(E_{p},\chi) = 
- k_{B}T_{s} ln [ g(E_{p},\chi )]. 
\label{w}
\end{equation}

In Figs. (10-12) we plot \(g(E_{p},\chi )\) for the
three sequences. The bottom panel in each of these figures shows the
contour plot of the histogram of states. For sequence G (Fig. (10))
it is clear that the NBA is the only
dominant maximum and consequently the kinetics on this surface is
expected to be two state-like. The plots in Fig. (10) for sequence G also show 
that after the NBA is located the chain only fluctuates in the
NBA. The free energy profile for sequence I (as suggested by Fig. (11)) 
has in addition to the
NBA at least one CBA. The presence of the CBA makes \(\Phi
\) smaller than that for sequence G, for which \(\Phi =
1.0\). Proteins with a larger \(\sigma \) would have several
CBA's. This is clearly indicated in Fig. (12) for sequence E 
which shows that there are
two discernible CBA's which makes this model protein only a moderate
folder. The profile of the potential of mean force for this sequence,
computed using
Eq. (\ref{w}), is shown in Fig. (13). This figure shows that in general
one has a complex structure for the free energy profile. It is also
clear that this multivalley structure naturally leads to the KPM
discussed in Sec. (IV.C). These figures also show that in special
cases (small values of \(\sigma \)) 
the  folding kinetics can be described in terms of only the
NBA or folding funnel \cite{Bryn95,Leop}.

\subsection{Dependence of \(\tau _{F}\) on \(\sigma \)}

It is clear from the results discussed above that the parameter
\(\sigma \) (for a given external condition, which in our case is the 
simulation temperature) may be used to predict approximate kinetic 
behavior of various
sequences. The folding time \(\tau _{F}\) , which is taken to be the mean 
first passage time \(\tau _{MFPT}\), is plotted as a function of \(\sigma \)
in Fig. (14a). This graph shows a remarkable correlation between these
\(\tau _{F}\) and \(\sigma \).  The sequences with small \(\sigma
\lesssim 0.4\) fold to the native conformation very rapidly, so that
\(\tau _{F}\) is less than about \(600\tau _{L} \). However, \(\tau
_{F}\) for the sequence with largest \(\sigma =0.65\) is as large as
\(875258\tau _{L}\). Thus, variation of the parameter \(\sigma \) from 0.14
to 0.65 results  in three orders of magnitude increase in the folding
time (from \(461\tau _{L}\) to \(875258\tau _{L} \)). It must be noted that
the correlation between \(\tau _{F}\) and \(\sigma \) should be
considered as statistical.  One can easily notice few pairs of closely
located data points in Fig. (14a), for which larger value of \(\sigma \)
does not correspond to larger \(\tau _{F}\).  Nevertheless, the general
conclusion following from Fig. (14a) remains apparent:  the parameter
\(\sigma \) allows us to predict the trend in the folding rate of the
sequences by knowing only its thermodynamic properties, such as
\(T_{\theta }\) and \(T_{F}\). 
It should also be pointed out that because of the difficulty in computing 
the low energy spectra of the off-lattice models
\cite{Honey90,Fukugita} correlations between 
folding times and other quantities (such as the energy gap or the 
relative value of the native energy compared to that of non-native 
conformations) 
were not tested. In addition, there appears to be no unambiguous 
way to determine the kinetic glass transition temperature, 
\(T_{g,kin}\). Therefore, we have not tested the proposal 
that foldable sequences 
have large values \(T_{F}/T_{g,kin}\) \cite{Gold92}.

	In order to study the dependence of the folding time on the parameter 
\(\sigma
\) we used the function \(P _{u}(t)\) and defined folding time as the
mean  first passage time \(\tau _{MFPT}\) (see Eq. (\ref{tfp})). It was 
already mentioned above that the alternative way is to analyze the overlap
function \(<\chi (t)>\) and take the largest exponent \(\tau _{2}\) in the
exponential fit to \(<\chi (t)>\) as an estimate for the folding time. 
Due to computational limitations we did this only for five
sequences and found the trend similar to that illustrated in
Fig.  (14a), i.e the folding time \(\tau _{2}\) correlates remarkably well
with the parameter \(\sigma \).

\subsection{Kinetics and folding times at moderate  friction}

	The results presented above have been obtained with the value of
friction coefficient fixed at \(\zeta_{L} =0.05\).  In order to study the
dependence of the folding kinetics on \(\zeta \) we have performed the
same study of nine sequences at a larger value of the
friction coefficient \(\zeta_{M} =5 = 100 \zeta _{L}\). The plot showing the
folding time \(\tau _{F} = \tau _{MFPT}\) as a function of the parameter 
\(\sigma \) at \(\zeta_{M} \) is
displayed in Fig. (14b). In accord with the results obtained at the lower
value of \(\zeta _{L}\) this figure also unambiguously demonstrates a good
correlation between \(\sigma \) and \(\tau _{F}\), so that the sequences
with small values of \(\sigma \) fold much faster than the sequences,
having large \(\sigma \).  Specifically, the sequence labeled I, which
has the smallest value of \(\sigma =0.14\), reaches the native
conformation very rapidly within \(\tau _{F} = 1554\tau _{L} \), while the
sequence labeled A with \(\sigma =0.65\) folds very slowly within \(\tau
_{F} = 2.4 \times 10^{6}\tau _{L}\). 
As one may expect the overall folding times in the
moderate friction limit are considerably larger than in the low friction
limit. The folding times vary almost linearly with \(\zeta \). 
For most sequences the ratio \(\tau _{F}(\zeta _{M})/\tau
_{F}(\zeta _{L})\) is 3 - 4. The largest value of this ratio is found
for the slow folding sequence labeled D and is equal to 5. 

In order
to compare the folding kinetics at \(\zeta _{M}\) with those obtained
at \(\zeta _{L}\) we analyzed 
several folding trajectories. Fig. (15) presents typical folding
trajectory (in terms of the overlap function \(\chi (t)\)) for the
sequence G, which displays two
state kinetics and is classified as a fast folder.  This figure shows
that after few tertiary native contacts are established the chain
rapidly reaches the native state. 
In Fig. (16) we plot \(\chi (t)\) for typical slow (upper panel) and fast
(lower panel) trajectories for the sequence E, which, in contrast to
sequence G, 
exhibits KPM and is 
classified as a moderate folder. It is seen that the fast trajectory for the
sequence E is very similar to a typical trajectory for the sequence
G. The reason for this is that the underlying mechanism for the fast
process, namely NCNC mechanism, is exactly identical. 
It is also very important to note
that  similar plots for these two sequence (Figs. (5-6,8)) obtained at
\(\zeta _{L}\) are virtually the same as those shown in Figs. (15,16). This
allows us to suggest that principal mechanisms of protein folding, such
KPM, nucleation collapse, appear to be independent of the
viscosity of surrounding medium. 
The time scales and the kinetic partition factor \(\Phi \), however,
depend critically on viscosity \cite{Thirum95}. 

	The classification of sequences into slow and fast folders
based on the parameter \(\sigma \) can also be carried out with the
larger  value of \(\zeta _{M}\). Fast folding sequences (4 out of 9) are
characterized by the values of \(\sigma \lesssim 0.4\). The mean 
first passage time for fast folders \(\tau _{MFPT} \) is below
\(3000\tau _{L}\). The function
\(P_{u}(t)\) for fast folders is adequately fit (apart from one 
sequence) with
single exponential just as in the low friction limit. Thus 
folding of these sequences proceeds via
nucleation collapse mechanism. The sequences with \(\sigma \gtrsim 0.4\) can be
classified as slow or moderate folders. These sequences have
significantly larger mean  first passage times \(\tau _{MFPT}\) 
ranging  from \(3285\tau
_{L}\) to \(2.4 \times 10^{6}\tau _{L}\).  Most importantly, the  fraction of
unfolded molecules \(P_{u}(t)\) is clearly two or three exponential (see Eq.
(\ref{Pufit})) which is an apparent manifestation of KPM.  As for the
low friction limit the fraction of fast folding trajectories \(\Phi \)
increases as the parameter \(\sigma \) decreases (Fig. (9b)). 
Specifically, for the sequence  A (\(\sigma =0.65\)) \(\Phi = 0.47\),
while for the fastest folding sequence I (\(\sigma =0.14\)) the
fraction of fast trajectories becomes as large as 0.93.

\subsection{Quantitative dependence of \(\tau _{F}\) on \(\sigma \)}

It is interesting to comment on the quantitative dependence of \(\tau
_{F}\) on \(\sigma \). Theoretical arguments suggest that, at least at
small values of \(\sigma \), \(\tau _{F}\) should scale algebraically
with \(\sigma \), i.e. \(\tau _{F} \sim \sigma ^{\theta }\) with
\(\theta = 3\) \cite{Thirum95}. The present simulations as well as
previous studies using lattice models suggest \cite{Klim,Camacho96} that the data can also be fit
with an exponential, i.e. 
\begin{equation}
\tau _{F} \simeq \tau _{0} F(N) \exp (\frac{\sigma }{\sigma _{0} })
\label{sigmafit}
\end{equation}
where \(\sigma _{0}\) depends on the value of friction and \(F(N)\) is
a function that depends on \(N\). It has been argued \cite{Thirum95}
that \(F(N) \sim N^{\omega }\) with \(3.8 \lesssim \omega \lesssim
4.2\) for \(\sigma \approx 0\) and   \(F(N) \sim \exp (\sqrt{N})\) for
larger \(\sigma \). 
The data in
Fig. (14) can be fit with Eq. (\ref{sigmafit}) with \(\sigma _{0}
\approx 0.06\) at \(\zeta _{L}\) and 
\(\zeta _{M}\). The fit of \(\tau _{F}\) to an algebraic power (\(\tau
_{F} \sim \sigma ^{\theta }\)) gives \(\theta \approx 3.9\) at \(\zeta
_{L}\)  and \(\zeta _{M}\). Further work
will be needed to fully quantify  the precise dependence of \(\tau
_{F}\) on \(\sigma \). It appears that both Eq. (\ref{sigmafit}) and
the algebraic behavior \cite{Thirum95} account adequately for the data
given here and elsewhere for lattice models. The fit given in Eq. 
(\ref{sigmafit}) appears to be a bit more accurate.

\section{Implications for Experiments}

The results presented here together with the time scale estimates
given in the Appendix have a number of implications for
experiments. Here we restrict ourselves to providing some comparisons
to the folding of chymotrypsin inhibitor 2 (CI2) which was probably the first
protein for which a kinetic two state transition was established
\cite{Jackson1,Jackson2}. 
These experiments established that the kinetics for the fast
phase, which corresponds to the molecules with proline residues 
in a trans conformation,
follows a two state behavior. Furthermore the thermodynamics also
displays a two state cooperative transition with the native
conformation being stable by about \(7 \ \  kcal/mol\) at \(T=25^{\circ }C
\), \(\text{pH}=6.3\) and at zero denaturant concentration. Although
not explicitly addressed here we have argued  elsewhere \cite{Thirum95}
that the marginal stability (relative to other structurally unrelated
conformations) of the native state of proteins satisfies
\begin{equation}
\Delta G \gtrsim k_{B}T\sqrt{N},
\label{G}
\end{equation}
where the unknown prefactor is assumed to be of the order of
unity. The CI2 examined by Jackson and Fersht has 83 residues and
consequently Eq. (\ref{G}) gives \(\Delta G \approx 5.5  \ \ kcal/mol\) at 
\(T=25^{\circ }C\). This is in fair agreement with the experimental
determination. It appears that  Eq. (\ref{G}) is consistent with the
marginal stability of proteins of varying size. The bound given above
seems to be a good estimate of the stability of biomolecules \cite{Thirum96}. 
We expect the scaling relation of the type given in Eq. (\ref{G})
to be accurate to only within a factor of two. Given that there is 
inherent experimental uncertainty in determining \(\Delta G \) the
agreement with the theoretical prediction within roughly 20 percent is
remarkable. 

The kinetics of folding of CI2 can be rationalized using the
ideas developed here. The time scale for native conformation
nucleation collapse according to Eq. (\ref{tauNCNC}) is \(\tau _{NCNC}
\approx 0.2 \ \  ms\) using the parameters specified in the Appendix and
with \(\sigma \approx 0.4\) (we have taken \(T_{\theta } \approx
60^{\circ }C\) and \(T_{F} \approx 37^{\circ }C\)). 
If we assume that the folding time
changes exponentially with \(\sigma \) (cf Eq. (\ref{sigmafit})) then
the estimate for the nucleation collapse time changes to about \(25 \
\ ms\), where we have used \(\sigma _{0} \approx 0.1\).  These estimates
give an interval (a relatively broad one) \(0.2 \ \ ms \lesssim \tau
_{NCNC}  \lesssim 25 \ \ ms\). Despite the uncertainties in the theoretical
estimates (unknown prefactors, errors in the estimates of \(\gamma,
a_{0}\) etc.) the estimated values of  \(\tau _{NCNC}\) are within
measured experimental values. The early experiments and more recent
ones on CI2 and a mutant of CI2 indicate that the folding time for
\(\tau _{NCNC}\) is in the range of \((1.5 - 18) \ \ ms\)
\cite{Otzen,Jackson1,Jackson2,Itzhaki}. 

The fastest folding time of \(1.5 \ \ ms\) is found for a 
mutant of CI2 \cite{OF}. Our theoretical
estimates show that even if the external conditions are constant and the
length of the polypeptide chain is fixed  \(\tau _{NCNC}\) can still
be altered if \(\sigma \) (see Eq. (\ref{tauNCNC})) is altered. Since
\(\sigma \) is very sensitive to sequence we suggest that the mutant of
CI2 has a different value of \(\sigma \) than the wild type. This can
readily explain the decrease in folding time for the mutant 
under otherwise similar 
external conditions. Further work is needed to quantify these ideas.

\section{Conclusions}

The folding of proteins  is a 
complex kinetic process involving scenarios that are not
ordinarily  encountered in simple chemical reactions. This complexity
arises due to the presence of several energy scales and the polymeric
nature of polypeptide chains. 
As a result, this complexity
leads to a  bewildering array of time scales that are only now
beginning to be  understood quantitatively in certain minimal models
of proteins \cite{Thirum95,Onuchic95}. Despite  this remarkable complexity  
it has been known
from the pioneering studies of  Anfinsen that 
the  specification of
the primary sequence determines the three dimensional  structure of
proteins, i.e. native state topology is encoded in the  primary
sequence. The study presented here as well as our earlier work on
lattice models \cite{Klim,Klimlong}  have shown clearly \underline{how} the kinetic
accessibility is also encoded  in the primary sequence itself.  Our
results suggest that a wide array of mechanisms that are encountered
in the folding process is, remarkably enough, determined by a simple
parameter expressible in terms of the properties that are intrinsic to the
sequence but affected by external conditions.  It appears  that the
two characteristic equilibrium temperatures \(T_{\theta }\) and \(T_{F}\)
determine the rate at which a given sequence reaches the native
conformation. \(T_{\theta }\) and \(T_{F}\) not only depend on
the  sequence but also can be dramatically changed by varying the
external  conditions such as pH, temperature etc. Thus, the
mechanism  for reaching the native conformation for a single domain protein
can change  dramatically depending on the external conditions. 
This implies that a protein that exhibits two state kinetics under
given external conditions does not necessarily follow the same
kinetics, if the ambient conditions (e.g., pH) are altered. 

Our results show that generically the polypeptide chain reaches the
native conformation by a kinetic partitioning mechanism (KPM). For a number
of  sequences studied here we have established that for given external
conditions (for the computational studies it is the temperature only)
a  fraction of molecules \(\Phi \) reaches the native conformation
directly  via nucleation collapse mechanism, while the
remainder follows  a complex three stage multipathway kinetics. For
both values of friction  coefficient studied here this general
scenario holds. 

It is clear from our results that once the external conditions are
specified \(\Phi \) is essentially determined by the interplay of
\(T_{\theta }\) and \(T_{F}\) as embodied in Eq. (\ref{sigma}).   
The folding time correlates  extremely well with the
dimensionless parameter \(\sigma = (T_{\theta }  -T_{F})/ T_{\theta
}\) independent of the value of the external friction.  The remarkable
correlation between \(\sigma \)  and several
kinetic properties lends credence  to the notion that in small
proteins at least a single collective coordinate description of
folding  may suffice \cite{Socci96}. It also follows from this study
that only when \(\sigma \) is small can folding be described in terms
of NBA. For moderate and slow folders it is important to consider the
interplay between NBA and CBA in determining folding kinetics. 
The independence of our general conclusions on
the type of  models (lattice versus off-lattice) \cite{Klim,Klimlong} 
and on the details of
the dynamics  (Langevin dynamics or Monte Carlo) seems to indicate that the 
kinetic partitioning mechanism 
(along with  \(\sigma \) determining the trends in folding times) may
describe in a concise fashion  the scenarios by which 
single domain proteins reach the native  conformation. 

There are quantitative differences between  the results obtained 
for lattice and
off-lattice models. For example using simulations of lattice models it
was concluded that fast folders (with \(\Phi \approx 1.0\)) have
values of \(\sigma \) less than about 0.15 \cite{Klimlong}. The
off-lattice models suggest that fast folders can have \(\sigma \) as
large as about 0.4. Since the estimates of \(T_{\theta }\) and
\(T_{F}\) using the off-lattice simulations appear to be in better
accord with experiments it is tempting to suggest that for
semi-quantitative comparison with experiments it is better to use
off-lattice simulations.

\acknowledgments
We are grateful to Alan Fersht for informing us of folding times for
CI2 and a mutant of CI2. 
This work was supported in part by grants from the National Science
Foundation (through grant numbers CHE93-07884 and CHE96-29845) 
and the Air Force Office of Scientific Research. 
T.V.\ gratefully acknowledges financial support from the Ecole Normale
Sup\'{e}rieure de Lyon, France.

\appendix 
\section{}

In this appendix we map the natural time units to real
times  so that an assessment of the folding times for these minimal
models as well as for small sized proteins can  be made. In addition
using a mapping between these models and proteins estimates for 
folding times  for proteins with small  number (\(\lesssim 70\)) 
of amino acids are also presented. 
We expect these estimates to be accurate to within an
order of magnitude due to large uncertainties in the estimates of
various quantities as well as a lack of theoretical understanding of
the conjectures. From the equation of motion (see Eq. (\ref{mx}))
it is clear  that when the inertial term dominates the natural unit of
time is \(\tau  _{L} = (m a^{2}/\epsilon
_{h})^{1/2}\). Typical values of \(m_{0}\)  and \(a_{0}\) for amino
acid residues are  \(3 \times 10^{-22} g\)  and \(5 \times 10^{-8} cm
\), respectively.  These are the masses and the Van     der Waals
radius of the amino acid residues. The hydrophobic interaction  energy
\(\epsilon _{h}\) is of the order of \(1 kcal/mol \) or \(7 \times
10^{-14} erg\). If these values are changed by factor of two or so
there  will be not a significant change in our conclusions. Assuming
that a bead in our model roughly represents one amino acid we evaluate
\(\tau _{L}\) as 
\begin{equation} 
\tau _{L} = \biggl(\frac{m_{0}a^{2}_{0}}{\epsilon _{h}}\biggr)^{1/2}
\approx 3  ps. 
\label{tau} 
\end{equation} 
The value of the low friction
coefficient used in our simulations \(\zeta  _{L} = 0.05 m/\tau
_{L} = 5 \times 10^{-12} g/s\) while the value  of \(\zeta _{M} = 100
\zeta _{L} = 5 \times 10^{-10} g/s\).  It is interesting to compare
these values for \(\zeta \) to that  obtained in water which has at
room temperature \(T = 25^{\circ }C\) a 
viscosity of \(0.01Poise\) with \(1 Poise\) being
equal to \(1\ \ g /(s \cdot  cm)\). The friction  on a bead of length
\(a_{0}\) may be estimated as  
\begin{equation} 
\zeta _{water} \simeq
6 \pi \eta _{water} a_{0} \approx 9 \times 10^{-9} g/s. 
\label{water}
\end{equation} 
From this we get \(\zeta _{L} / \zeta _{water} \approx
10^{-3}\), while  \(\zeta _{M} / \zeta _{water} \approx 0.1\). The low
friction would  correspond to the energy diffusion regime in the
Kramer's description of  the unfolding to folding reaction. In the
moderate friction there could  be a competition between inertial 
and viscous damping terms leading  perhaps to the Kramer's turnover regime
familiar in the literature on  simple reactions. 

In the overdamped limit the inertial term can be ignored and the natural
measure for time is  
\begin{equation} 
\tau _{H} \simeq \frac{\zeta
a^{2}}{k_{B}T_{s}} \simeq \frac{6 \pi \eta   a^{3}}{k_{B}T_{s}} = 
\alpha \tau _{L} \frac{\epsilon _{h}}{k_{B}T_{s}},
\label{tauH}
\end{equation} 
where \(\alpha \) is a constant. In our simulations \(\alpha =0.05\)
for \(\zeta _{L}\) and \(\alpha =5.0\) for \(\zeta _{M}\). The typical
value of \(\epsilon _{h}/k_{B}T_{s}\) is about 2, where once again we
have used \(\epsilon _{h} = 1 kcal/mole \) and taken \(T_{s}\) to be the
room temperature. For water at room temperature \(\tau _{H} \approx 3
ns \) with \(\alpha  \approx 100\).  
If we
assume that the  higher value of friction used in this study is in the
slightly overdamped  limit   then  the approximate time unit becomes \(\tau
_{M} \approx 0.3 ns\). Since  the higher value of friction is more
realistic we can estimate the folding times for small proteins  
using the computed time scale. 
For \(\alpha =5.0\) our simulation results give
the folding time ranging from 100 \(\tau _{M}\) to \(10^{6} \tau
_{M}\). 
The folding time for the case  of higher 
friction (with \(\alpha \) exceeding \(5\)) also ranges 
from \(10^{3} \tau _{M}\) (for the  smallest \(\sigma
\)) to \(10^{6} \tau _{M}\) (for the largest  \(\sigma \)) (Klimov and
Thirumalai, unpublished). A naive
estimate using these results would suggest that the folding time for  these sequences
ranges from \(10^{-6} s\) to \(10^{-4} s\). 

A better estimate of these times can perhaps be made by recognizing
that  each bead in the minimal model corresponds to a blob containing \(g\) number of amino
acids \cite{DeGenes}. If  the structure within a blob is represented by roughly
spherical size \(a\)  than we can use \(a \approx g^{\nu}a_{0}\),
where \(g^{\nu}\) is the  "swelling factor" mapping the minimal model to
real proteins. Then the  natural time unit for the motion of such a
blob in the overdamped limit  becomes  
\begin{equation} 
\tau _{H}^{R}
\simeq \frac{6 \pi \eta g^{3\nu} a^{3}_{0} }{k_{B}T} =  g^{3\nu} \tau
_{H}  
\label{tauHR} 
\end{equation}   
The range of \(\nu \) is \(\frac{1}{3} -
1\), with \(\nu = \frac{1}{3}\) corresponding to globular structure
within a blob and \(\nu = 1\) corresponding to maximum repulsion among
the residues in a blob. This should be viewed as a guess  and is not expected to be
correct given that \(g\) is small. Realistic  values of \(g\) are
expected to be between 2 and 3 \cite{Onuchic95,Bryn96} making the 22-mer  minimal model to
(perhaps) correspond with 44-66 amino acid residue  proteins. For \(g =2
\), \(\tau _{H}^{R} \) ranges from \(0.4 ns\) to  \(1.6 ns\), while for
\(g =3 \) \(\tau _{H}^{R} \) ranges from \(0.4 ns\) to \(5.4
ns\). Assuming that \(g =3 \) and \(\nu =1 \) (which would give the
largest time scales ) the folding estimates for small proteins (number
of  amino acids smaller than 70) range from \(8 \times 10^{-6} s\)
(for small  \(\sigma \)) to \(10 ms \) (for large \(\sigma \)). 

This exercise suggests that no matter how the mapping is done the
most  relevant time scale for folding kinetics of small proteins under
normal folding conditions (around room temperature and low denaturant
concentration) is  between microseconds to milliseconds. In
particular, for those proteins  that reach the native conformation
predominantly via the nucleation collapse process
(characterized by relatively small \(\sigma  \)) the time scale for
folding  is between microseconds to  milliseconds
for small proteins. One of us has argued \cite{Thirum95} that the  
time scale for the NCNC  process is given  by
\begin{equation} 
\tau _{NCNC}  \simeq  \frac{\eta a_{0}}{\gamma }
\sigma ^{3} N^{\omega } 
\label{tauNCNC} 
\end{equation} 
where \(\omega
\) is in the range of \(3.8\) to \(4.2\). There is usually  a large
uncertainty in the surface tension \(\gamma \) between the
hydrophobic residues and water. The range for \(\gamma \) is between
\(25  - 75 \ \ cal / (\text{\AA}^{2} \cdot mol)\). The largest time scale
to Eq. (\ref{tauNCNC}) emerges when \(w \simeq 4.2\) and \(\gamma
\simeq 25 \ \ cal / (\text{\AA}^{2} \cdot mol)\).  Using  \(\eta \simeq
0.01\ \ Poise\), \(\sigma \approx  0.2\), and \(a_{0} \approx 5 \times
10^{-8} \ \ cm
\) \(\tau _{NCNC}\)  ranges from \(10^{-6} s\) to \(0.1
ms \) as \(N\) varies from 22 to 66.  The values based on theoretical
arguments (cf. Eq. (\ref{tauNCNC})) are  consistent with the numerical
estimates based on the  simulations. 

It is interesting to compare the estimates for the fast process,
corresponding to the NCNC mechanism, obtained using
Eqs. (\ref{tauNCNC},\ref{tauHR}) and simulation results with
experimental results. All the theoretical estimates yield \(\tau
_{NCNC} \approx 0.1 \ \ ms\). The recent experiments on
chymotrypsin inhibitor 2 suggest that the time scale for the
nucleation collapse process is in the range of \(1.5 - 15 \ \ ms\)
\cite{OF} depending on external conditions. The experimental times are not
inconsistent with our simulation results and theoretical
estimates given the uncertainty in the values of the various
parameters. Our studies further underscore the importance of processes
relevant for folding of proteins in submillisecond time scale
especially for the NCNC process. 
Further experiments on these  
time scales are needed for an explicit  experimental demonstration of
the native conformation  nucleation collapse mechanism 
\cite{Jones,Gruebele,Gray}.

\newpage
\centerline{FIGURE CAPTIONS}

Fig. 1. The \(\beta \)-type native structure of the sequence  
\(LB_{9}(NL)_{2}NBLB_{3}LB\) labeled G. In the turn region chain backbone 
adopts {\em gache}-conformations. The hydrophobic beads are given
by a blue color, the hydrophilic beads are represented by red, 
and the neutral beads are shown in grey. 

Fig. 2. The temperature dependence of the thermodynamic quantities for 
sequence G calculated using slow-cooling method: 
(a) overlap function \(<\chi  (T)>\); (b) fluctuations of the overlap 
function \(\Delta \chi (T)\); (c) energy \(<E(T)>\); (d) specific heat 
\(C_{v}(T)\). The peaks in the graphs of \(\Delta \chi (T)\) and 
\(C_{v}(T)\) correspond to the folding transition and collapse transition 
temperatures, \(T_{F}\) and \(T_{\theta }\). 

Fig. 3. The dependence of the simulation temperature \(T_{s}\), as defined by 
Eq. (\ref{026}), on the parameter \(\sigma = (T_{\theta } - 
T_{F})/T_{\theta }\). 

Fig. 4. The fraction of the unfolded molecules 
\(P_{u}(t)\) as a function of time for the sequences G (a) and A (b). Time is
measured in the units of \(\tau _{L}\) (cf. Eq. (\ref{tau})). 
The solid line in Fig. (4a) is a single exponential 
fit to the data. This implies that for this sequence folding is 
kinetically a two state process (\(\Phi = 1.0\)). 
The solid line in Fig. (4b) is a three exponential
fit to the data.  The multiexponential process is indicative of the
kinetic partitioning mechanism with \(\Phi = 0.43\) (see Eq. (\ref{Pufit})).  

Fig. 5. Dynamics of a typical fast-folding trajectory as measured  
by \(\chi (t)\) 
(sequence G) at \(\zeta _{L}\). It is seen in Fig. (5a) 
that on a very short time 
\(380\tau 
_{L} \) the native conformation is reached. 
After the native conformation is reached \(\chi (t) \)
fluctuates around the equilibrium value \(<\chi >\). 
(b) Another trajectory for this sequence, for which inherent
structures at the times (1-6) were determined. Horizontal arrow in
this plot 
indicates the region of native basin of attraction (NBA). It is seen
that the chain approaches NBA but spends finite amount of time there
before reaching the native state at \(1525 \tau _{L} \). 
Dashed line 
indicates  \(<\chi > =0.26 \) at \(T_{s}\).  Vertical arrows indicates
the first passage time.

Fig. 6. Examples of two trajectories that reach the native
conformation by an indirect off-pathway process. These trajectories
are for sequence E at \(\zeta _{L}\). The kinetics exhibited by the
off-pathway process suggests that the native state is reached by a
three state multipathway mechanism. Fig. (6a) 
shows that after 
initial rapid collapse on the time scale of \((100-200) \tau _{L}\)
the chain gets trapped in misfolded compact structure 
(indicated by nearly constant value of \(\chi \) for long times). In
this case the
native state is eventually reached at \(\approx 3026 \tau
_{L}\). Fig. (6b) which is given for a different slow trajectory shows that 
the chain samples at least two distinct misfolded
structures before the first passage time is attained at \(\approx 1970
 \tau _{L}\). In both figures numbers indicate the points where
inherent structures were determined. The time course of \(\chi (t)\) 
reveals that the chain samples a number of kinetic and equilibrium
intermediates. It was found that inherent structures (1-5,7-20)
(Fig. (6a)), (11-17) (Fig. (6b)), and (1-5) (Fig. (8)) are almost identical and
are accessible {\em before} and {\em after} first passage time. For
this  they 
are classified as native-like equilibrium intermediates. However, the
inherent structures (1-2) and (3-7) (Fig. (6b)) are examples of kinetic
intermediates. 
Dashed
lines in these plots indicate the equilibrium value of  \(<\chi >
=0.26 \) at \(T_{s}\). Horizontal arrows indicate the regions of
CBA's. Vertical arrows indicate the first passage time. 

Fig. 7. Dynamics of one of the few off-pathway trajectories for sequence I
(\(\Phi = 0.95\)). Inherent structures are determined at the points
marked by the numbers. Analysis shows that structures (1-6) and (7-12)
are identical that allows us to refer to them as equilibrium
native-like intermediates. Note that the vast majority (\(\approx
0.95\)) of trajectories fold via NCNC mechanism. The native state is
reached at \(2384 \tau _{L}\). Horizontal arrows indicate the regions of
CBA's. Vertical arrow indicates the first passage time.

Fig. 8. This figure shows the dynamics of \(\chi (t) \) for one
trajectory for sequence E (with \(\Phi = 0.72\)) that reaches the
native conformation rapidly. In this example the native conformation
is attained at \(\approx 325 \tau _{L}\). Comparison of this figure
with Fig. (5) (for sequence G with \(\Phi > 0.9\)) shows that the
dynamics is very similar. This implies that the underlying mechanism
(NCNC mechanism) of fast folding trajectories of sequences with large
\(\sigma \) (or equivalently small \(\Phi \)) is similar to that by
which the molecules reach the native state in kinetically two
state manner. Horizontal arrow indicates the regions of
CBA's. Vertical arrow indicates the first passage time.

Fig. 9. Correlation between the fraction of fast folding trajectories 
\(\Phi \) and the parameter \(\sigma = (T_{\theta } - T_{F})/T_{\theta
}\). Most sequences with small \(\sigma \) have 
\(\Phi \approx 1.0\). Vertical dashed line shows classification of
sequences with respect to \(\Phi \). (a) is for low friction value.  
(b) corresponds to moderate friction. These sequences with \(\Phi
\gtrsim  0.9\) are classified as fast folders. The classification of
sequences into slow category is somewhat arbitrary. The
classification does not seem to depend on the value of \(\zeta \).

Fig. 10. (a) Histogram of states \(g \) as a 
function of two variables,
\(E_{p}\) and \(\chi \), is given for sequence G. (b) The contour plot
of the histogram of states \(g\) for this sequence. Lighter areas
correspond to peaks of  \(g\). 
Single peak of the histogram of states suggests that at equilibrium 
the chain is completely 
confined to a native basin of attraction. 

Fig. 11. Histogram of states \(g \) and contour plot of
\(g\) for sequence I. These plots reveal two peaks of the histogram of
states that manifests the presence of competing
basin of attraction which makes the partition factor
\(\Phi \) less than unity. 

Fig. 12. Histogram of states \(g \) and contour plot of
\(g\) for sequence E. One can clearly see at least three maximums of
\(g\). These plots illustrates the existence of several competing
basins of attractions (intermediates) 
that gives rise to complex folding kinetics which features a
combination of three stage
multipathway and nucleation collapse mechanisms. 

Fig. 13. The profile of the potential of the mean force \(W\) in terms of
two variables, \(E_{p}\) and \(\chi \), for sequence E. This further
illustrates that the free energy landscape of this sequence features
multiple funnels (basins of attractions). The plane at \(W=3.8\) is
given for eye reference.

Fig. 14. Dependence of the folding time \(\tau _{F}\) on 
the parameter \(\sigma = (T_{\theta } - T_{F})/T_{\theta
}\). It is seen that  \(\tau _{F}\) correlate 
remarkably well with \(\sigma \), so that sequences with small value of \(\sigma \) reach 
native state very rapidly, whereas those characterized by large \(\sigma 
\) fold slowly. Solid lines indicate exponential fit the the data. 
The actual fit to
the data is discussed in the text. Fig. (14a) is for the low friction
value and  Fig. (14b) shows data for the moderate friction limit. 

Fig. 15. An example of a fast folding trajectory that reaches the
native conformation by a NCNC process. This is for sequence G at
\(\zeta _{M}\). This figure shows that the dynamics of the fast
folding trajectory is qualitatively similar to that obtained at 
\(\zeta _{L}\) (see Fig. (5)). In both cases the native conformation
is reached rapidly following the formation of a critical number of
contacts (nucleus) and collapse. The first passage time for this
trajectory is \(603 \tau _{L}\). Dashed line gives the equilibrium
value of \(<\chi > =0.26 \) at \(T_{s}\). 
Vertical arrow indicates the first passage time.

Fig. 16. (a) An example of a slow folding trajectory as recorded by  
\(\chi (t)\)
(sequence E) at \(\zeta _{M}\). After initial rapid collapse on the
time scale 
\(\lesssim 1000\tau _{L}\) the chain  
samples various compact conformations and finally 
reaches the native state at \(12919\tau _{L} \) as indicated by an 
arrow. This trajectory shows that at least four distinct kinetic
structures are sampled as the chain navigates to the native
conformation. 
Dashed line in both panels  indicates the equilibrium value of
\(<\chi > = 0.26 \) at \(T_{s}\). 
(b)  Typical fast folding trajectory for this sequence. This 
trajectory is similar  to those characteristic of fast folders (see
Fig. (15)). The 
native conformation is found very rapidly at \( 563 \tau _{L}\) as 
indicated by an arrow. The results displayed in Figs. (5-8) and
Figs. (15,16) show that the qualitative aspect of the kinetic
partitioning mechanism is not dependent on the friction coefficient. 
Vertical arrow indicates the first passage time.

\newpage

\begin{table}
\caption{The list of sequences and their parameters studied in simulations}

\begin{tabular}{cccccc}
Sequence label & Sequence & \(\Lambda \) & \(T_{F}\) & \(T_{\theta }\)
& \(\sigma \)  \\ \hline

A & \(B_{9}N_{3}(LB)_{5}\)                 &  0  & 0.20 & 0.58 & 0.65 \\
B & \(B_{8}(NL)_{2}NBLB_{3}(LB)_{2} \)     &  0  & 0.30 & 0.62 & 0.51 \\
C & \(B_{8}NLN_{2}LBLB_{3}(LB)_{2} \)      & 0.3 & 0.38 & 0.72 & 0.47 \\
D & \(B_{9}N_{2}LNB(LB)_{4}   \)           & 0.17& 0.36 & 0.62 & 0.41 \\
E & \(LB_{7}NBNLN(BL)_{2}B_{3}LB \)        & 0.1 & 0.40 & 0.66 & 0.39 \\
F & \(LB_{9}(NL)_{2}NBLB_{3}LB  \)         & 0.3 & 0.46 & 0.76 & 0.39 \\
G & \(LB_{9}(NL)_{2}NBLB_{3}LB  \)         & 0.3 & 0.62 & 0.78 & 0.20 \\
H & \(LB_{9}(NL)_{2}NBLB_{3}LB   \)        & 0.3 & 0.59 & 0.80 & 0.26 \\
I & \(LBNB_{3}LB_{3}N_{2}B_{2}LBLB_{3}LB\) & 0.3 & 0.54 & 0.62 & 0.14 \\ \hline

\end{tabular}
\end{table}

\newpage

\hspace{10cm}
\begin{center}
\begin{minipage}{18.5cm} 
\[
\psfig{figure=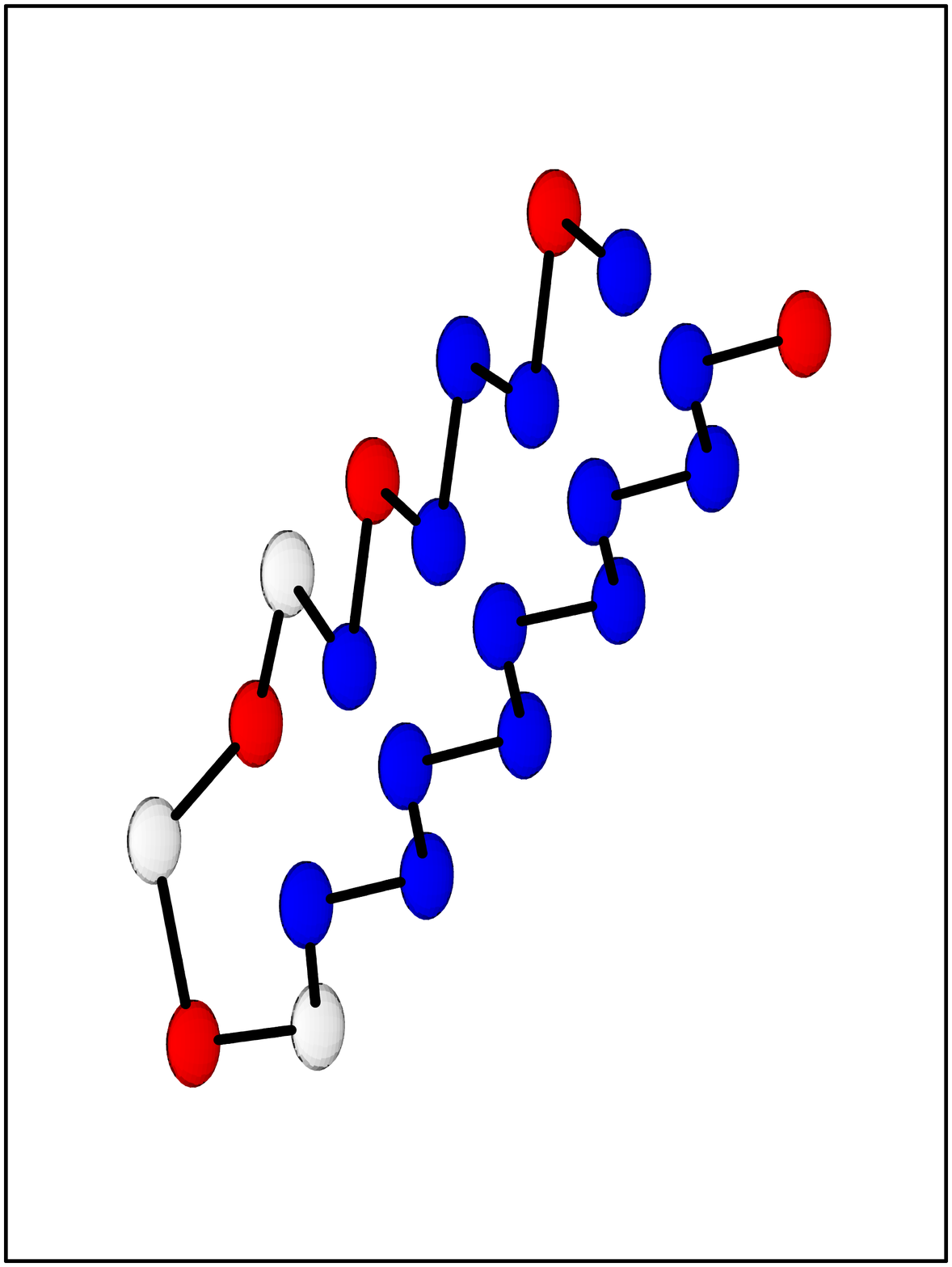,height=13.5cm,width=18.5cm}
\]
\end{minipage}   
{\bf \large Fig. 1}\\
\end{center}

\newpage

\begin{center}
\begin{minipage}{15cm}
\[
\psfig{figure=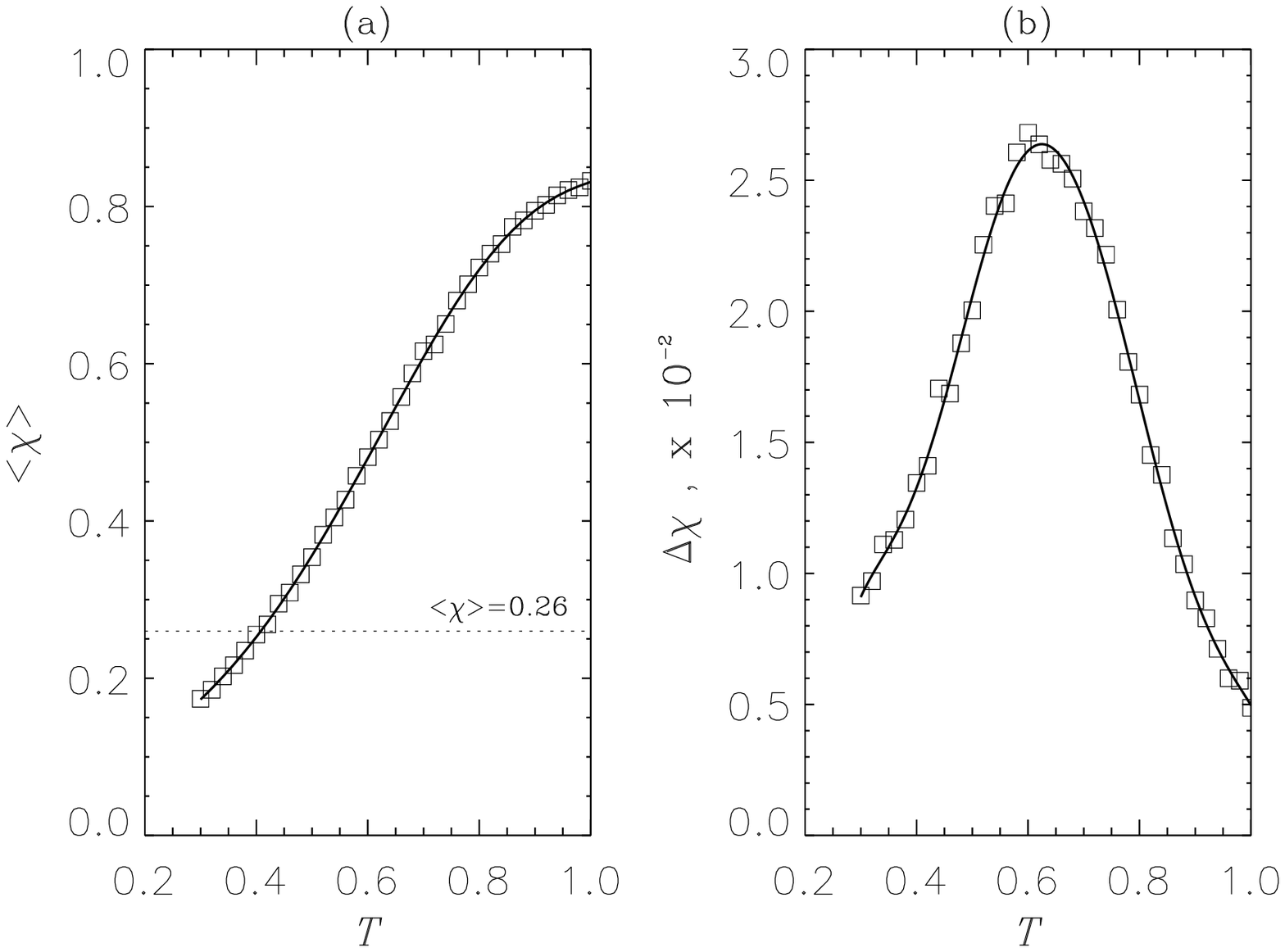,height=10cm,width=14cm}
\]
\end{minipage}
\end{center}

\begin{center}
\begin{minipage}{15cm}
\[
\psfig{figure=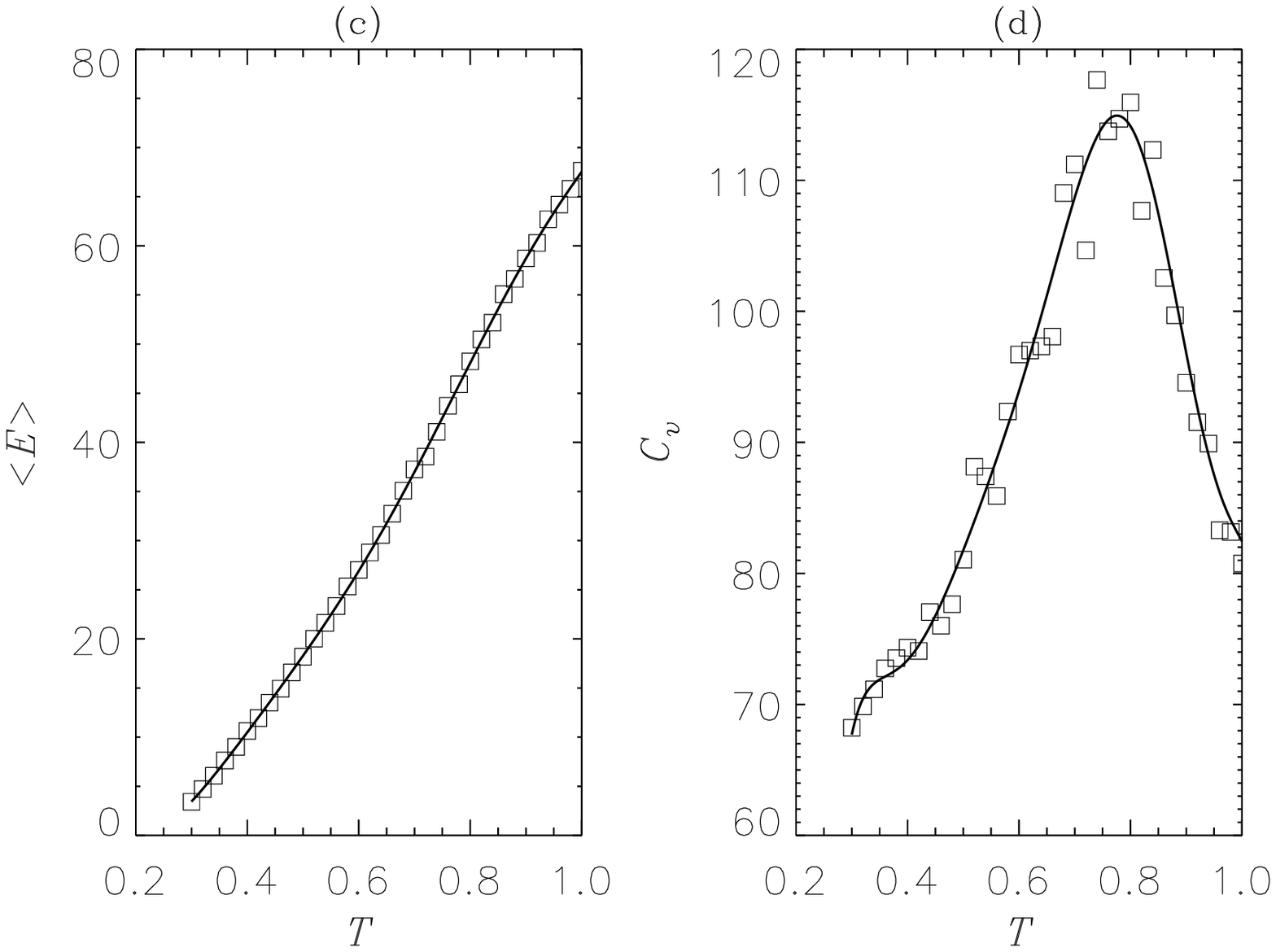,height=10cm,width=14cm}
\]
\end{minipage}

{\bf \large Fig. 2}\\
\end{center}

\newpage

\begin{center}
\begin{minipage}{15cm}
\[
\psfig{figure=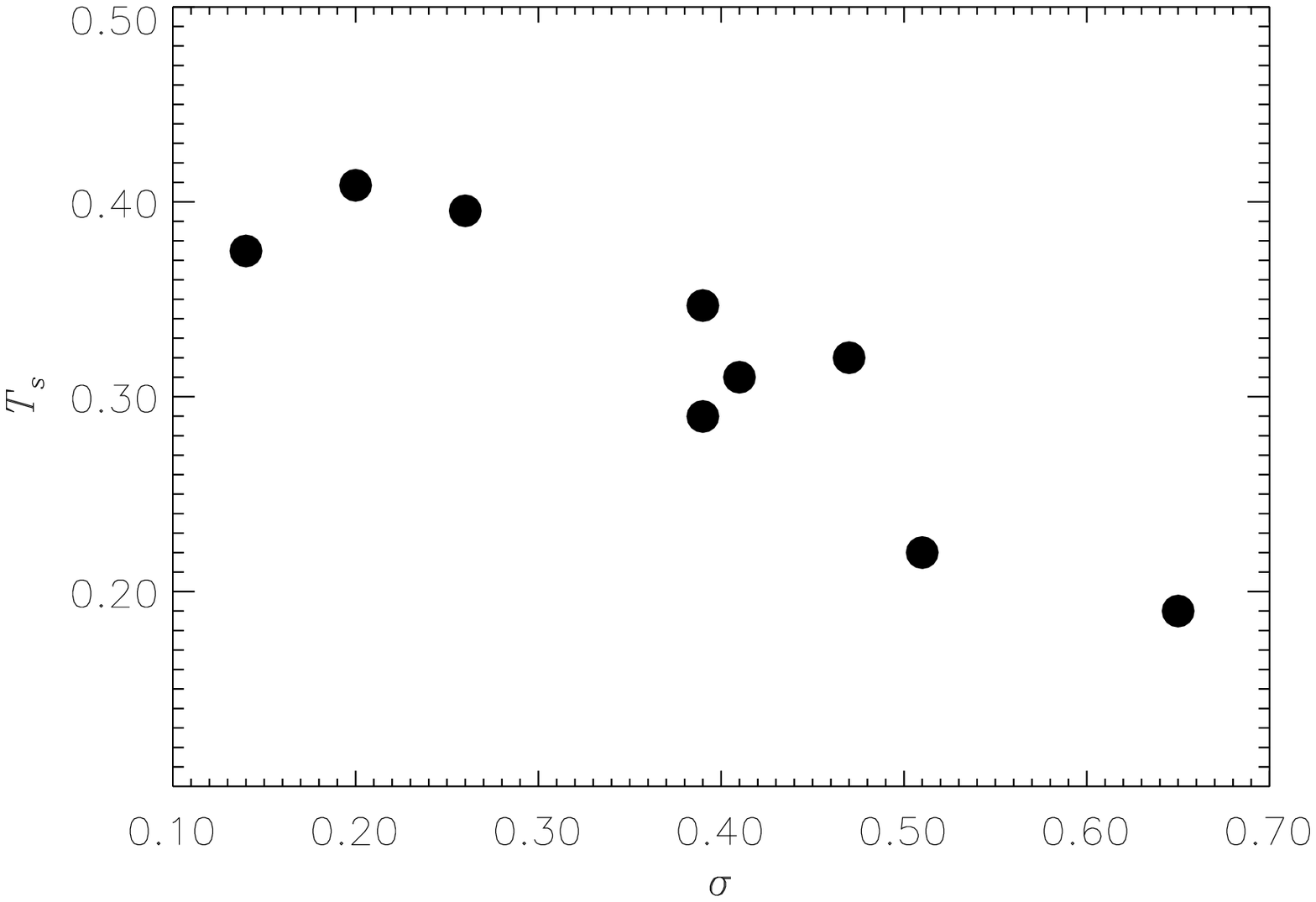,height=10cm,width=14cm}
\]
\end{minipage}

{\bf \large Fig. 3}\\

\end{center}

\newpage 

\begin{center}
\begin{minipage}{15cm}
\[
\psfig{figure=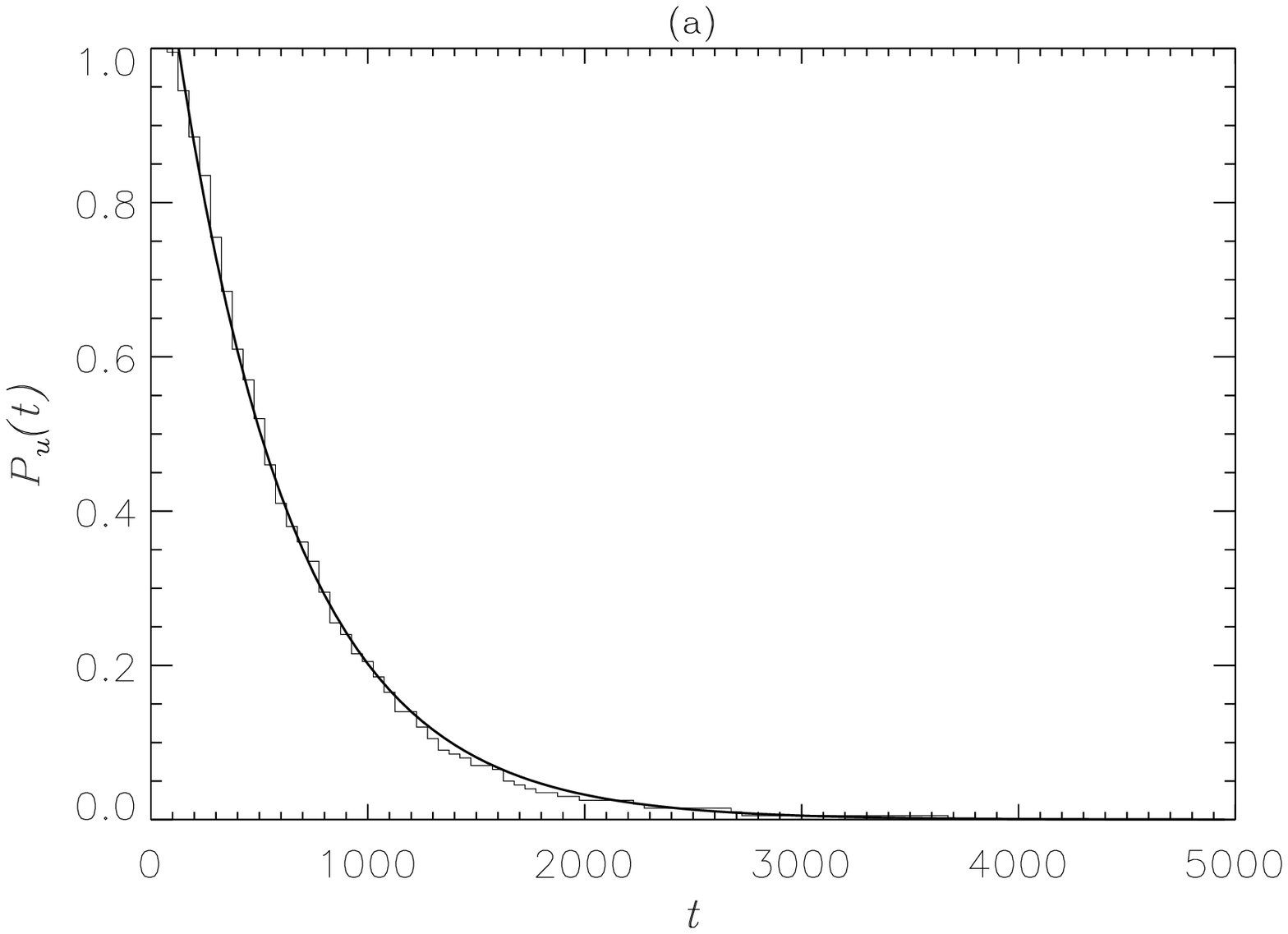,height=10cm,width=14cm}
\]
\end{minipage}

\end{center}

\begin{center}
\begin{minipage}{15cm}
\[
\psfig{figure=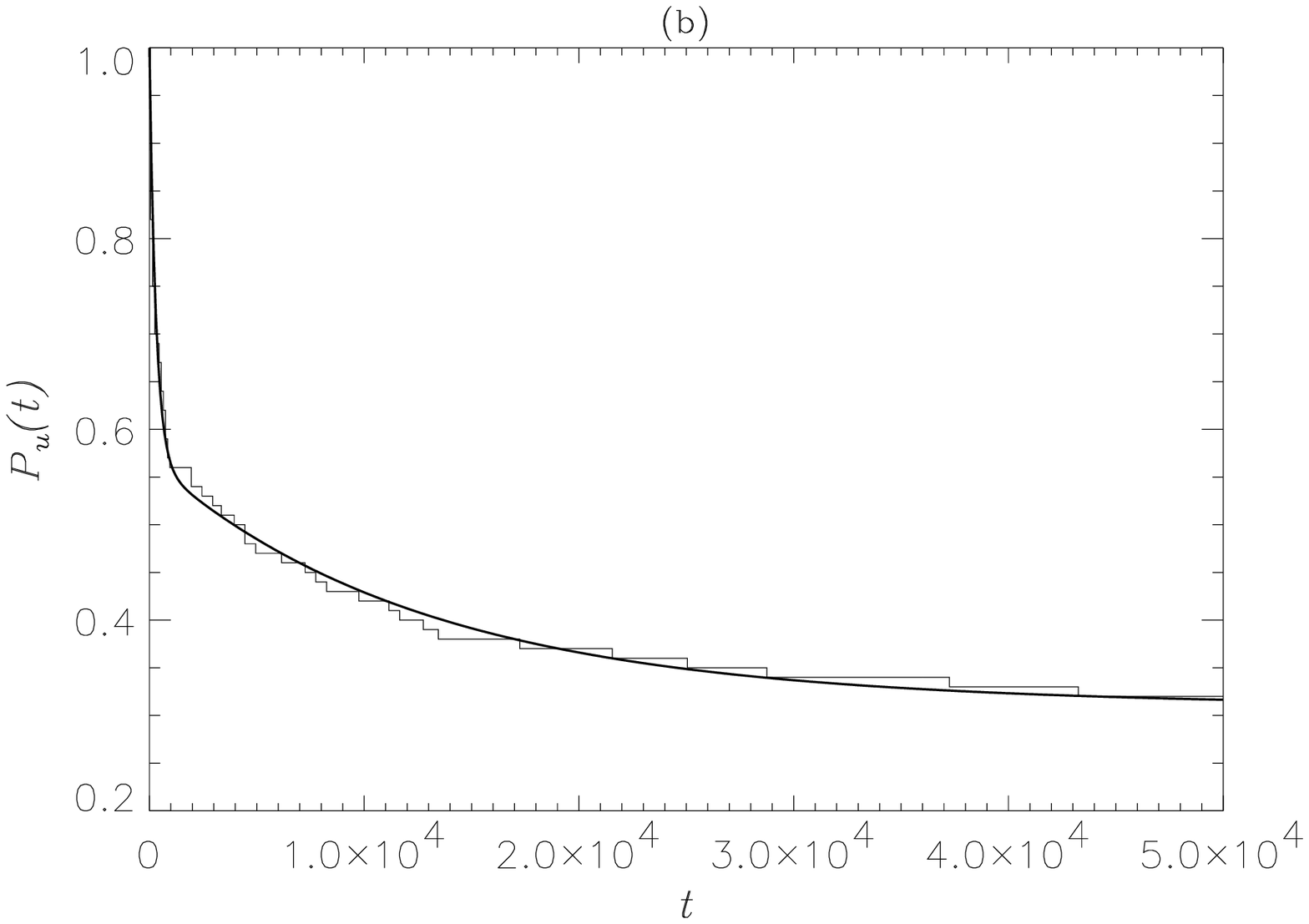,height=10cm,width=14cm}
\]
\end{minipage}
  
{\bf \large Fig. 4}\\
\end{center}
 
\newpage

\begin{center}
\begin{minipage}{15cm}
\[
\psfig{figure=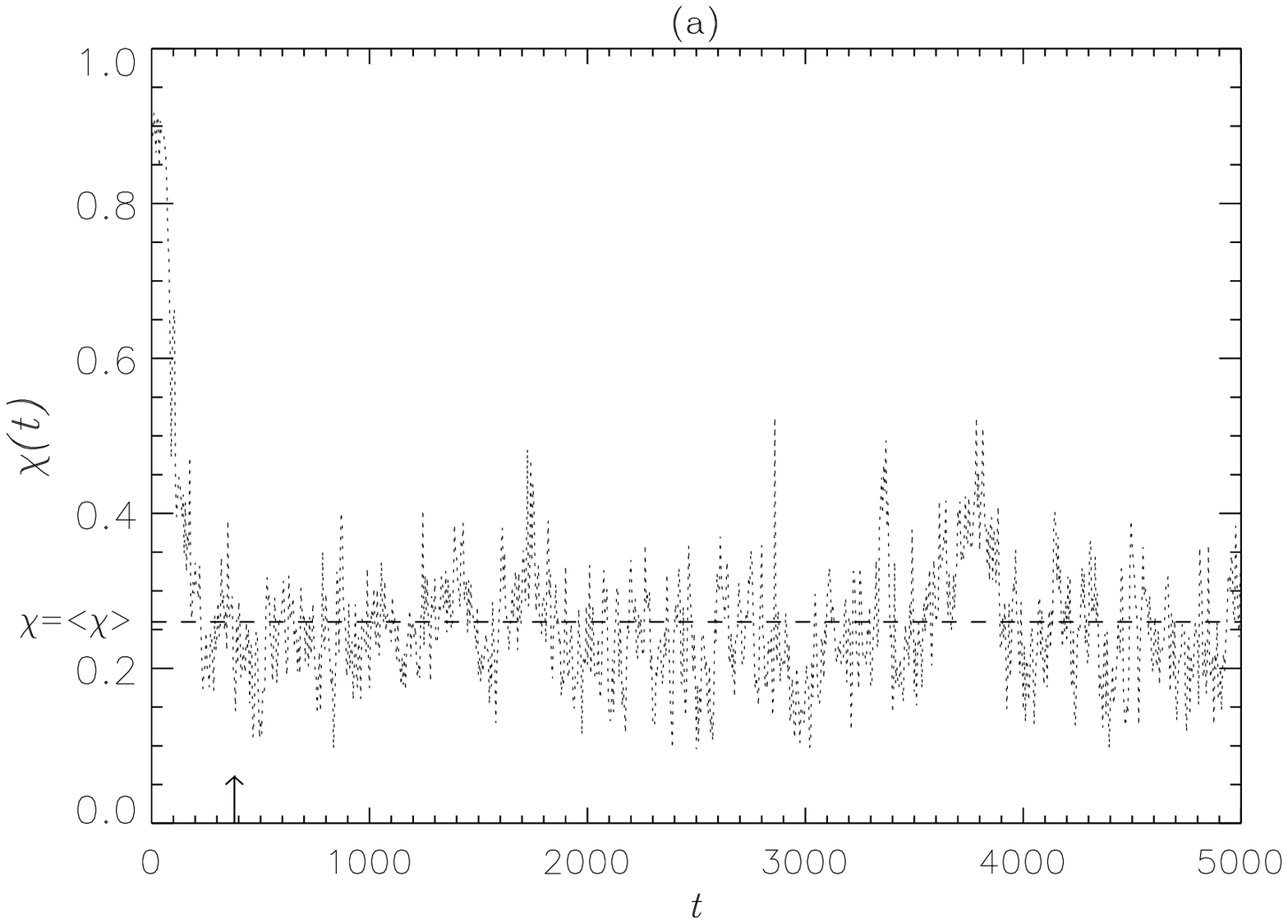,height=10cm,width=14cm}
\]
\end{minipage}

\end{center}

\begin{center}
\begin{minipage}{15cm}
\[
\psfig{figure=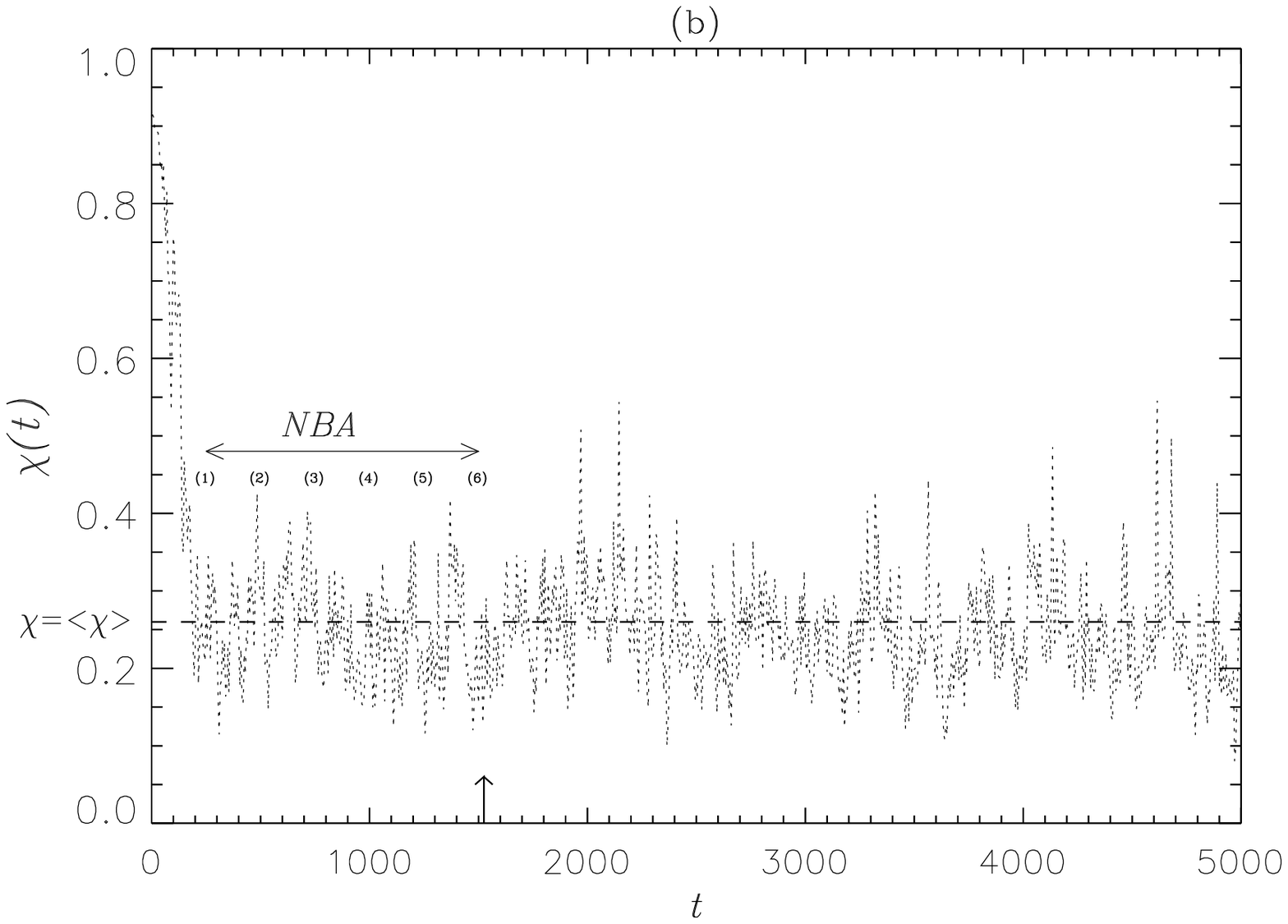,height=10cm,width=14cm}
\]
\end{minipage}

{\bf \large Fig. 5}\\

\end{center}

\newpage

\begin{center}
\begin{minipage}{15cm}
\[
\psfig{figure=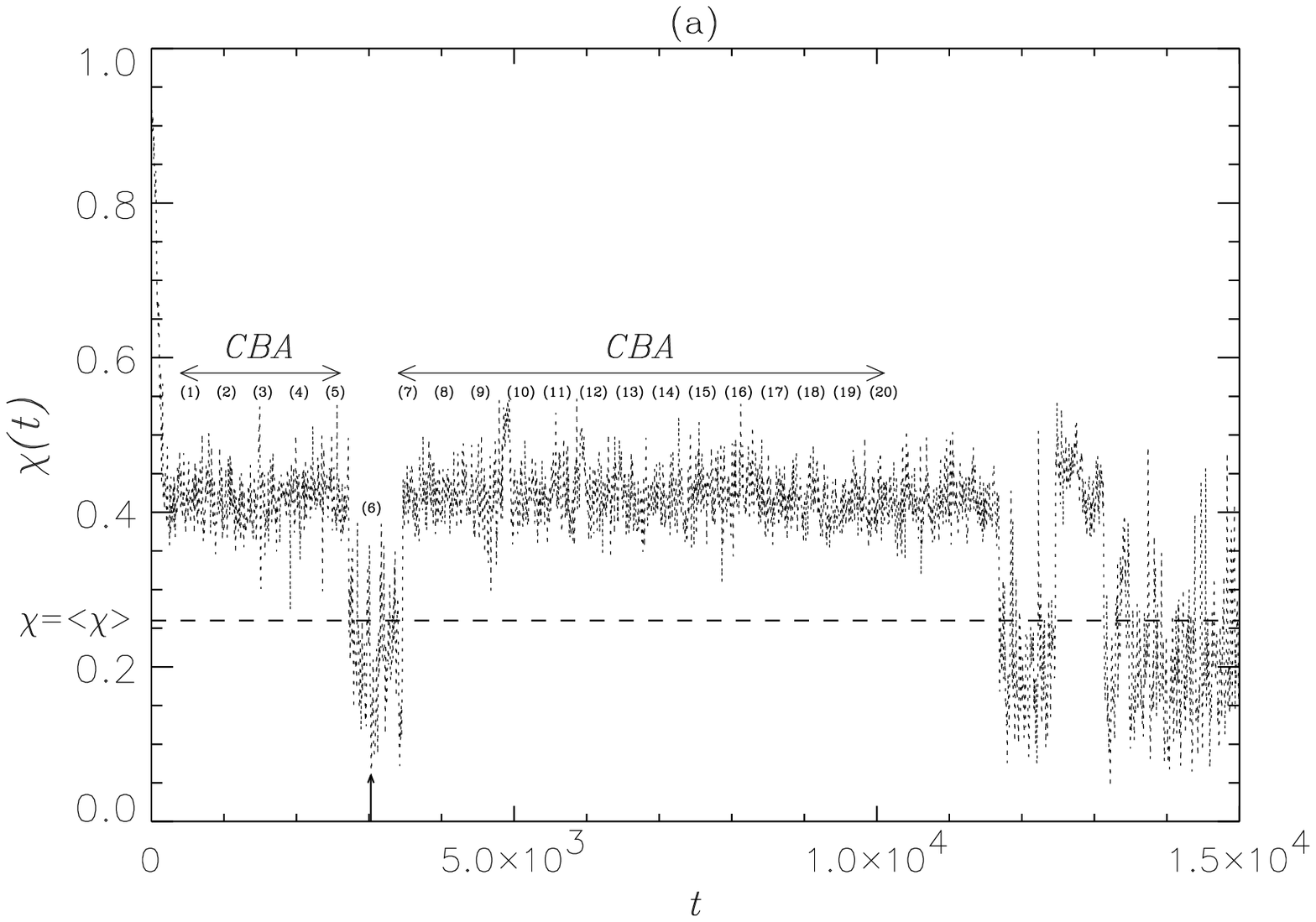,height=10cm,width=14cm}
\]
\end{minipage}

\end{center}

\begin{center}
\begin{minipage}{15cm}
\[
\psfig{figure=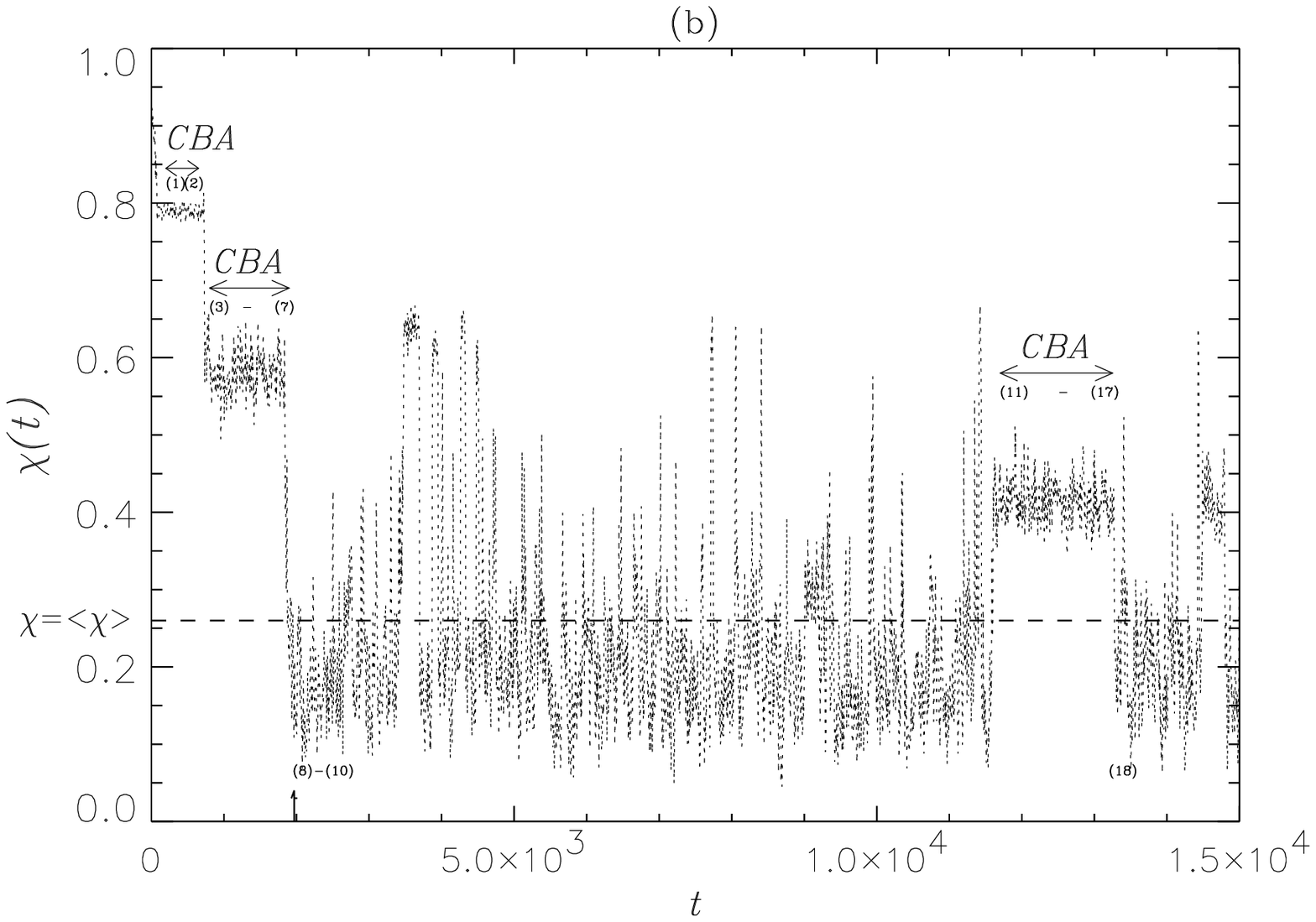,height=10cm,width=14cm}
\]
\end{minipage}

{\bf \large Fig. 6}\\

\end{center}

\newpage
\begin{center}
\begin{minipage}{15cm}
\[
\psfig{figure=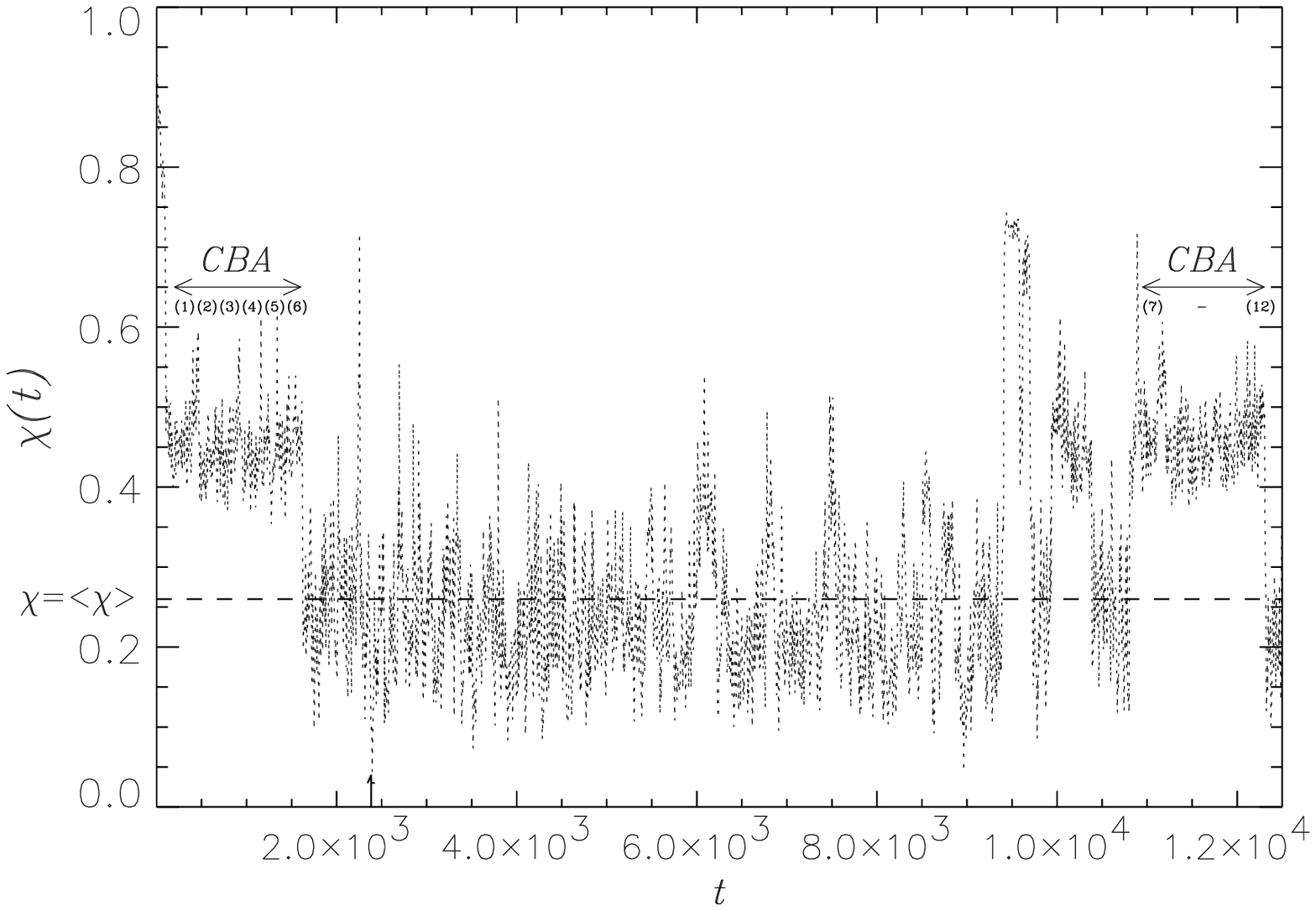,height=10cm,width=14cm}
\]
\end{minipage}

{\bf \large Fig. 7}\\

\end{center}

\begin{center}
\begin{minipage}{15cm}
\[
\psfig{figure=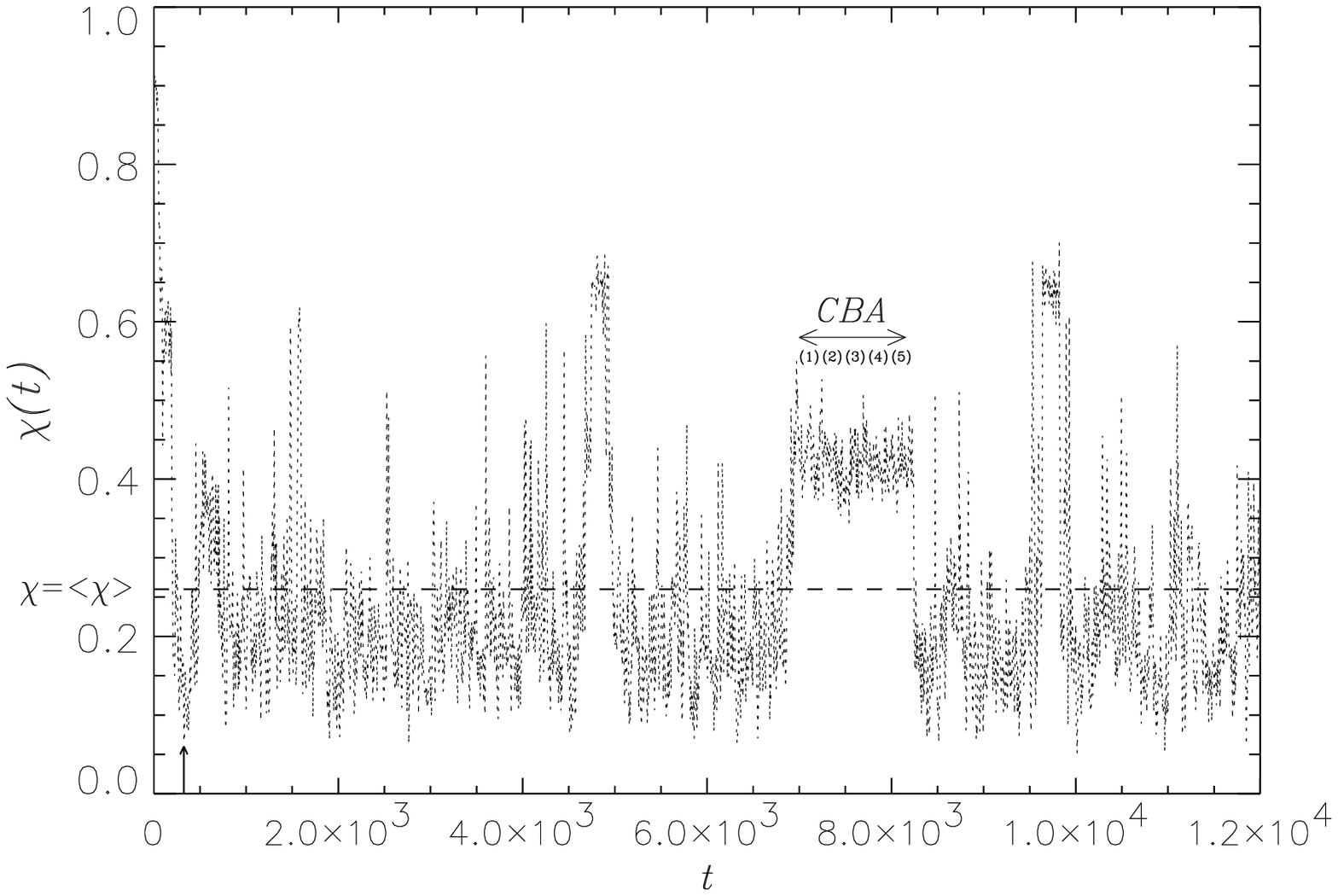,height=10cm,width=14cm}
\]
\end{minipage}

{\bf \large Fig. 8}\\

\end{center}

\newpage

\begin{center}
\begin{minipage}{15cm}
\[
\psfig{figure=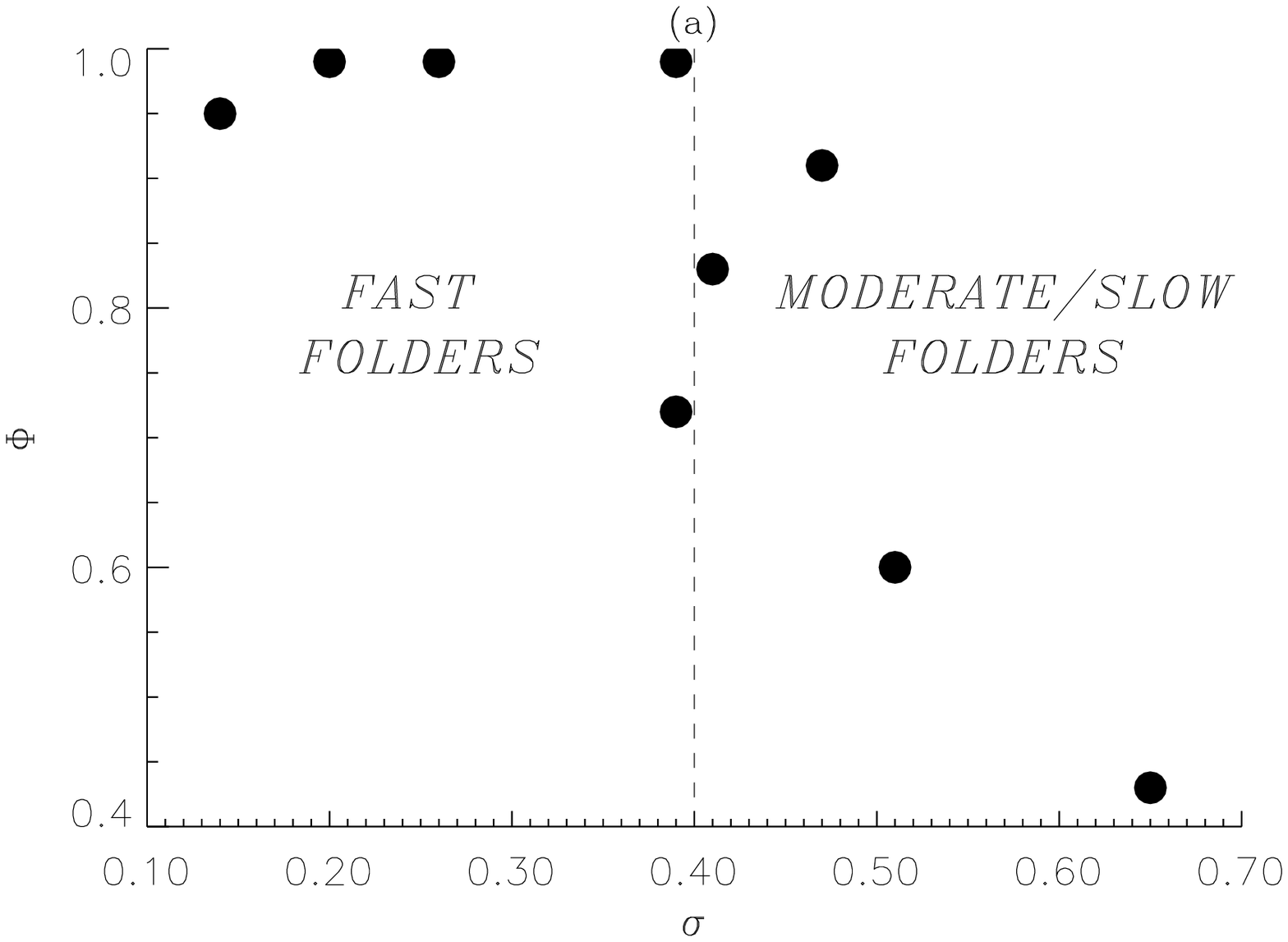,height=10cm,width=14cm}
\]
\end{minipage}
\end{center}

\begin{center}
\begin{minipage}{15cm}
\[
\psfig{figure=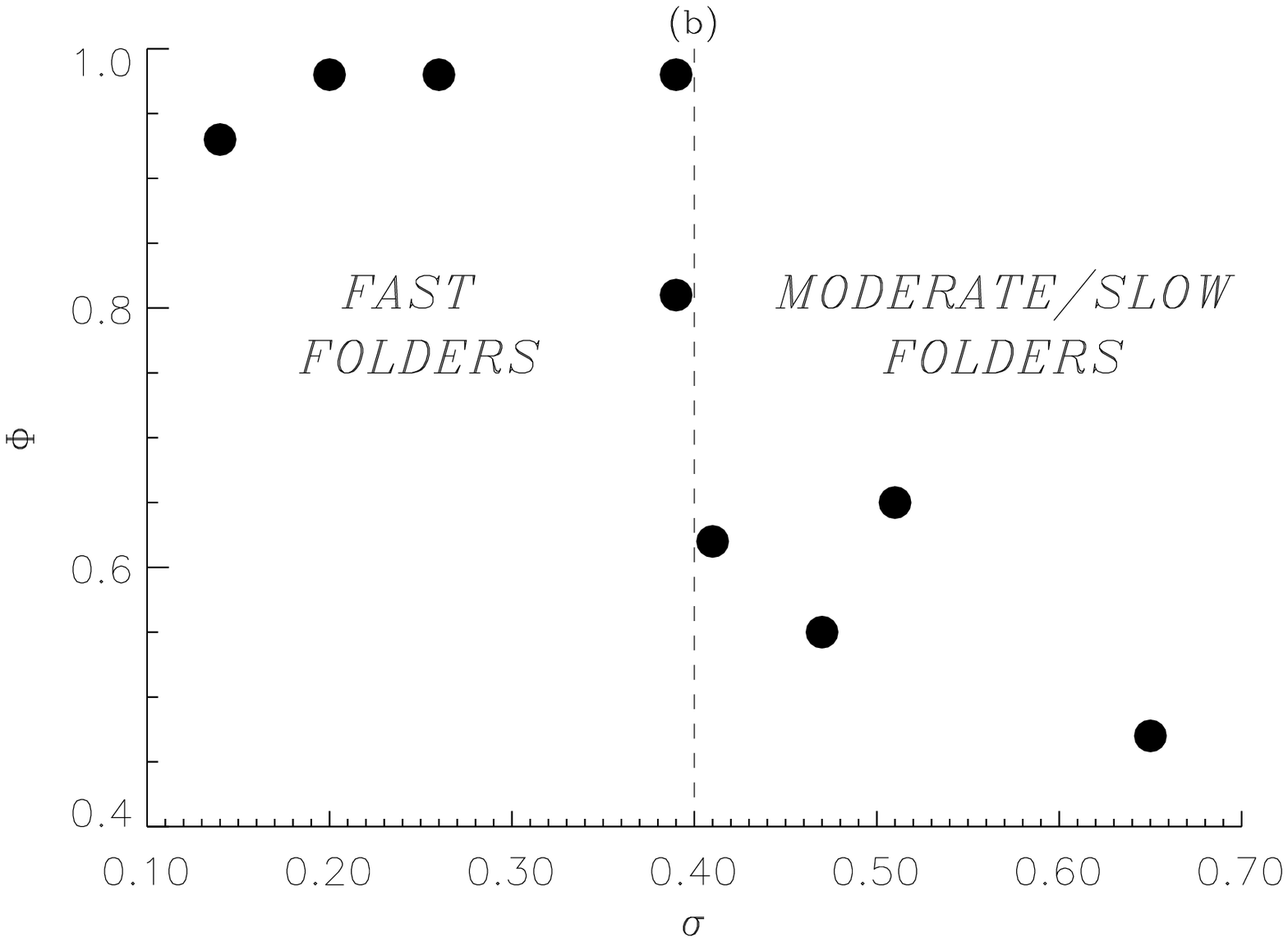,height=10cm,width=14cm}
\]
\end{minipage}

{\bf \large Fig. 9}\\
\end{center}

\newpage

\begin{center}
{\bf \large (a)}\\

\begin{minipage}{15cm}
\[
\psfig{figure=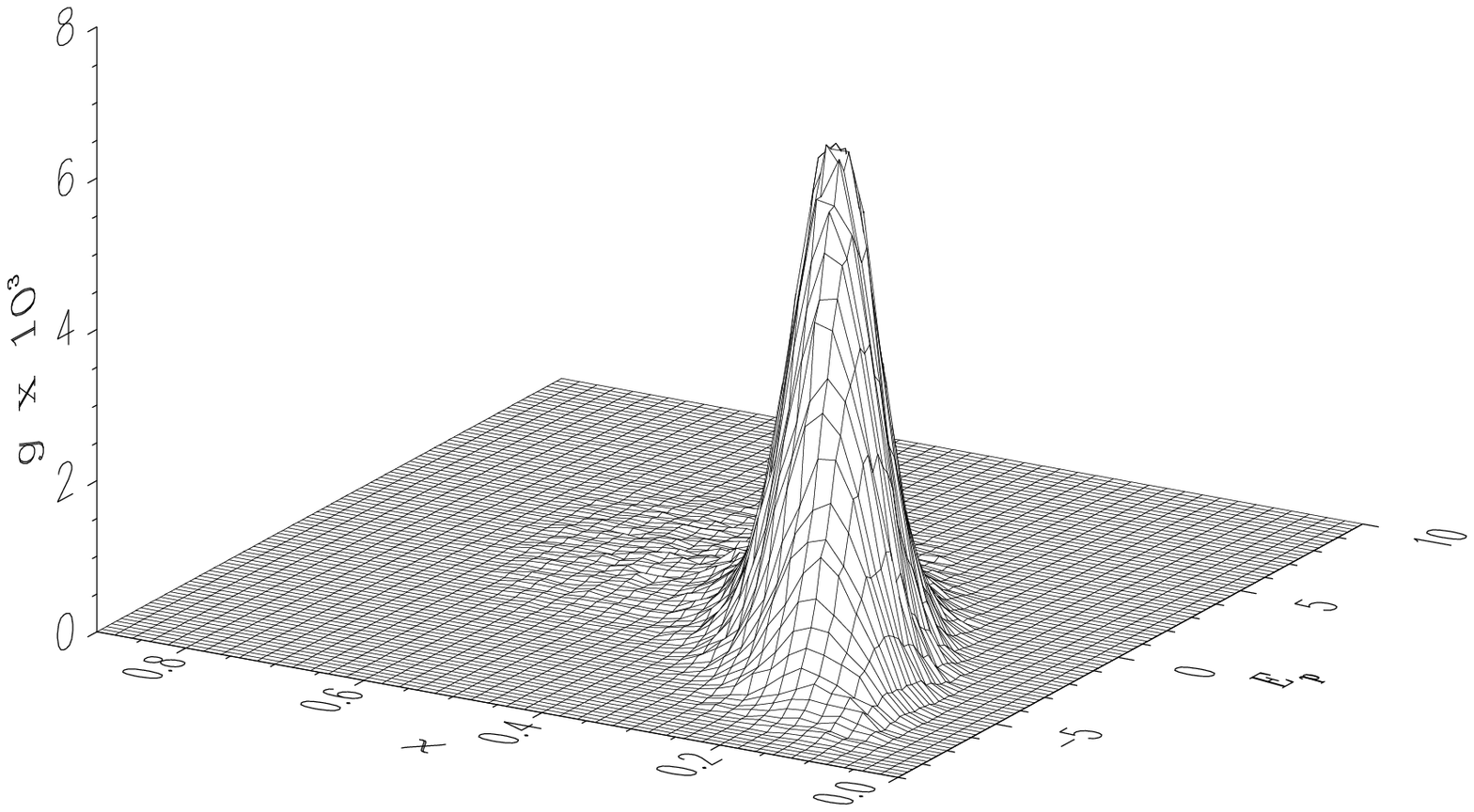,height=9.8cm,width=14cm}
\]
\end{minipage}
\end{center}

\begin{center}
{\bf \large (b)}\\

\begin{minipage}{15cm}
\[
\psfig{figure=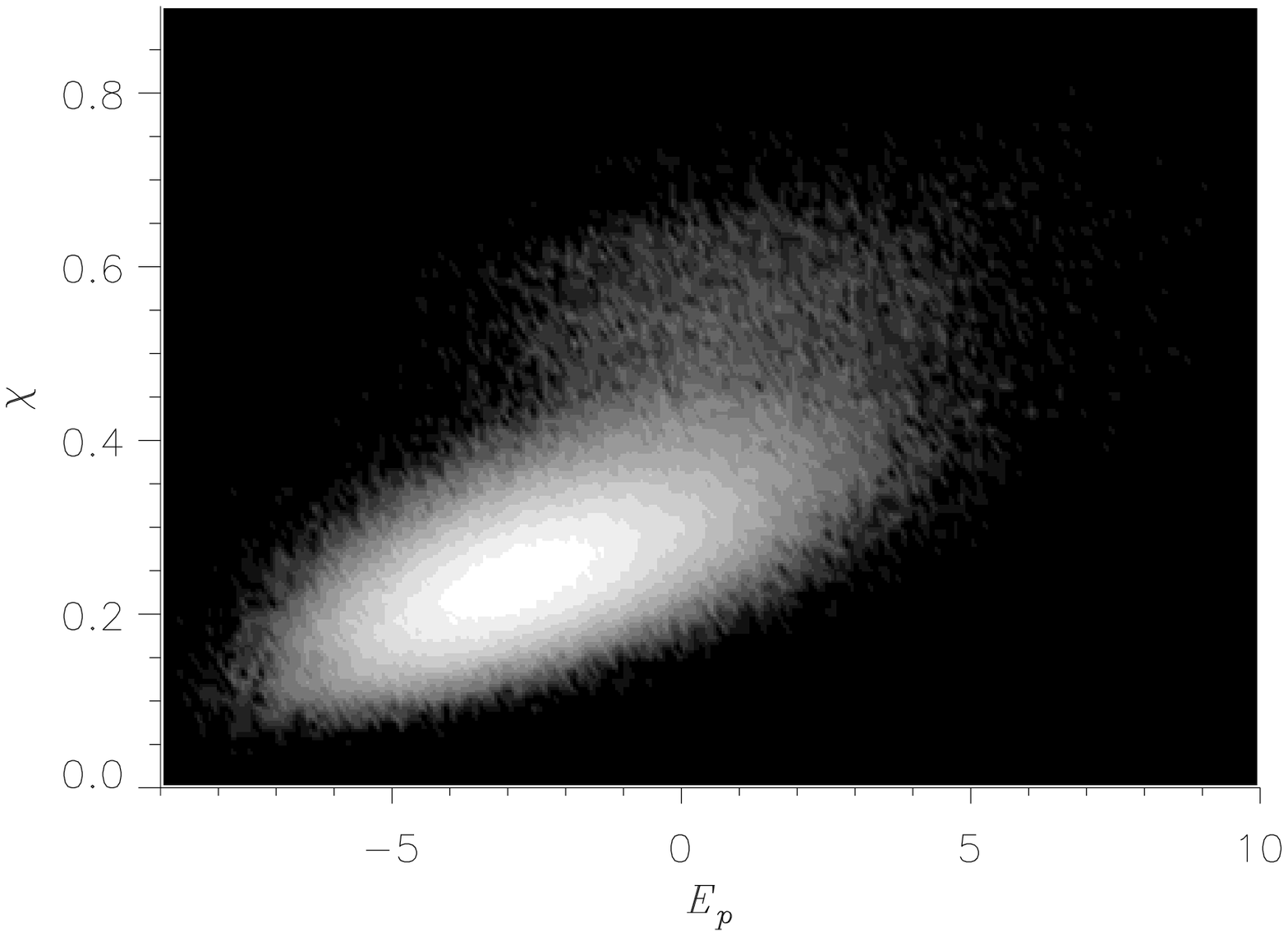,height=9.8cm,width=14cm}
\]
\end{minipage}

{\bf \large Fig. 10}\\
\end{center}

\newpage
\begin{center}
{\bf \large (a)}\\

\begin{minipage}{15cm}
\[
\psfig{figure=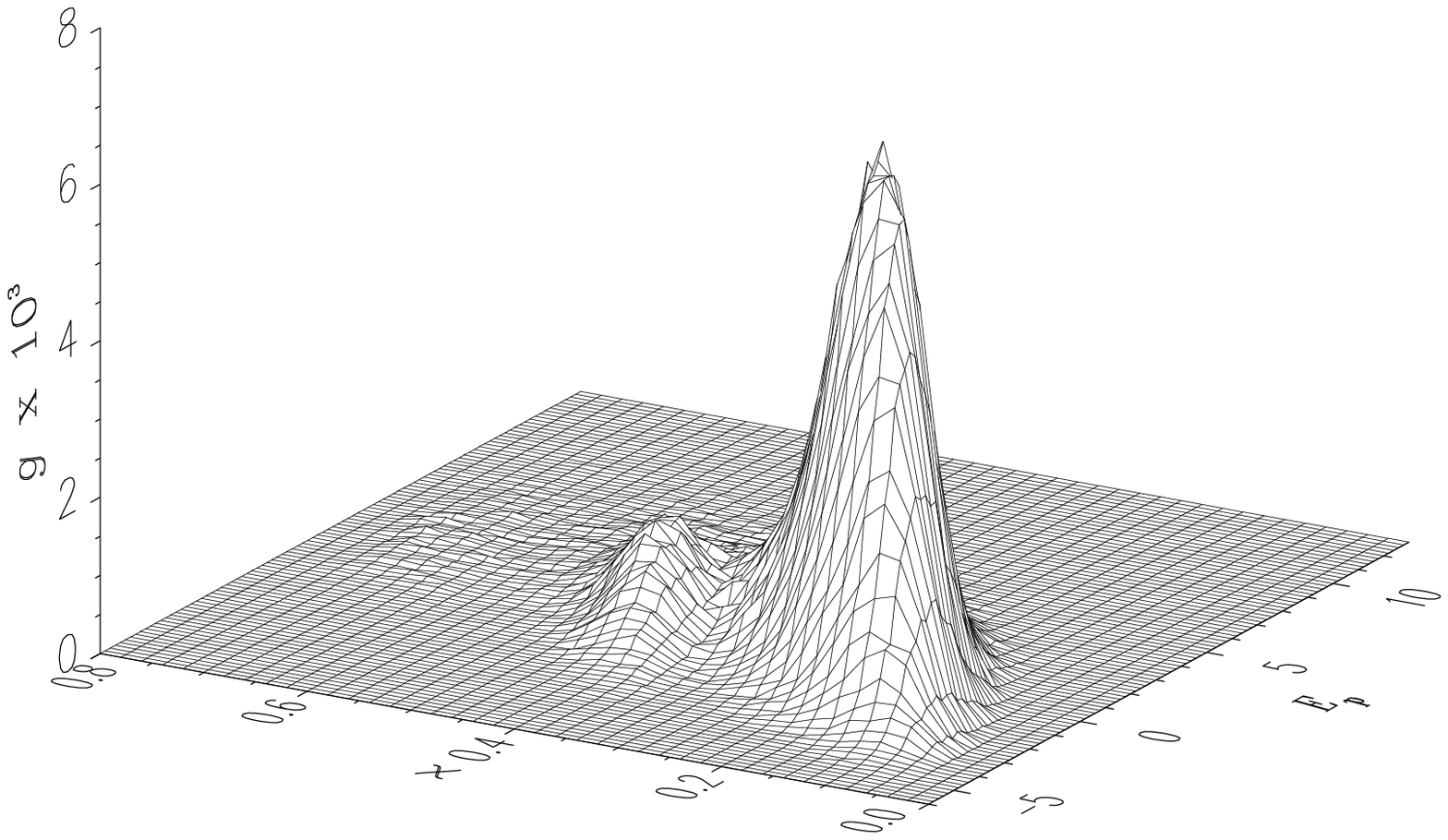,height=9.8cm,width=14cm}
\]
\end{minipage}
\end{center}

\begin{center}
{\bf \large (b)}\\

\begin{minipage}{15cm}
\[
\psfig{figure=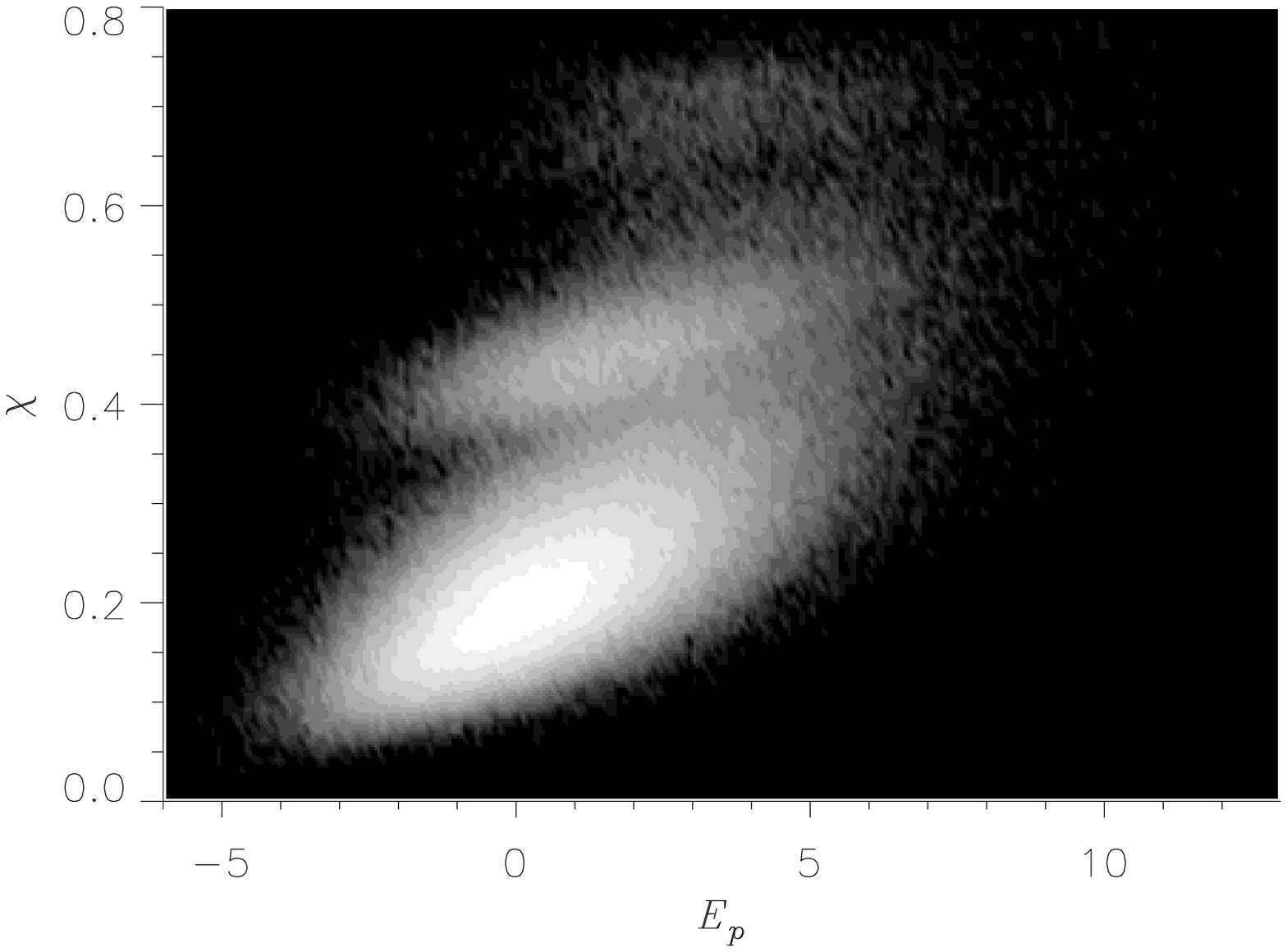,height=9.8cm,width=14cm}
\]
\end{minipage}

{\bf \large Fig. 11}\\
\end{center}

\newpage

\begin{center}
{\bf \large (a)}\\

\begin{minipage}{15cm}
\[
\psfig{figure=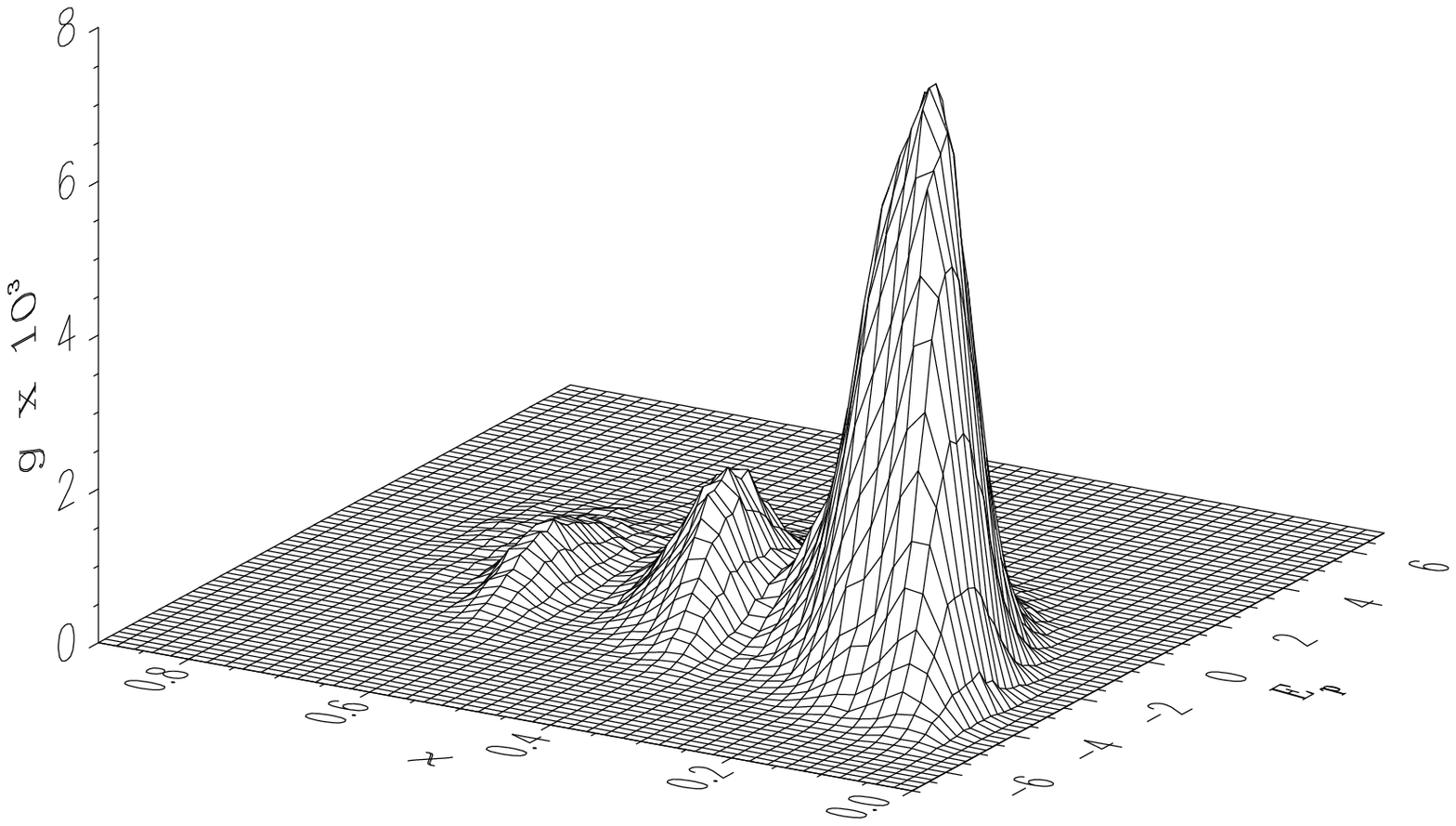,height=9.8cm,width=14cm}
\]
\end{minipage}
\end{center}

\begin{center}
{\bf \large (b)}\\

\begin{minipage}{15cm}
\[
\psfig{figure=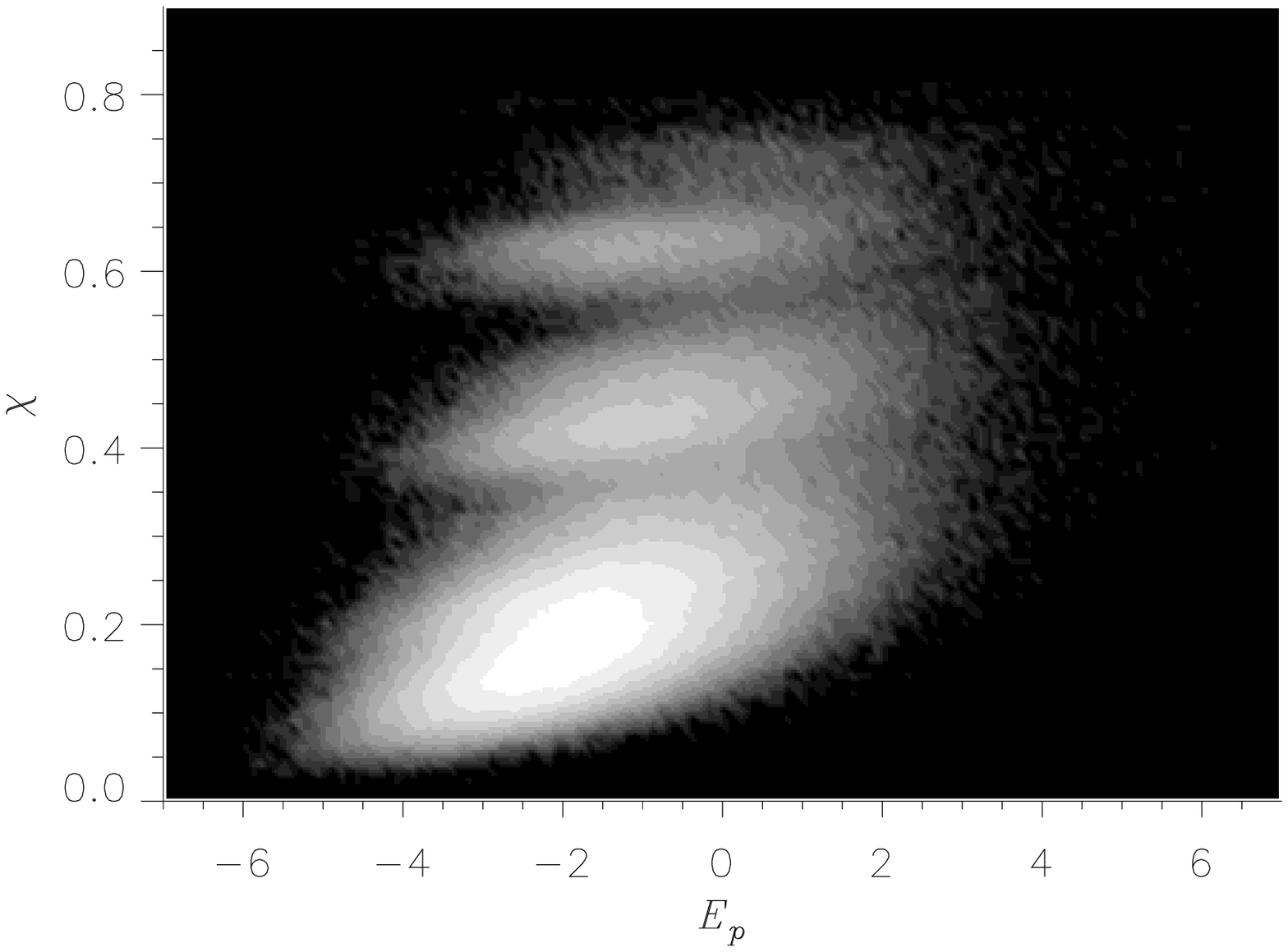,height=9.8cm,width=14cm}
\]
\end{minipage}

{\bf \large Fig. 12}\\
\end{center}

\newpage

\begin{center}

\begin{minipage}{15cm}
\[
\psfig{figure=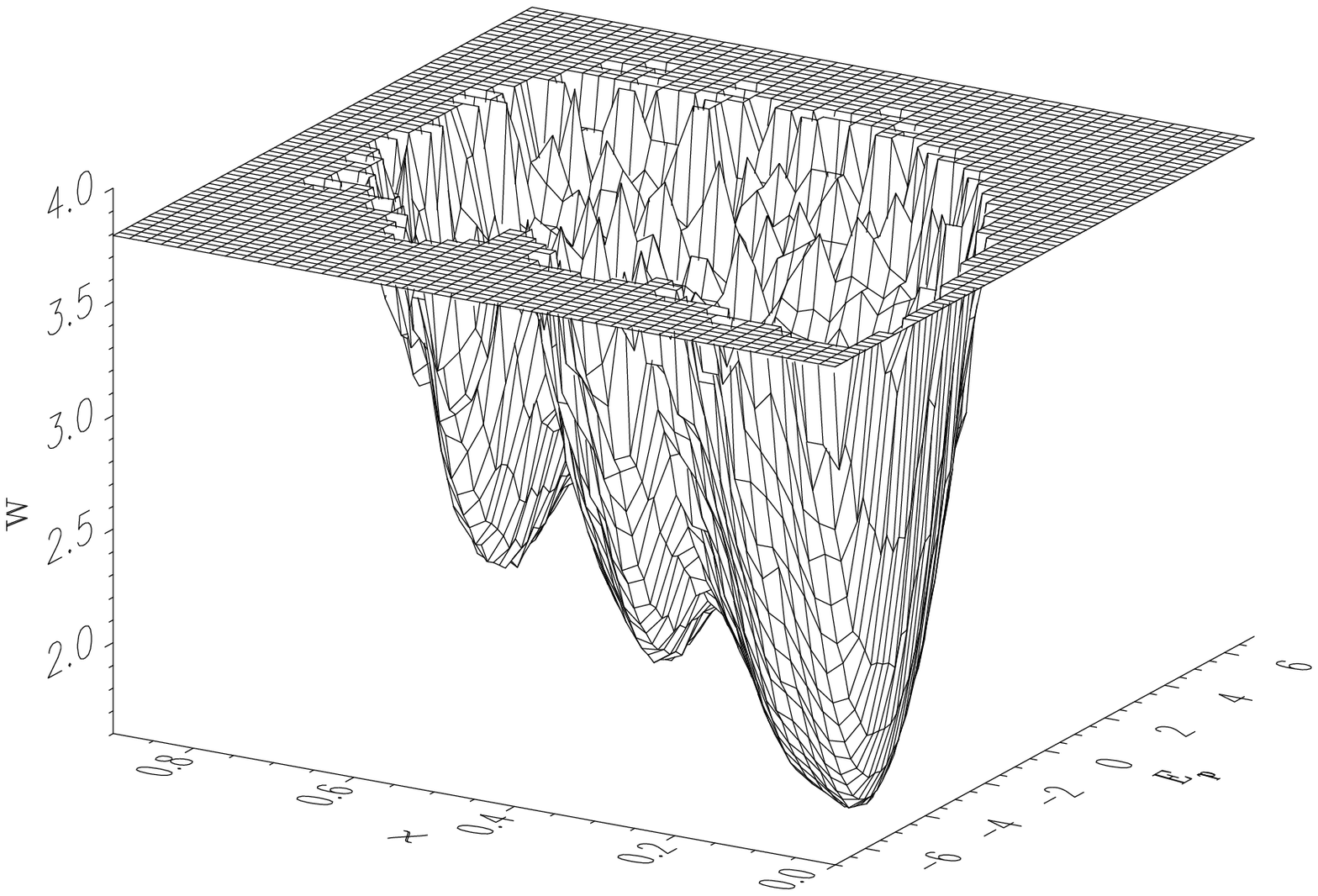,height=10cm,width=14cm}
\]
\end{minipage}

{\bf \large Fig. 13}\\
\end{center}

\newpage

\begin{center}
\begin{minipage}{15cm}
\[
\psfig{figure=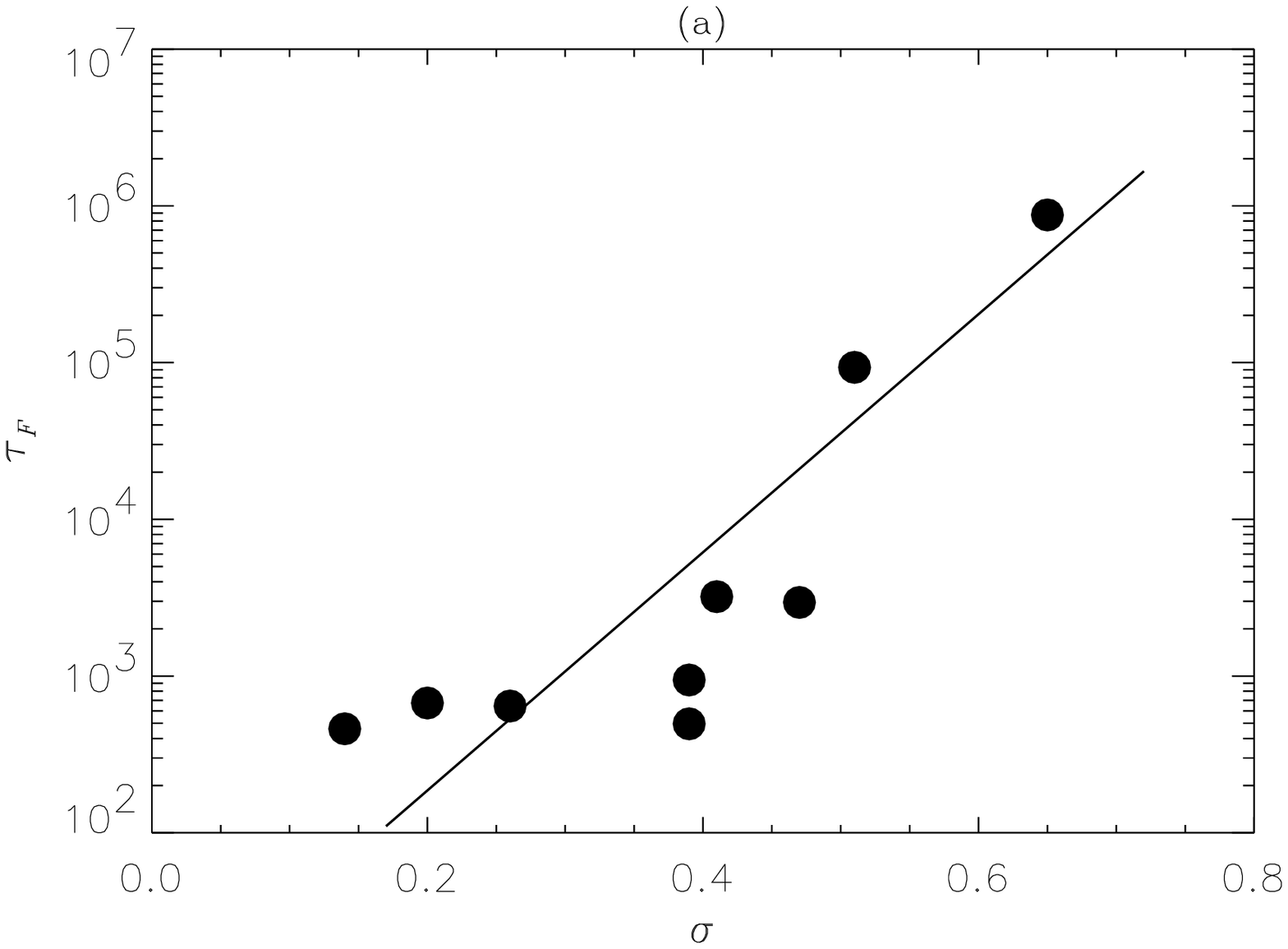,height=10cm,width=14cm}
\]
\end{minipage}

\end{center}

\begin{center}
\begin{minipage}{15cm}
\[
\psfig{figure=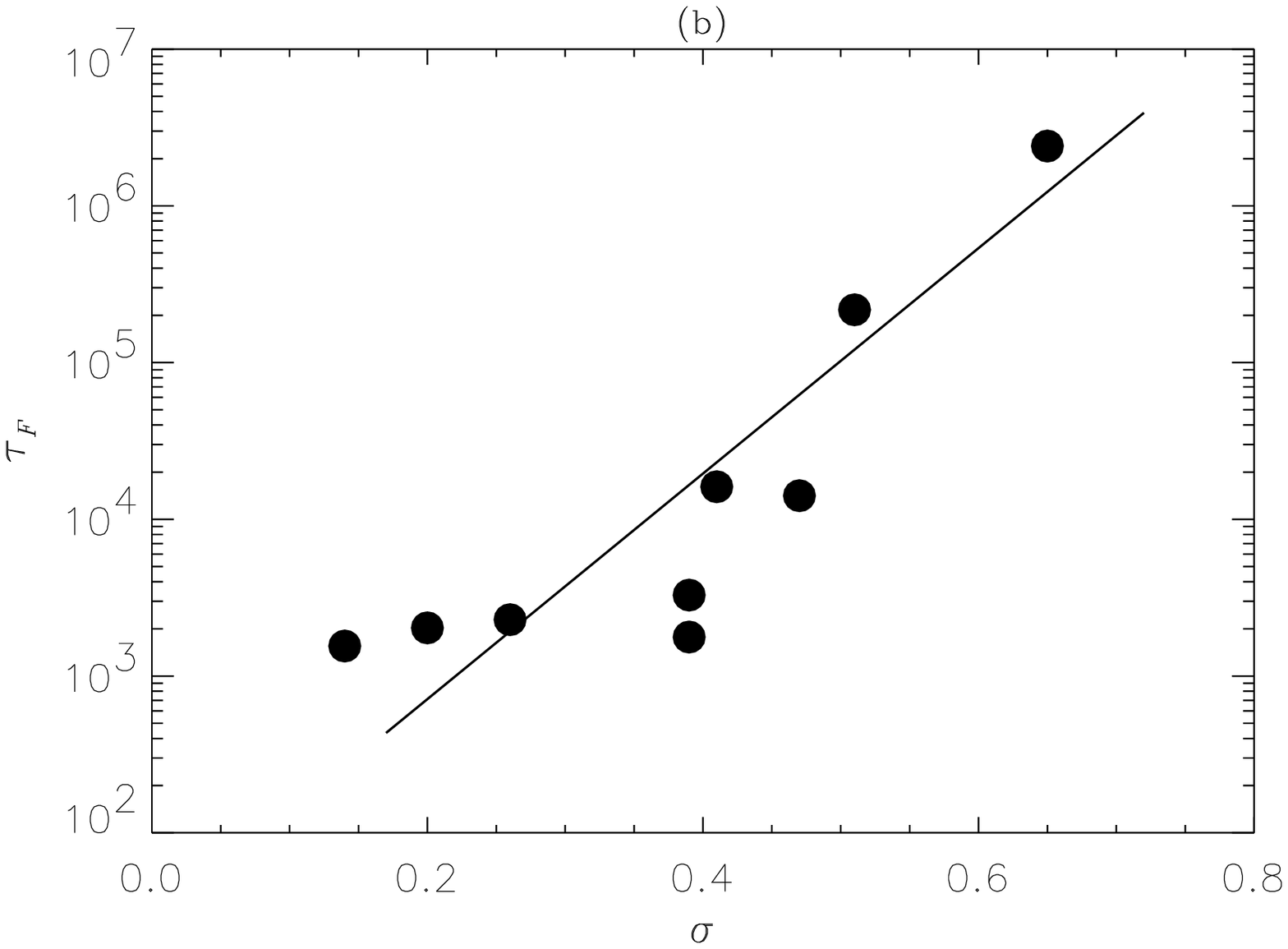,height=10cm,width=14cm}
\]
\end{minipage}

{\bf \large Fig. 14}\\
\end{center}

\newpage

\begin{center}
\begin{minipage}{15cm}
\[
\psfig{figure=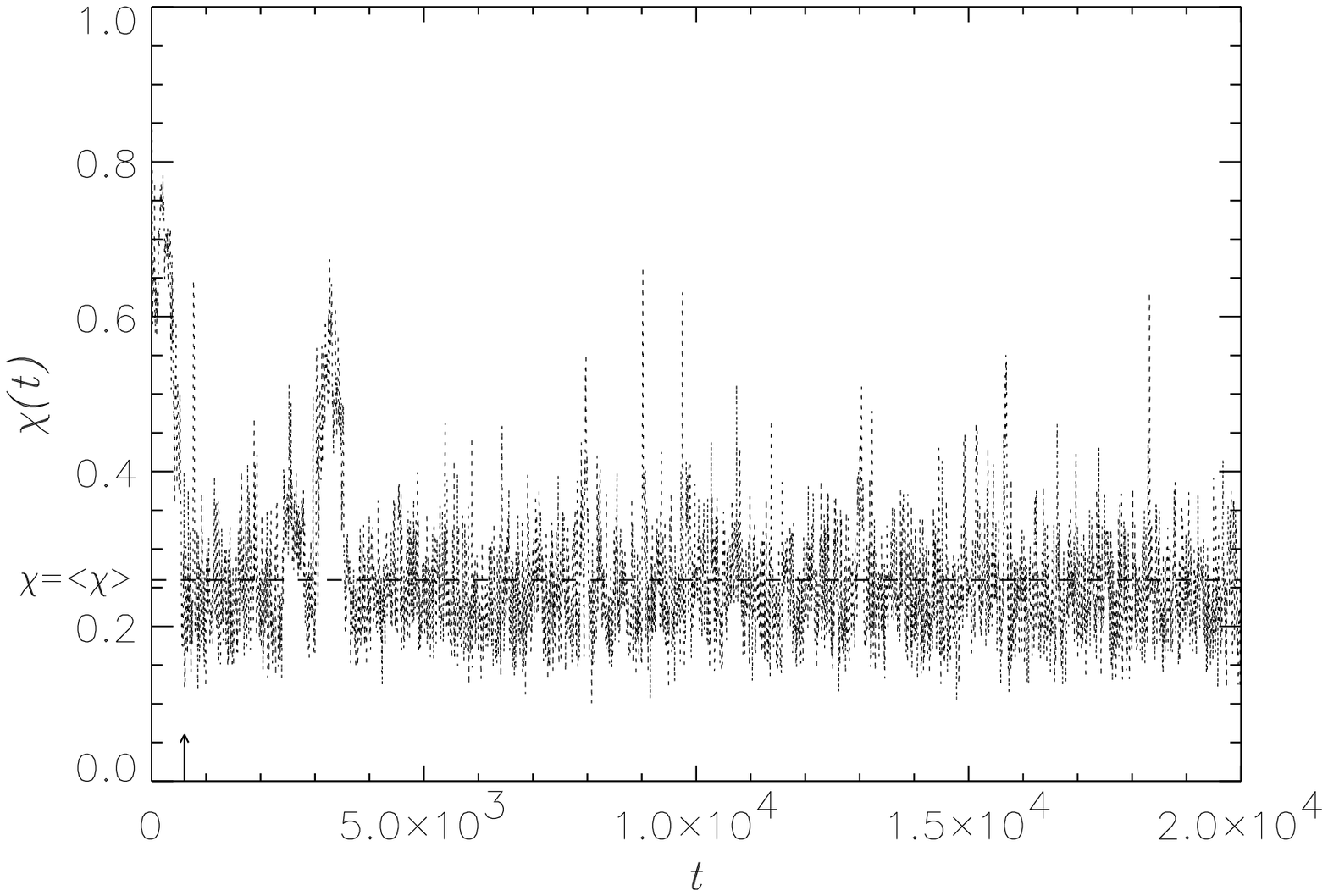,height=10cm,width=14cm}
\]
\end{minipage}

{\bf \large Fig. 15}\\

\end{center}

\newpage

\begin{center}
\begin{minipage}{15cm}
\[
\psfig{figure=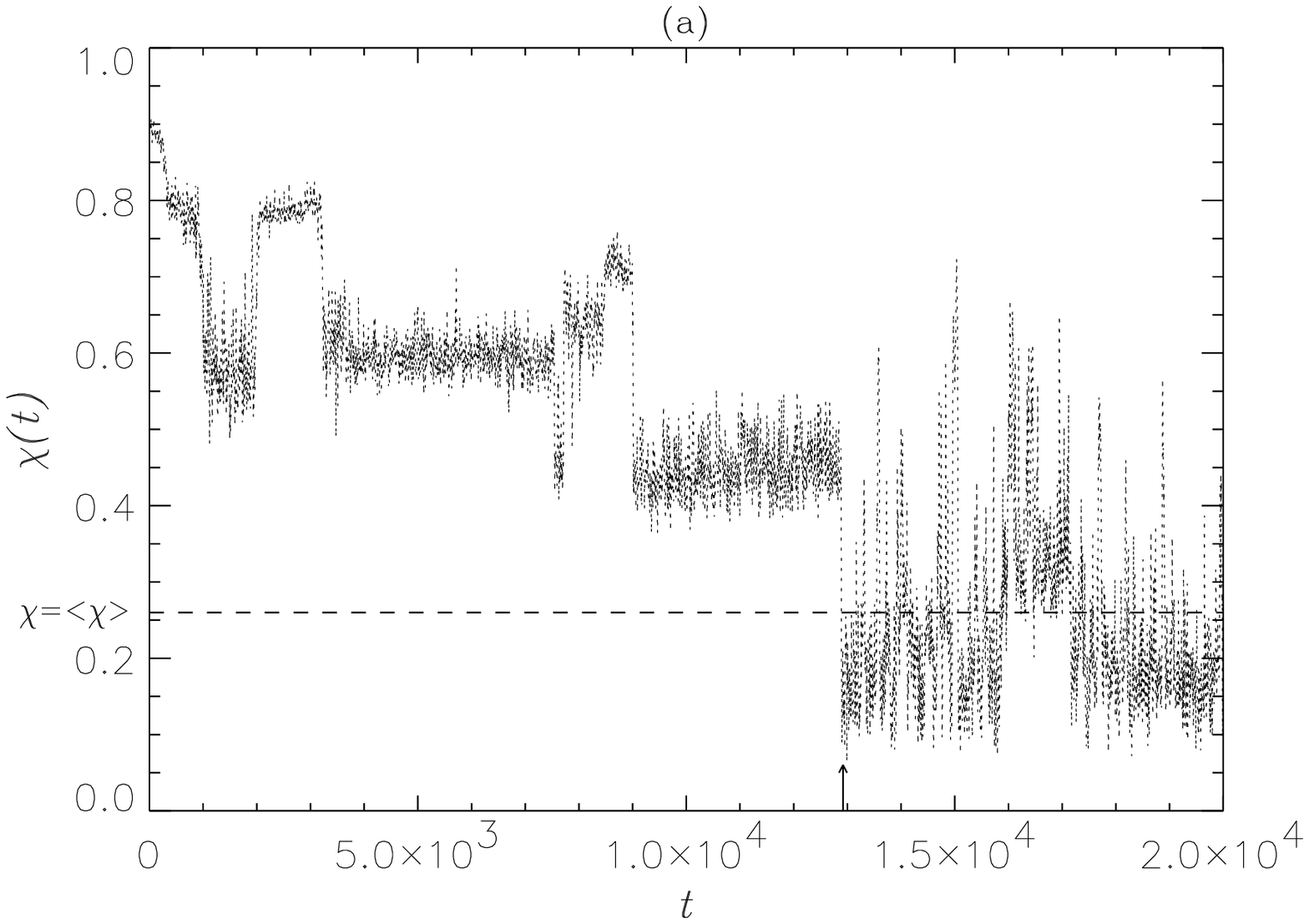,height=10cm,width=14cm}
\]
\end{minipage}

\end{center}

\begin{center}
\begin{minipage}{15cm}
\[
\psfig{figure=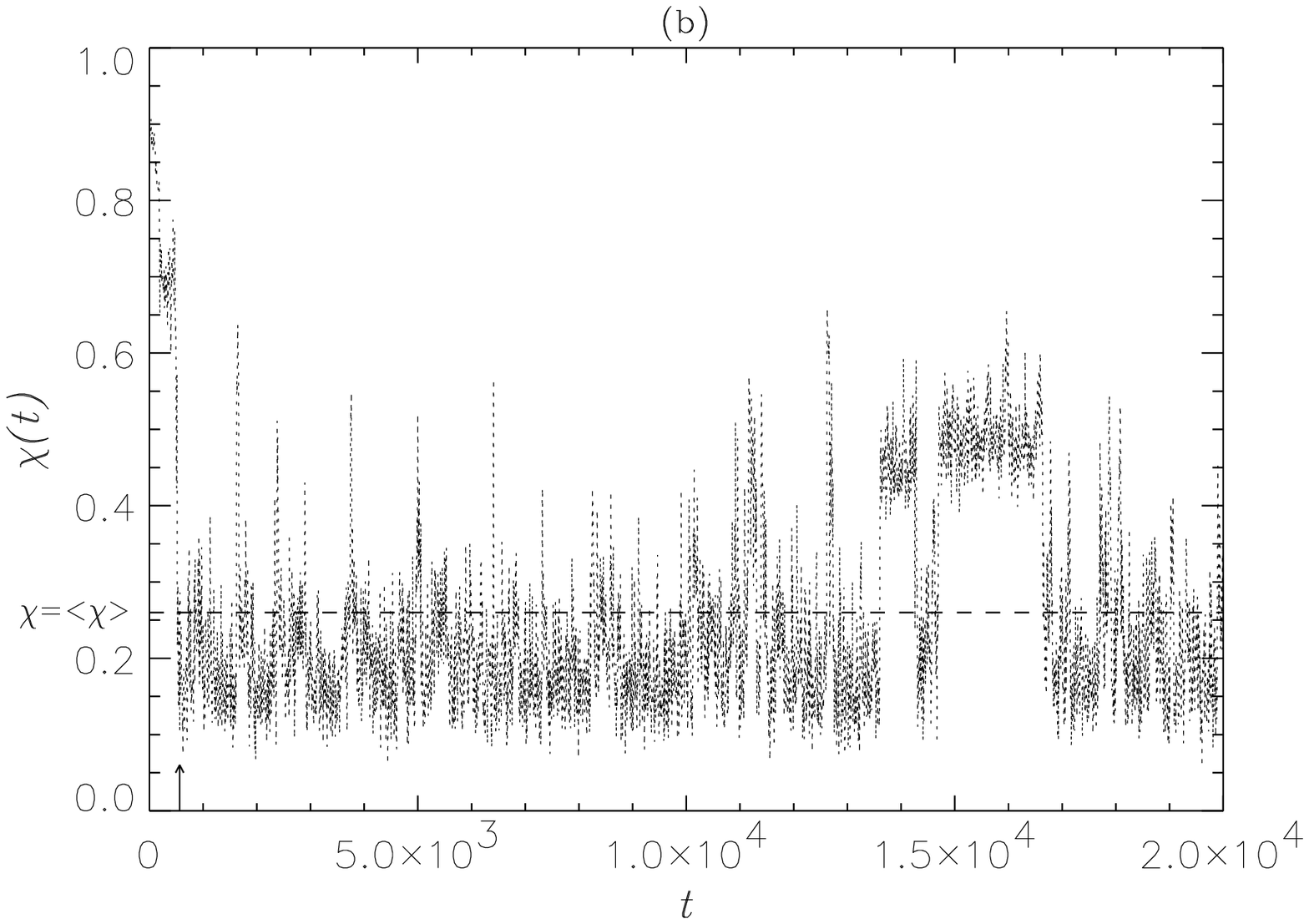,height=10cm,width=14cm}
\]
\end{minipage}

{\bf \large Fig. 16}\\
\end{center}


\begin{references}

\bibitem{Dill} Dill, K.A., Bromberg, S., Yue, K., Fiebig, K.M., Yee, D.P.,
Thomas, P.D. \& Chan, H.S. (1995). Principles of protein folding
 - A perspective from simple exact models. {\em Protein Sci. } {\bf 4},
561-602. 

\bibitem{Bryn95} Bryngelson, J.D., Onuchic, J.N., Socci, N.D. \& Wolynes,
P.G. (1995). Funnels, pathways and the energy landscape of protein folding: A
synthesis. {\em Proteins Struct. Funct. Genet. } {\bf 21 }, 167-195. 

\bibitem{Wol} Wolynes, P.G., Onuchic, J.N. \& Thirumalai, D. (1995). 
Navigating the folding routes. {\em Science } {\bf 267},  1619-1620.


\bibitem{Chan94} Chan, H.S. \& Dill, K.A. (1994). 
Transition states and folding
dynamics of proteins and heteropolymers. {\em J. Chem. Phys. } 
{\bf 100}, 9238-9257. 


\bibitem{Thirum94} Thirumalai, D. (1994) Theoretical perspectives on in
vitro and in vivo protein folding. In: {\em Statistical Mechanics,
Protein Structure, and Protein Substrate Interactions}. (Doniach, S.,   
ed.), pp. 115-133,  Plenum Press, New York. 

\bibitem{Abk1} Abkevich, V.I., Gutin, A.M. \&  Shakhnovich,
E. (1994). 
Specific nucleus as the transition state for protein folding: Evidence
from the lattice model.  {\em Biochemistry} {\bf 33}, 10026-10036.

\bibitem{Socci95} Socci, N.D.  \& Onuchic, J.N. (1995). 
Kinetic and thermodynamic
analysis of protein-like heteropolymers: Monte Carlo histogram technique.
{\em J. Chem. Phys. } {\bf 103}, 4732-4744. 

\bibitem{Thirum96} Thirumalai, D. \& Woodson, S.A. (1996). 
Kinetics of folding of proteins and RNA. 
{\em Acc. Chem. Res.} (to be published). 


\bibitem{Otzen} Otzen, D.E., Itzhaki, L.S. \&  Fersht, A.R.  (1994). 
Structure of the
transition state for the folding/unfolding of the barley chymotrypsin
inhibitor 2 and its implications for mechanisms of protein folding. {\em Proc.
Natl. Acad. Sci. USA} {\bf 91}, 10422-10425. 

\bibitem{Fer} Fersht, A.R. (1995). Optimization of rates of protein   
folding: The nucleation - condensation mechanism and its implications.
{\em Proc. Natl. Acad. Sci. USA}{\bf 92}, 10869-10873.

\bibitem{Rad} Radford, S.E. \&  Dobson, C.M. (1995). Insights into protein
folding using physical techniques: Studies of lysozyme and
alpha-lactalbumin. {\em Phil. Trans. Roy. Soc. Lond. B} {\bf 348}, 17-25.

\bibitem{Englander} Sosnick, T. R., Mayne, L.,  Hiller, R. \& 
 Englander, S. W. (1994).  
The barriers in protein folding. {\em Nature Struct. Biol.} {\bf 1}, 149-156. 

\bibitem{Schindler} Schindler, T., Herrler, M., Marahiel, M.A. \& 
Schmid, F.X.  (1995). Extremely rapid
folding in the absence of intermediates. {\em Nature Struct. Biol.} {\bf 2},
663-673. 

\bibitem{Kief} Kiefhaber, T.  Kinetic traps in lysozyme
folding. (1995). {\em Proc. Natl. Acad. Sci. USA} {\bf 92}, 9029-9033. 

\bibitem{Guo} Guo, Z. \&  Thirumalai, D. (1995). Kinetics of
protein folding: Nucleation mechanism, time scales, and
pathways. {\em Biopolymers} {\bf 36}, 83-103.

\bibitem{ThirumGuo} Thirumalai, D. \&  Guo, Z.  (1995). 
Nucleation mechanism for
protein folding and theoretical predictions for hydrogen-exchange labeling
experiments. {\em Biopolymers} {\bf 35}, 137-140. 


\bibitem{CamProx}  Camacho, C.J. \&  Thirumalai, D. (1995). Theoretical
predictions of folding pathways using the proximity rule with
applications to BPTI.  {\em Proc. Natl. Acad. Sci. USA }{\bf 92}, 1277-1281. 

\bibitem{Dadlez} Dadlez, M. \&  Kim, P.S.  (1995). 
A third native one-disulphide 
intermediate in the
folding of bovine pancreatic trypsin inhibitor. {\em Nature
Struct. Biol. } 
{\bf 2}, 674-679. 

\bibitem{Mirny} Mirny, L.A., Abkevich, V. \&  Shakhnovich,
E.I. (1996). Universality and diversity of the protein folding scenarios: a
comprehensive analysis with the aid of a lattice model. {\em Folding \&
Design} {\bf 1}, 103-116. 

\bibitem{Jackson1} Jackson, S.E. \&  Fersht, A.R. (1991). Folding of
chymotrypsin inhibitor 2. 1. Evidence for a two-state
transition. {\em Biochemistry} {\bf 30}, 10428-10435. 

\bibitem{Jackson2} Jackson, S.E. \&  Fersht, A.R. (1991).  
Folding of chymotrypsin inhibitor 2. 2. Influence of
proline isomerization on the folding kinetics and 
thermodynamic characterization of the transition state of folding. 
{\em Biochemistry} {\bf 30}, 10436-10443. 


\bibitem{Thirum95} Thirumalai, D. (1995). From minimal models to real   
proteins: Time scales for protein folding kinetics. {\em J. Physique (Paris)
I }{\bf 5}, 1457-1467.

\bibitem{Onuchic95} Onuchic, J.N., Wolynes, P.G., Luthey-Schulten,
Z.A. \&  Socci, N.D. (1995). 
Toward an outline of the topography of a realistic protein-folding
funnel. {\em Proc. Natl. Acad. Sci. USA} {\bf 92}, 3626-3630. 

\bibitem{Klim} Klimov, D.K. \&  Thirumalai, D. (1996). A Criterion that
determines  the foldability of proteins. {\em Phys. Rev. Lett. } 
{\bf 76}, 4070-4073. 

\bibitem{Cam93} Camacho, C.J. \&  Thirumalai, D. (1993). Kinetics
and thermodynamics of folding in model proteins. {\em Proc. Natl. Acad. Sci.
USA }{\bf 90}, 6369-6372.

\bibitem{Alex} Alexander, P., Fahnestock, S., Lee, T., Orban, J. \& 
Bryan, P. (1992). Thermodynamic analysis of the folding of the streptococcal
protein G IgG-binding domains B1 and B2: Why small proteins tend to
have high denaturation temperatures.  {\em Biopolymers }{\bf 31},  3597-3603. 

\bibitem{Leop} Leopold, P.E., Montal, M. \&  Onuchic, J.N. (1992). 
Protein folding funnels: A kinetic approach to the sequence-structure
relationship. {\em Proc. Natl. Acad. Sci. USA} {\bf 89}, 8721-8725.

\bibitem{Honig} Honig, B. \&  Cohen, F.E. (1996). Adding backbone to protein
folding: Why protein are polypeptides. {\em Folding \& Design} 
{\bf 1}, R17-R20. 

\bibitem{Rey} Rey, A. \&  Skolnick, J. (1991). A comparison of lattice Monte
Carlo dynamics and Brownian dynamics folding pathways of \(\alpha
\)-helical hairpins. {\em Chem. Phys.} {\bf 158},  199-219. 

\bibitem{Kastella} Garrett, D. G., Kastella, K. \&  Ferguson,
D. M. (1992). 
New results on protein folding from simulated
annealing. {\em J. Am. Chem. Soc. }{\bf 114}, 6555-6556. 

\bibitem{Helix} Guo, Z. \&  Thirumalai, D. (1996). 
Kinetics and thermodynamics of
folding of a {\em de novo} designed 
four-helix bundle protein. {\em J. Mol. Biol.} {\bf 263
}, 000-000 (to be published).   


\bibitem{Straub} Straub, J. E. \&  Thirumalai, D. (1996). On the approximate
incorporation of side chains in the minimal off-lattice models of
proteins (unpublished). 

\bibitem{Itzhaki} Itzhaki, L.S., Otzen, D.E. \&  Fersht, A.R. (1995). 
The structure of the
transition state for folding of chymotrypsin inhibitor 2 analyzed by
protein engineering methods: Evidence for a nucleation-condensation
mechanism for protein folding. {\em J. Mol. Biol.} {\bf 254}, 260-288. 

\bibitem{Sosnick} Sosnick, T.R., Mayne, L. \&  Englander,
S.W. (1996). Molecular
collapse: The rate limiting step in two-state cytochrome C folding. 
{\em Proteins Struct. Funct. Genet.} (to be published). 

\bibitem{Honey92} Honeycutt, J.D. \&  Thirumalai, D. (1992). 
The nature
of folded states of globular proteins. {\em Biopolymers }{\bf 32}, 695-709.

\bibitem{Pangali} Pangali, C., Rao, M. \&  Berne, B.J. (1979). 
A Monte Carlo simulation of the hydrophobic 
interaction. {\em J. Chem. Phys.} {\bf 71}, 2975-2981. 

\bibitem{Andersen} Andersen, H.C. (1983). 
Rattle: A "velocity" version of the
shake algorithm for molecular dynamics calculations. 
{\em J. Comp. Phys.} {\bf 52}, 24-34. 

\bibitem{Creightonbook} Creighton, T.E. (1993). {\em 
Proteins: Structures and
Molecular Properties}, W.H. Freeman \& Co., New York. 

\bibitem{McCammon} McCammon, J.A. \&  Harvey, S.C. (1988). 
{\em Dynamics of Proteins
and Nucleic Acids}, Cambridge University Press, Cambridge. 

\bibitem{Verlet} Swope, W.C., Andersen, H.C., Berens, P.H. \&  Wilson,
K.R. (1982). A computer simulation method for the calculation of equilibrium
constants for the formation of physical clusters of molecules:
Application to small water clusters. {\em J. Chem. Phys.}  
{\bf 76}, 637-649. 

\bibitem{Amara} Amara, P. \&  Straub, J.E. (1995). Folding model
proteins using kinetic and thermodynamic annealing of the classical   
density distribution. {\em J. Phys. Chem. }{\bf 99}, 14840-14853.

\bibitem{DeGenes} De Gennes, P.G. (1979). {\em Scaling Concept in Polymer
Physics}, Cornell University Press, New York. 

\bibitem{Klimlong} Klimov, D.K. \&  Thirumalai, D. (1996). 
Factors governing the 
foldability of proteins.  {\em Proteins Struct. Funct. Genet. } 
(to be published). 


\bibitem{Str2} Straub, J. \&  Thirumalai, D. (1993). Theoretical probes of
conformational fluctuations in S-peptide and RNase A/3'-UMP enzyme
product complex. {\em Proteins Struct. Funct. Genet.} {\bf 15},  360-373. 

\bibitem{Bryn89} Bryngelson, J.D. \&  Wolynes, P.G. (1989). 
Intermediates and barrier crossing in a random energy model (with
application to protein folding). {\em J. Phys. Chem. }{\bf 93}, 6902-6915.

\bibitem{Cam95} Camacho, C.J. \&  Thirumalai, D. (1995). Modeling
the role of disulfide bonds in protein folding: Entropic barriers and
pathways. {\em Proteins Struct. Funct. Genet.} {\bf 22}, 27-40.

\bibitem{Sali94b} Sali, A., Shakhnovich, E. \&  Karplus, M. (1994). 
Kinetics of protein folding: A lattice model study of the requirements
for folding to the native state. {\em J. Mol. Biol. }
{\bf 235}, 1614-1636.

\bibitem{Camacho96} Camacho, C.J. \&  Thirumalai, D. (1996). A
criterion that determines fast folding of proteins: A model study. 
{\em Europys. Lett.} (to be published). 

\bibitem{Shakh93} Shakhnovich, E. \&  Gutin, A.M. (1993). 
A new approach to the
design of stable proteins. {\em Protein Eng. }{\bf 6}, 793-800. 

\bibitem{Shakh94} Shakhnovich, E. (1994). 
Proteins with selected sequences fold
into unique native conformation. {\em Phys. Rev. Lett. }{\bf 72}, 3907-3910. 


\bibitem{Deutsch} Deutsch, J.M. \&  Kurosky, T. (1996). 
New algorithm for protein
design. {\em Phys. Rev. Lett. }{\bf 76}, 323-326. 

\bibitem{Garel} Garel, T., Orland, H. \&  Thirumalai, D. (1996). 
Analytical
theories of protein folding.  In: {\em New Development in Theoretical Studies
of Proteins}. (Elber, R., ed.), World Scientific, Singapore. 

\bibitem{Gold92} Goldstein, R.A., Luthey-Schulten, Z.A. \&  Wolynes, P.G. 
(1992). Optimal protein-folding codes from spin-glass theory. {\em Proc. 
Natl. Acad. Sci. USA }{\bf 89}, 4918-4922.

\bibitem{Socci96} Socci, N.D., Onuchic, J.N. \&  Wolynes,
P.G. (1996). 
Diffusive dynamics of the reaction coordinate for protein 
folding funnels. {\em J. Chem. Phys. }{\bf 104}, 5860-5871. 

\bibitem{Bryn96} Bryngelson, J.D. (1996). In: {\em Physics of
Biological 
Systems}. (Flyvberg, H., ed.), Springer Verlag, New York 
(to be published). 

\bibitem{OF}  Otzen, D.E. \&  Fersht, A.R. (unpublished). 

\bibitem{Go83} Go, N. (1983). Theoretical studies of protein folding.  
{\em Ann. Rev. Biophys. Bioeng. }{\bf 12},  183-210. 

\bibitem{Still} Stillinger, F.H. \&  Weber, T.A. (1982). 
Hidden structure in 
liquids. {\em Phys. Rev. A }{\bf 25}, 978-989. 

\bibitem{Honey90} Honeycutt, J.D. \&  Thirumalai, D. (1990). 
Metastability of the folded states of globular proteins. {\em Proc. Natl. 
Acad. Sci. USA }{\bf 87}, 3526-3529.

\bibitem{Fukugita} Fukugita, M., Lancaster, D. \&  Mitchard,
M.G. (1996). {\em 
Biopolymers }(to be published). 

\bibitem{Jones} Jones, C.M., {\em et al.}, \& Eaton, W.A. (1993). 
Fast events in protein folding initiated by nanosecond laser
photolysis. {\em Proc. Natl. Acad. Sci. USA }{\bf 90}, 11860-11864. 

\bibitem{Gruebele} Ballew, R.M., Sabelko, J., \& Gruebele, M.
(1996). Direct observation of fast protein folding: 
The initial collapse of apomyoglobin.  {\em Proc. Natl. Acad. Sci. USA
} {\bf 93}, 5759-5764. 

\bibitem{Gray} Pascher, T., Chesick, J.P.,  Winkler, J.R., \& Gray,
H.R. (1996). Protein folding triggered by electron transfer. 
{\em Science } {\bf 271}, 1558-1560. 


\end{references}
\end{document}